\documentclass[pdflatex]{article}
\usepackage{graphicx,grffile,wrapfig}
\usepackage{amsmath,amssymb}
\usepackage{braket,mathtools,cancel,bm}
\usepackage[colorlinks,linkcolor=mediumtealblue,urlcolor=blue,citecolor=Ao(English),bookmarksopenlevel=4]{hyperref}
\usepackage[usenames]{color}
\usepackage{ulem}
\usepackage[left=30truemm,right=30truemm,top=30truemm,bottom=30truemm]{geometry}
\usepackage[percent]{overpic}
\usepackage{here}
\usepackage{cite}

\newcommand{\dmeasure}[2]{\mathrm{d}^{#1} #2 \,}
\newcommand{\tr}{\mathrm{tr}\,}
\newcommand{\psibar}{\bar{\psi}}
\newcommand{\Gammaf}[1]{\Gamma\left(#1\right)}
\newcommand{\Minkowski}[1]{\mathcal{M}^{#1}}
\newcommand{\Rspace}[1]{\mathbb{R}^{#1}}
\newcommand{\Unitary}[1]{\mathrm{U} \left( #1 \right)}
\newcommand{\sphere}[1]{S^{#1}}
\newcommand{\sgn}{\mathrm{sgn}\,}
\newcommand{\Sym}[2]{\mathsf{{}^{#1}_{#2} S}}
\newcommand{\Bro}[2]{\mathsf{{}^{#1}_{#2} B}}
\newcommand{\sfbtcpg}[2]{\textcolor{my-purple}{\bf #1}/\textcolor{my-green}{\bf #2}}

\renewcommand{\Re}{\mathrm{Re}\,}

\graphicspath{{graph/}}
\definecolor{blue}{rgb}{0.0, 0.0, 1.0}
\definecolor{Ao(English)}{rgb}{0.0, 0.5, 0.0}
\definecolor{mediumtealblue}{rgb}{0.0, 0.33, 0.71}
\definecolor{my-red}{RGB}{148,0,27}
\definecolor{my-yellow}{RGB}{237,185,,24}
\definecolor{my-green}{RGB}{0,100,0}
\definecolor{my-purple}{RGB}{128,0,128}

\title{Precise Phase Structure in Four-fermion Interaction Model on Torus}

\author{
Tomohiro Inagaki${}^{1, 2, 3}$
$\footnote{\href{mailto:inagaki@hiroshima-u.ac.jp}{inagaki@hiroshima-u.ac.jp}}$,
Yamato Matsuo${}^{4}$
$\footnote{\href{mailto:ya-matsuo@hiroshima-u.ac.jp}{ya-matsuo@hiroshima-u.ac.jp}}$,
and Hiromu Shimoji${}^{4}$
$\footnote{\href{mailto:h-shimoji@hiroshima-u.ac.jp}{h-shimoji@hiroshima-u.ac.jp}}$}

\date{\textit{
${}^1$Information Media Center, Hiroshima University, Higashi-Hiroshima, 739-8521, Japan,\\
${}^2$Core of Research for the Energetic Universe, Hiroshima University, Higashi-Hiroshima, 739-8526, Japan,\\
${}^3$Lab. Theor. Cosmology, Tomsk State University of Control Systems and Radioelectronics (TUSUR), 634050 Tomsk, Russia,\\
${}^4$Graduate School of Science, Hiroshima University, Higashi-Hiroshima, 739-8526, Japan}}

\begin{document}

\maketitle

\begin{abstract}
 We investigate finite-size effects on chiral symmetry breaking in a four-fermion interaction model at a finite temperature and a chemical potential.
 Applying the imaginary time formalism, the thermal quantum field theory is constructed on an $\sphere{1}$ in the imaginary time direction.
 In this paper, the finite-size effect is introduced by a compact $\sphere{1}$ spatial direction with a $\Unitary{1}$-valued boundary condition.
 Thus, we study the model on a $\Rspace{D-2} \times \sphere{1} \times \sphere{1}$ torus.
 Phase diagrams are obtained by evaluating the local minima of the effective potential in the leading order of the $1/N$ expansion.
 From the grand potential, we calculate the particle number density and the pressure, then we illustrate the correspondence with the phase structure.
 We obtain a stable size for which the sign of the pressure flips from negative to positive as the size decreases.
 Furthermore, the finite chemical potential expands the parameter range that the stable size exists.
\end{abstract}

\section{Introduction}
Spontaneous symmetry breaking is a key concept in modern physics.
In the early years, Y.~Nambu and G.~Jona Lasinio made a significant contribution through a four-fermion interaction model with chiral symmetry, now called the Nambu--Jona Lasinio model \cite{Nambu:1961tp, Nambu:1961fr}.
Another four-fermion interaction model is the Gross--Neveu model that is constructed in two dimensions and invariant under a discrete chiral transformation \cite{Gross:1974jv}.
These models have often been used as tools to study chiral symmetry breaking in quantum chromodynamics (QCD) \cite{Rosenstein:1990nm, Klevansky:1992qe, Hatsuda:1994pi, Inagaki:1997kz, Miransky:2015ava}.

It is well known that the chiral phase structure of a four-fermion interaction model is determined by its conditions, for example, a finite temperature and a chemical potential.
The structure is also affected by the size of the system and boundary conditions (topology of the system).
Boundary conditions are often imposed to be periodic and antiperiodic \cite{Kim:1987db, Ishikawa:1998uu, Gamayun:2004wv, Ebert:2010eq, Abreu:2017lxf, Ishikawa:2018yey, Ishikawa:2019dcn}.
In the Matsubara formalism, the finite temperature and chemical potential can be regarded as the size and boundary conditions for the imaginary time direction.
It is also pointed out that inhomogeneous phases can be energetically favored in a finite size system in Ref \cite{Flachi:2012pf}.

There are some works that have previously studied an $\Unitary{1}$-valued boundary condition \cite{Higuchi:1988qa, HUANG1994644, Song:1993da, Flachi:2013iia, Flachi:2013zaa, Yoshii:2014fwa, Inagaki:2019kbc}.
In an Abelian gauge theory, such a boundary condition is introduced as the Aharonov--Bohm effect \cite{Aharonov:1959fk}.
The effect is observed in a superconducting ring with an Aharonov--Bohm magnetic flux.
Another finite-size phenomena that occurs in quantum systems, is observed as the Casimir effect \cite{Casimir:1948dh, Milton:1978sf, Milton:1979yx, BREVIK198236, BREVIK1982179, PLUNIEN198687}, which is caused by a non-trivial quantum vacuum.
The Casimir effect has been studied in a four-fermion interaction model \cite{Flachi:2013bc,Flachi:2017cdo} and in the context of lattice simulations \cite{Chernodub:2018aix, Chernodub:2019nct, Ishikawa:2020ezm, Ishikawa:2020icy}.

In our previous work \cite{Inagaki:2019kbc}, the finite-size effect, which comes from the size of the system and $\Unitary{1}$-valued boundary condition, is discussed in a four-fermion interaction model on $\Rspace{D-1} \times \sphere{1}$ without considering any thermal effects.
It is assumed that the fermion field acquires a phase, $e^{i\pi\delta}$, after going around the spatially compact $\sphere{1}$ direction.
The ground state evaluated under a constant expectation value breaks the chiral symmetry for a strong coupling at the large $\sphere{1}$ size limit\footnote{
Because the continuous chiral symmetry cannot be broken in two dimensions, the Gross--Neveu model is employed, and the discrete chiral symmetry is evaluated in two dimensions.}.
In the periodic ($\delta=0.0$) boundary condition, as the size decreases, chiral symmetry breaking tends to be enhanced, and at the small size limit the symmetry is always broken for any finite coupling constant.
On the other hand, the antiperiodic ($\delta=1.0$) boundary condition has opposite effects on the chiral symmetry.
The chiral symmetry breaking tends to be suppressed as the size decreases, and is restored for an arbitrary coupling constant at the small size limit.
The system with the antiperiodic boundary condition is regarded to mimic the finite temperature system of the fermion field.
Sign-flip boundaries of the string tension, which is given by the derivative of the effective potential with respect to the size, are obtained in $0.4 < \delta \leq 0.5$ for any finite size.
In particular, we have found a stable size that appears for a limited phase region, for which the string tension is zero and becomes negative (positive) for the larger (smaller) size.
It seems difficult to realize the limited phase region, but there is a possibility to relax the condition for the stable size at a finite temperature and a chemical potential.
Thus, we launched a plan to study the finite-size effect for a four-fermion interaction model at a finite temperature and chemical potential.

This paper is organized as follows.
In Sec. \ref{sec:ffi-model} we begin with a brief review of a four-fermion interaction model on a $D$-dimensional Minkowski spacetime at a finite temperature with a compact direction.
In the imaginary time formalism, the effective potential for the model is identical with one on a Euclidean torus, $\Rspace{D-2} \times \sphere{1} \times \sphere{1}$.
We obtain the explicit expression of the effective potential in the leading order of the $1/N$ expansion.
In Sec. \ref{sec:phase-structure} we show the phase bifurcation diagrams on $\mu$-$T$ and $L$-$T$ planes and the behavior of the dynamically generated fermion mass.
In Sec. \ref{sec:therm-quant} we calculate the grand potential, then we evaluate the particle number density and the pressure.
The chiral phase transition and the stability of the system are also discussed.
Finally, we summarize our results in Sec. \ref{sec:conclusions}.

\section{Four-fermion interaction model on torus}\label{sec:ffi-model}
A Dirac fermion $\psi$ in even dimensions is composed of two components, the left- and right-handed fermions $\psi_L$ and $ \psi_R$, that are eigenstates of the chirality operator $\gamma^5$: $\gamma^5 \psi_L = - \psi_L$ and $\gamma^5 \psi_R = \psi_R$.
The chiral symmetry is defined as the invariance of a theory under the chiral transformation $\psi \to e^{i\gamma^5 \theta} \psi$.
Explicit mass terms are prohibited by the symmetry.
The simplest interaction between fermions and anti-fermions that preserves the chiral symmetry is the four-fermion interaction.

To investigate the phase structure that emerges by spontaneous chiral symmetry breaking, we employ a four-fermion interaction model.
The action on a $D$-dimensional Minkowski spacetime, $\Minkowski{D} $($2 \leq D < 4$), is written as
\begin{align}
 S = \int \hspace{-0.3em} \dmeasure{D}{x} \hspace{-0.3em}\left[ \sum_{a=1}^N \psibar_a(x) i \gamma^\mu \partial_\mu \psi_a(x) + \frac{\lambda_0}{2 N} \left( \left( \sum_{a=1}^N \psibar_a(x) \psi_a(x)\right)^2 \hspace{-0.5em} + \left( \sum_{a=1}^N \psibar_a(x) i \gamma^5\psi_a(x)\right)^2 \right)\right],
 \label{eq:action:original}
\end{align}
where the index, $a$, denotes a species of the fermion field, $\psi(x)$, $N$ is the number of species and $\lambda_0$ represents a bare coupling constant of the four-fermion interactions.
For simplicity, we omit the index, $a$, and the summation symbol in all subsequent equations.

By introducing the auxiliary fields $\sigma(x)$ and $\pi(x)$, the action can be rewritten as
\begin{align}
 S = \int \dmeasure{D}{x} \left[ \psibar\left(x\right) \left( i \gamma^\mu \partial_\mu - \sigma\left(x\right) - i \gamma^5 \pi\left(x\right) \right) \psi\left(x\right) - \frac{N}{2 \lambda_0} \left( \sigma\left(x\right)^2 + \pi\left(x\right)^2 \right) \right].
 \label{eq:action:auxiliary}
\end{align}
The original action \eqref{eq:action:original} is reproduced by substituting the solutions of the equations of motion: $\sigma(x) = -N^{-1} \lambda_0 \psibar(x)\psi(x)$ and $\pi(x) = -N^{-1} \lambda_0 \psibar(x)i\gamma^5\psi(x)$.
When $\sigma(x)$, or $\psibar(x)\psi(x)$, develops a non-vanishing vacuum expectation value, the fermion fields dynamically acquire mass and the chiral symmetry is spontaneously broken.

In the following, we assume that the expectation values are homogeneous and set $\sigma(x) = \sigma$ and $\pi(x) = \pi$.
Performing the path integrals of the fermion field, we obtain the effective potential in the leading order of the $1/N$ expansion \cite{Inagaki:1997kz, Inagaki:2019kbc},
\begin{align}
 V_D(\sigma, \pi) = \frac{\sigma^2 + \pi^2}{2\lambda_0} - \int \frac{\dmeasure{D}{k}}{i(2\pi)^D} \tr \ln \frac{\gamma^\mu k_\mu - \sigma - i\gamma^5\pi}{-M},
 \label{eq:eff-pot:two-auxiliary}
\end{align}
where $M$ is an arbitrary mass scale and the trace, $\tr$, denotes the sum over the spinor indices.
By the chiral transformation we can set $\pi = 0$ without loss of generality.

The expectation value of $\sigma$ is regarded as an order parameter of the chiral symmetry breaking and the value is determined by observing the minimum of the effective potential. Thus, the dynamically generated fermion mass, $m_0$, is found as a solution of the gap equation,
\begin{align}
 \left. \frac{\partial V_D \left( \sigma \right)}{\partial \sigma} \right|_{\sigma = m_0}
 = 0.
 \label{eq:gap-equation:RD}
\end{align}
The effective potential \eqref{eq:eff-pot:two-auxiliary} is divergent because of the infinite momentum integral.
Without changing the solution of the gap equation, the infinite zero-point energy, $V_D(\sigma = 0)$, is removed from the potential $V_D(\sigma) \to V_D(\sigma) - V_D(\sigma = 0)$. The divergence induced by the interaction is subtracted by introducing the renormalized four-fermion coupling, $\lambda_r$,
\begin{align}
 \left. \frac{\partial^2 V_D \left( \sigma\right)}{\partial \sigma^2} \right|_{\sigma = \mu_r}
 = \frac{\mu_r^{D-2}}{\lambda_r},
 \label{eq:renormalization-proc}
\end{align}
where $\mu_r$ is the renormalization scale.

On $\Minkowski{D}$ the gap equation \eqref{eq:gap-equation:RD} has a non-trivial solution if the renormalized coupling, $\lambda_r$, is larger than the critical one, $\lambda_c$,
\begin{align}
 \lambda_c
 = \frac{ \left(4\pi\right)^{\frac{D}{2}}}{\tr I \cdot \left(1-D\right)\Gammaf{1-\frac{D}{2}}}.
 \label{eq:critical-coupling}
\end{align}
The mass, $m_0$, is derived as a function of the dimension, $D$, the renormalized coupling, $\lambda_r$ and the renormalization scale, $\mu_r$,
\begin{align}
 m_0 = \left[\frac{(4\pi)^{D/2}}{\tr I\Gamma(1-D/2)}\left(\frac{1}{\lambda_r}-\frac{1}{\lambda_c}\right)\right]^{1/(D-2)}\mu_r
 \label{eq:mass:m0}
\end{align}
In the present paper we focus on the strong coupling theory, $\lambda_r > \lambda_c$.

We introduce finite-temperature and finite-size effects to the action \eqref{eq:action:original}.
According to the Matsubara formalism we assign the antiperiodic boundary condition to the imaginary-time direction,
\begin{align}
 \psi(x^1, \dots, x^{D-1}, x^D + \beta)
 = - \psi(x^1, \dots, x^{D-1}, x^D), \label{eq:boundary-cond:temp}
\end{align}
where $\beta$ denotes the inverse temperature.
A finite particle number density and a chemical potential, $\mu$, are introduced by adding a term, $ - \mu \int \dmeasure{D}{x} \psi(x)^\dagger \psi(x)$, to the action \eqref{eq:action:original}.
We compactify the spacial direction, $x^{D-1}$, by imposing a $\Unitary{1}$-valued boundary condition,
\begin{align}
 \psi(x^1, \dots, x^{D-1} + L, x^D)
 = e^{- i \pi \delta} \psi(x^1, \dots, x^{D-1}, x^D),
 \label{eq:boundary-cond:space}
\end{align}
where $L$ is the length of the compactified space and $\delta$ is a $\Unitary{1}$ phase from the boundary condition.
Therefore, we assign boundary conditions for the two directions, $x^{D-1}$ and $x^{D}$, and study the model on a torus like topology, $\Rspace{D-2}\times S^1 \times S^1$.

Under the boundary conditions \eqref{eq:boundary-cond:temp} and \eqref{eq:boundary-cond:space} the momentum takes the discrete values,
\begin{align}
 k_{\delta, n}
 = \frac{2\pi}{L} \left( n + \frac{\delta}{2}\right),
 \qquad
 \omega_{\mu, n^\prime}
 = \frac{2\pi}{\beta} \left( n^\prime + \frac{1}{2} \right) - i \mu,
\end{align}
where $k_{\delta, n}$ denotes a discrete momentum for the compactified space, and $\omega_{\mu, n^\prime}$ a Matsubara frequency with chemical potential, $\mu$.
Thus, the effective potential \eqref{eq:eff-pot:two-auxiliary} is rewritten as
\begin{align}
 V_{D} \left(\sigma; L, \delta, \beta, \mu \right)
 = \frac{\sigma^2}{2\lambda_0}
 - \frac{\tr I}{2\beta L} \sum_{n, n^\prime = -\infty}^{\infty} \int\frac{\dmeasure{D-2}{\bm{k}}}{(2\pi)^{D-2}} \ln \frac{\bm{k}^2 + k_{\delta, n}^{2} + \omega_{\mu, n^\prime}^2 + \sigma^2}{M^2}.
 \label{eq:eff-pot:main}
\end{align}
The momentum integrals are replaced by summations over the integers $n$ and $n^\prime$.

Because the ultraviolet divergences are not modified by the compactification, a finite expression is obtained by subtracting the divergent zero-point energy, $V_{D} (\sigma=0; L=\infty, \delta=0, \beta=\infty, \mu=0 )$, and substituting the renormalized coupling, $\lambda_r$, defined by \eqref{eq:renormalization-proc} on $\Minkowski{D}$.
By applying the zeta function regularization to calculate the Matsubara frequency summation \cite{Inagaki:2019kbc}, we obtain
\begin{align}
 \frac{V_{D} \left(\sigma; L, \delta, \beta, \mu \right)}{\mu_r^D}
 =& \frac{1}{2}\left(\frac{1}{\lambda_r}-\frac{1}{\lambda_c}\right)\left(\frac{\sigma}{\mu_r}\right)^2
 - \frac{\tr I \cdot \Gamma\left(1-\frac{D}{2}\right)}{(4\pi)^{\frac{D}{2}}D}
 \left( \frac{\sigma}{\mu_r}\right)^{2 \cdot \frac{D}{2}} \notag\\
 & \hspace{-2em}- \dfrac{\tr I }{C(D+1) L\mu_r} \int_0^\infty \frac{\dmeasure{}{q}}{\mu_r} \left(\frac{q}{\mu_r}\right)^{D-2}\hspace{-0.5em} \ln \left(2 \frac{ \cosh{ L\sqrt{q^2+\sigma^2}} - \cos\pi\delta }{\exp{L\sqrt{q^2+\sigma^2}}} \right) \notag\\
 & \hspace{-5em} -  \dfrac{\tr I }{C(D) \beta L \mu_r^2} \sum_{n = -\infty}^\infty \int_0^\infty \frac{\dmeasure{}{q}}{\mu_r} \left(\frac{q}{\mu_r}\right)^{D-3} \hspace{-0.5em} \ln\left( 2\frac{\cosh{\beta \sqrt{q^2 + k^2_{\delta, n} + \sigma^2}} + \cosh \beta \mu}{\exp{\beta \sqrt{q^2 + k^2_{\delta, n} + \sigma^2}} }\right).
 \label{eq:eff-pot:MatsubaraFrequencySummation}
\end{align}
where we set $C(D)=(4\pi)^{\frac{D-2}{2}}\Gammaf{\frac{D-2}{2}}$.
The renormalized coupling $\lambda_r$ and the renormalization scale $\mu_r$ are eliminated from Eq. \eqref{eq:eff-pot:MatsubaraFrequencySummation} by using the dynamically generated fermion mass $m_0$ \eqref{eq:mass:m0} on $\Minkowski{D}$.
Therefore, equation \eqref{eq:eff-pot:MatsubaraFrequencySummation} is simplified to
\begin{align}
 \frac{V_{D} \left(\sigma; L, \delta, \beta, \mu \right)}{m^D_0}
 =& \frac{\tr I \cdot \Gamma\left(1-\frac{D}{2}\right)}{(4\pi)^{\frac{D}{2}}}
 \left[ \frac{1}{2}\left(\frac{\sigma}{m_0}\right)^2 -\frac{1}{D}\left( \frac{\sigma}{m_0}\right)^{2 \cdot \frac{D}{2}} \right] \notag\\
 & \hspace{-2em}- \dfrac{\tr I }{C(D+1) Lm_0} \int_0^\infty \frac{\dmeasure{}{q}}{m_0} \left(\frac{q}{m_0}\right)^{D-2}\hspace{-0.5em} \ln \left(2 \frac{ \cosh{ L\sqrt{q^2+\sigma^2}} - \cos\pi\delta }{\exp{L\sqrt{q^2+\sigma^2}}} \right) \notag\\
 & \hspace{-6em} -  \dfrac{\tr I }{C(D) \beta L m_0^2} \sum_{n = -\infty}^\infty \int_0^\infty \frac{\dmeasure{}{q}}{m_0} \left(\frac{q}{m_0}\right)^{D-3} \hspace{-0.5em} \ln\left( 2\frac{\cosh{\beta \sqrt{q^2 + k^2_{\delta, n} + \sigma^2}} + \cosh \beta \mu}{\exp{\beta \sqrt{q^2 + k^2_{\delta, n} + \sigma^2}} }\right).
 \label{eq:eff-pot:evaluable}
\end{align}
The first term is equivalent to the effective potential on $\Minkowski{D}$ and the others show contributions from finite size, a finite temperature, and a finite chemical potential.
As is known, the four-fermion interaction model is non-renormalizable in four dimensions, because the first term diverges at $D=4$.

\begin{figure}[H]
 \begin{center}
  \begin{tabular}{cc}
   \begin{minipage}{0.47\hsize}
    \begin{center}
     \begin{tabular}{cc}
      \begin{minipage}{0.48\hsize}
       \begin{overpic}[width=1\hsize,clip]
        {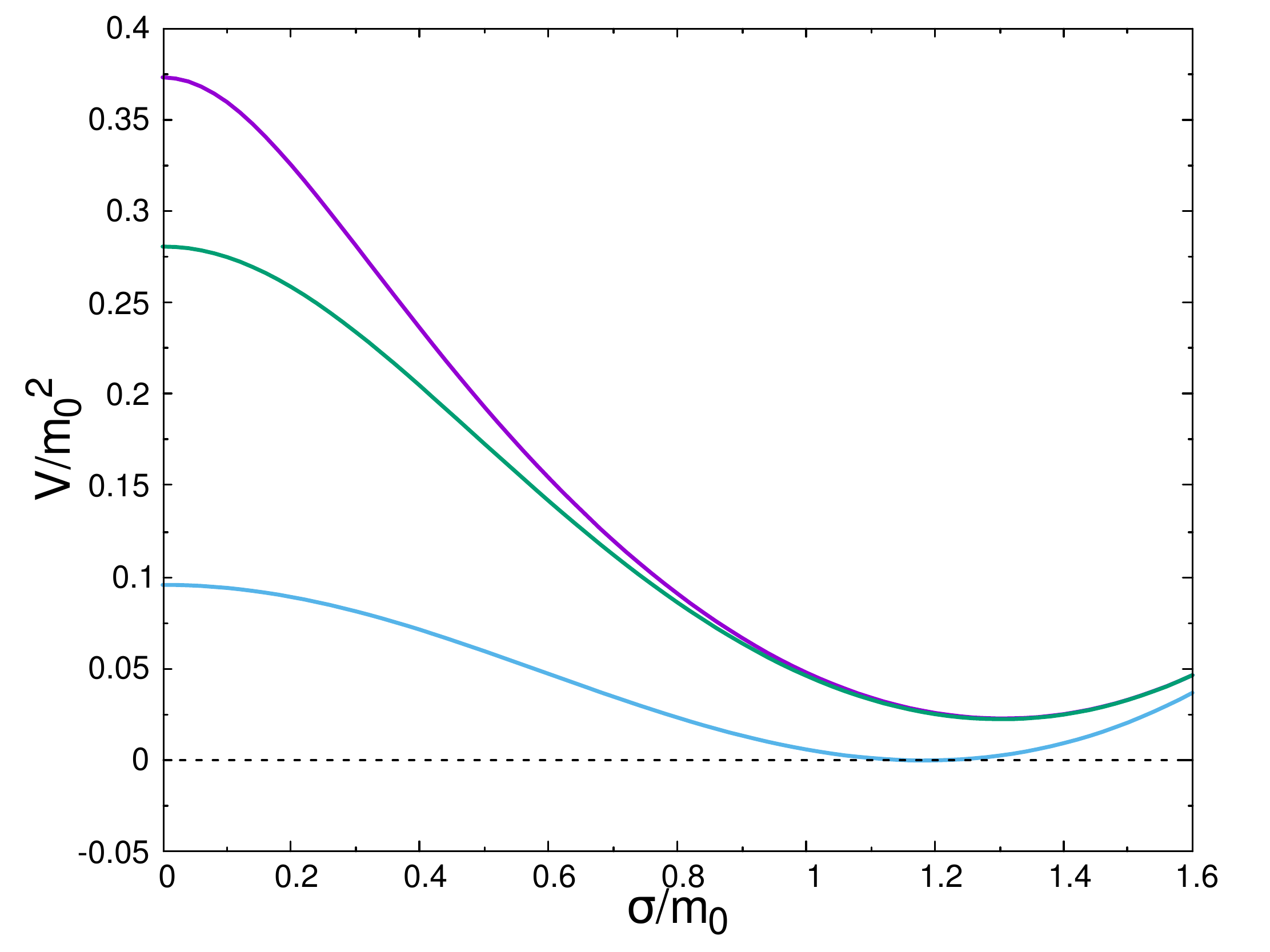}
        \put(50,66){\scriptsize $\mu/m_0=0.0$}
       \end{overpic}
      \end{minipage}
      &
      \begin{minipage}{0.48\hsize}
       \begin{overpic}[width=1\hsize,clip]
        {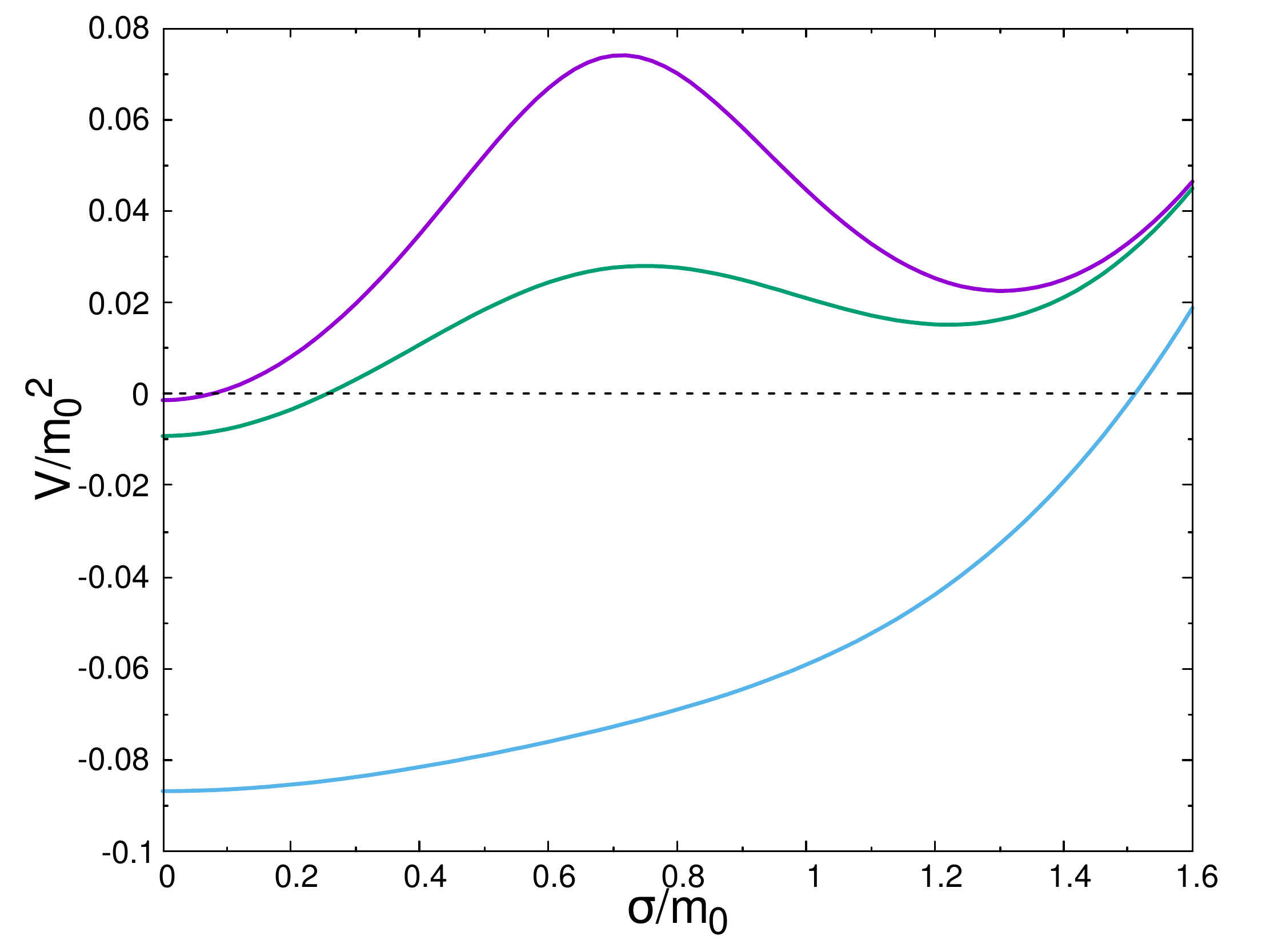}
        \put(76,66){\scriptsize $0.7$}
       \end{overpic}
      \end{minipage}
     \end{tabular}
     {\footnotesize $D=2.0$}
    \end{center}
   \end{minipage}
   &
   \begin{minipage}{0.47\hsize}
    \begin{center}
     \begin{tabular}{cc}
      \begin{minipage}{0.48\hsize}
       \begin{overpic}[width=1\hsize,clip]
        {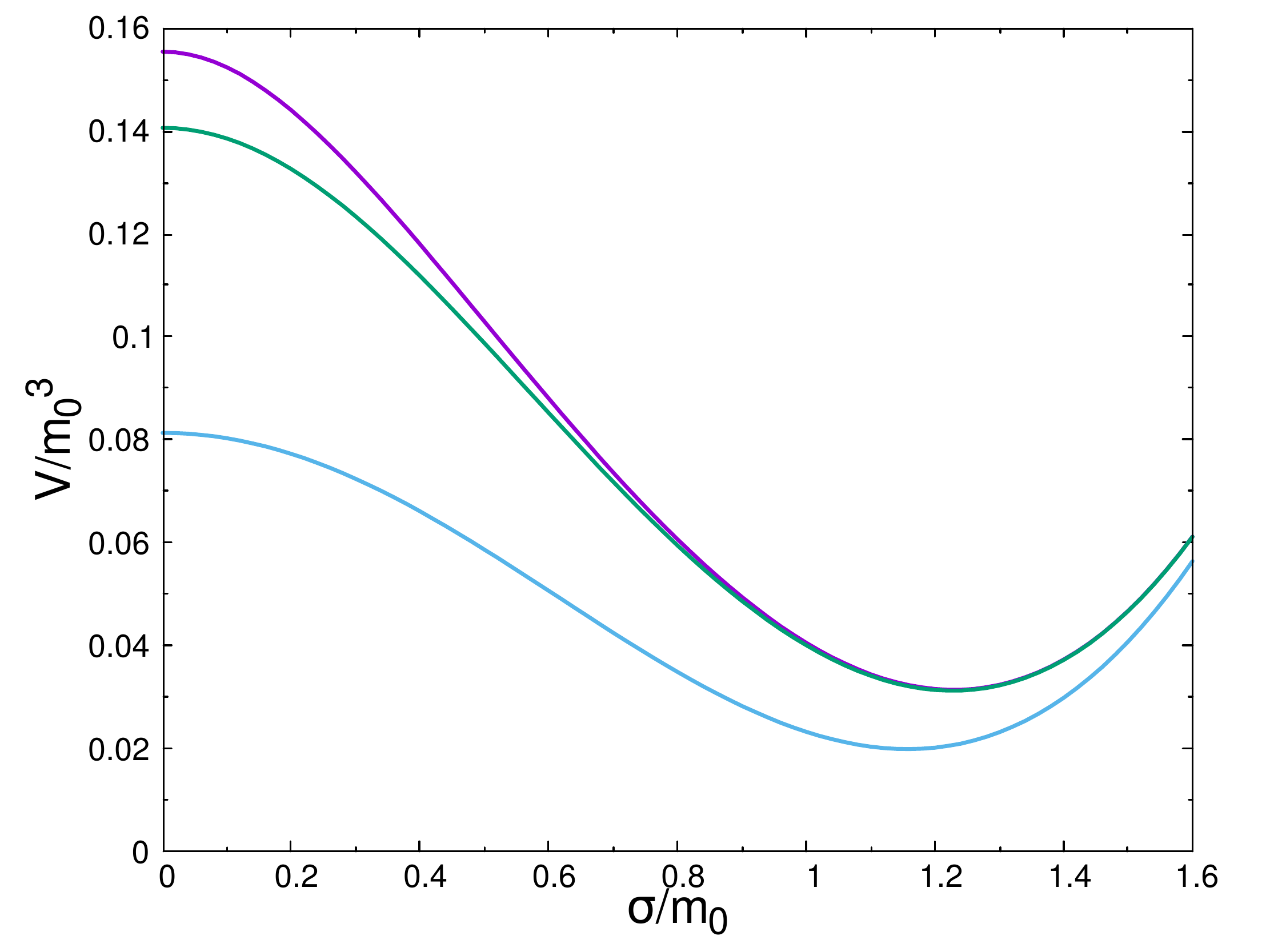}
        \put(50,66){\scriptsize $\mu/m_0=0.0$}
       \end{overpic}
      \end{minipage}
      &
      \begin{minipage}{0.48\hsize}
       \begin{overpic}[width=1\hsize,clip]
        {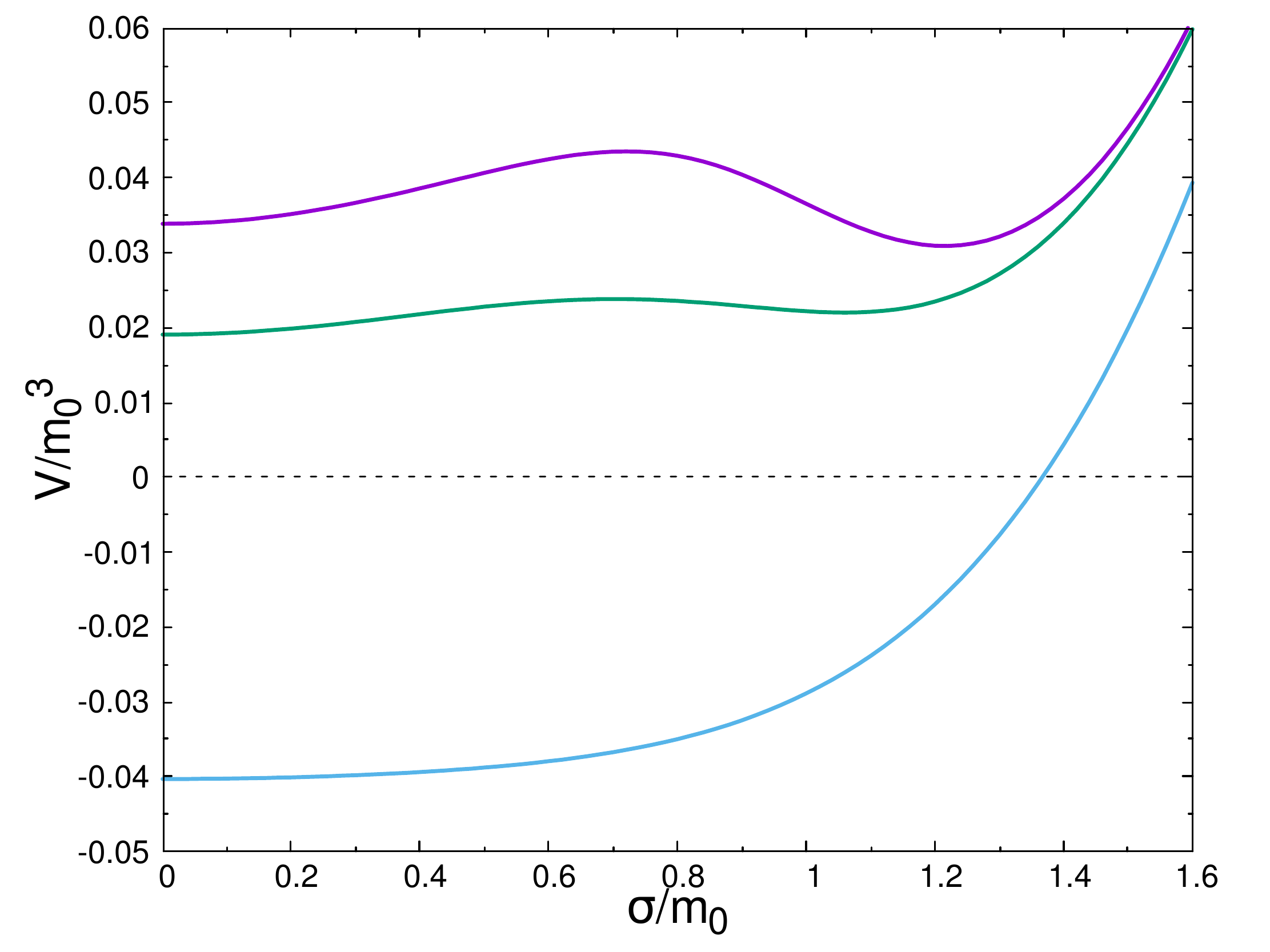}
        \put(76,66){\scriptsize $0.9$}
       \end{overpic}
      \end{minipage}
     \end{tabular}
     {\footnotesize $D=3.0$}
    \end{center}
   \end{minipage}
  \end{tabular}
  \vspace{-1em}
 \end{center}
 \caption{Behavior of the effective potential (purple: $T/m_0=0.1$, green: $T/m_0=0.2$ and light blue: $T/m_0=0.4$)
 for $Lm_0=1.5$ and $\delta=0.0$.}
 \label{fig:typ-eff-pot}
\end{figure}

In the following, we numerically evaluate the effective potential in the two- and three-dimensional instances.
The trace in the spinor space is taken as $\tr I = 2^{D/2}$.
Typical behavior of the effective potential is shown in Fig. \ref{fig:typ-eff-pot}, where we do not normalize the effective potential as $V_D(\sigma=0;L,\delta,\beta,\mu)=0$, because the $L, \delta, \beta, \mu$ dependence of $V_D(\sigma;L,\delta,\beta,\mu)$ has a decisive contribution to the thermodynamic properties discussed in Sec. \ref{sec:therm-quant}.
It is observed that the chiral symmetry tends to be restored for a high temperature and chemical potential;
in particular, a first-order phase transition takes place via the effect of the chemical potential.
Because the third term in \eqref{eq:eff-pot:evaluable} behaves similar to a step function, $\theta\left( \left|\mu\right| - \sqrt{q^2 + k_{\delta, n}^2 + \sigma^2} \right)$, at $T=0$, the chemical potential has no significant contribution for $\mu \lesssim \left|\sigma\right|$.

\section{Phase structure}\label{sec:phase-structure}
An order parameter of chiral symmetry breaking is the expectation value of the composite operator, $\Braket{\bar{\psi}\psi}$.
It is proportional to the auxiliary field, $\sigma$, in the ground state.
The state is obtained by evaluating the minimum of the effective potential, $V_{D} (m; L, \delta, \beta, \mu ) \leq V_{D} (\sigma; L, \delta, \beta, \mu )$ for any $\sigma$.
The value at the minimum, $\sigma = m$, corresponds to a dynamically generated fermion mass.
The local minima of the effective potential indicate the existence of metastable states.
Not only the ground state but also the metastable states have some phenomenological consequences.
In this paper, we assume a spatially constant ground state and evaluate the number of extrema.
In that sense, our results do not directly indicate the presence of the inhomogeneous phase, but the ground state can become inhomogeneous in the presence of multiple minima in certain region of the parameter space.

\subsection{Phase diagrams}
\begin{figure}[H]
 \begin{center}
  \begin{tabular}{c}
   \begin{minipage}{1\hsize}
    \begin{center}
     \begin{overpic}[width=0.9\hsize,clip]
      {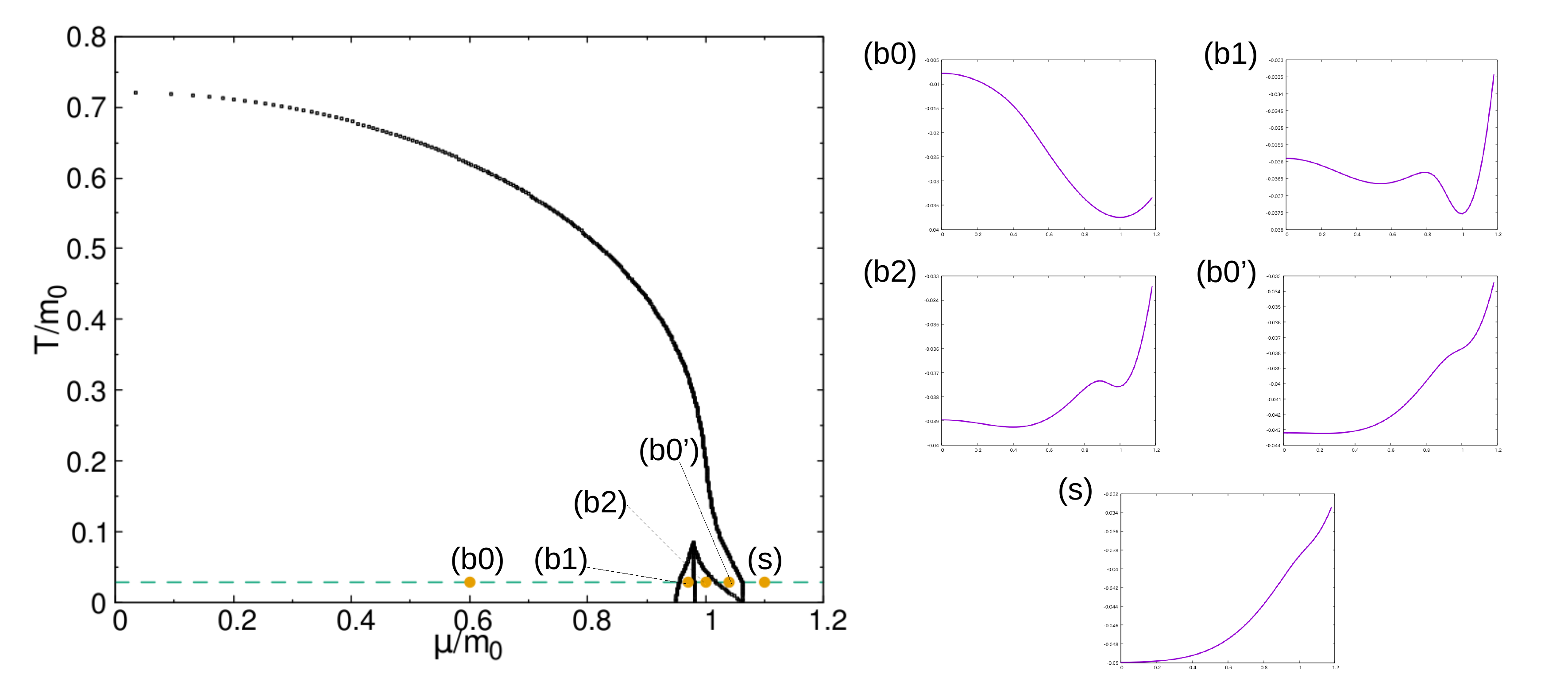}
      \put(70,37){$\Bro{2}{2}$}
      \put(90,37){$\Bro{4}{4}$}
      \put(70,23){$\Bro{4}{2}$}
      \put(90,23){$\Bro{2}{2}$}
      \put(80,9){$\Sym{1}{1}$}
     \end{overpic}
     \vspace{-2em}
    \end{center}
   \end{minipage}
  \end{tabular}
 \end{center}
 \caption{Phase diagram on the $\mu$-$T$ plane ($D =3.0$, $Lm_0 = 8.0$ and $\delta=1.0$) and behavior of the effective potential along the dashed line ($T/m_0=0.03$).}
 \label{fig:behavior-eff-pot-phase-diagram}
\end{figure}
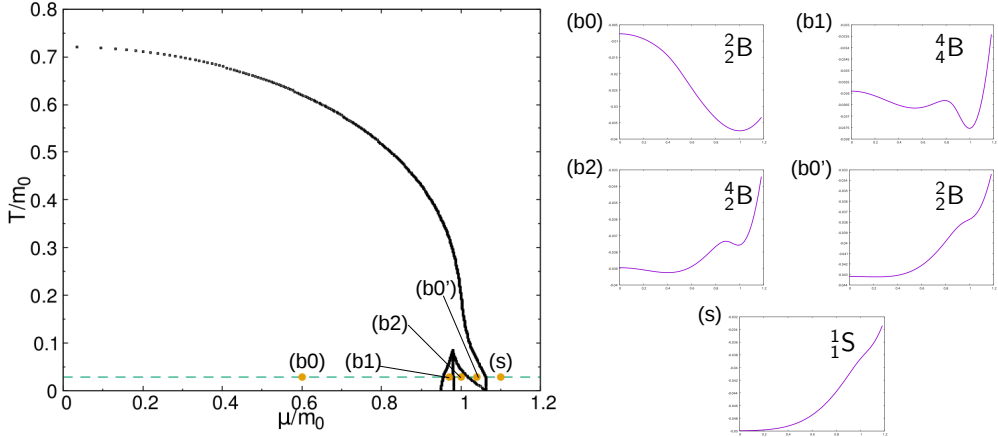

To find the precise phase structure of the four-fermion interaction model, we evaluate the local minima of the effective potential and plot bifurcation diagrams.
Because of the chiral symmetry, the potential is an even function of $\sigma$.
The states are classified by the number of extrema and the position of the minimum of the effective potential for $\sigma \geq 0$.
The class of states is described by introducing two symbols, $\Sym{a}{b}$ and $\Bro{a}{b}$.
The former and the latter represent a symmetric phase and a broken phase, respectively.
The upper index $\mathsf{a}$ is the number of extrema for $\sigma \geq 0$ and the lower index $\mathsf{b}$ is the number of the extrema from the origin to the minimum, i.e. the $\mathsf{b}$th extremum is the minimum and $\mathsf{b} \leq \mathsf{a}$.
As an example, the phase diagram on a $\mu$-$T$ plane is divided into four regions for $D =3.0$, $Lm_0 = 8.0$ and $\delta=1.0$.
The correspondences between the symbol and behavior of the effective potential are shown in Fig. \ref{fig:behavior-eff-pot-phase-diagram}.
The symmetric phase $\Sym{1}{1}$ is realized for a high temperature and chemical potential, outside the outer boundary.
At the outer boundary, a second-order phase transition takes place.
The broken phase contains three regions identified by the number of the extrema and the position of the minimum.
A discontinuous change of the dynamical mass is observed on the boundary between (b1), $\Bro{4}{4}$, and (b2), $\Bro{4}{2}$.

\begin{figure}[H]
 \begin{center}
  \begin{tabular}{cc}
   \begin{minipage}{0.48\hsize}
    \begin{center}
     \begin{overpic}[width=0.96\hsize,height=13.5em]
      {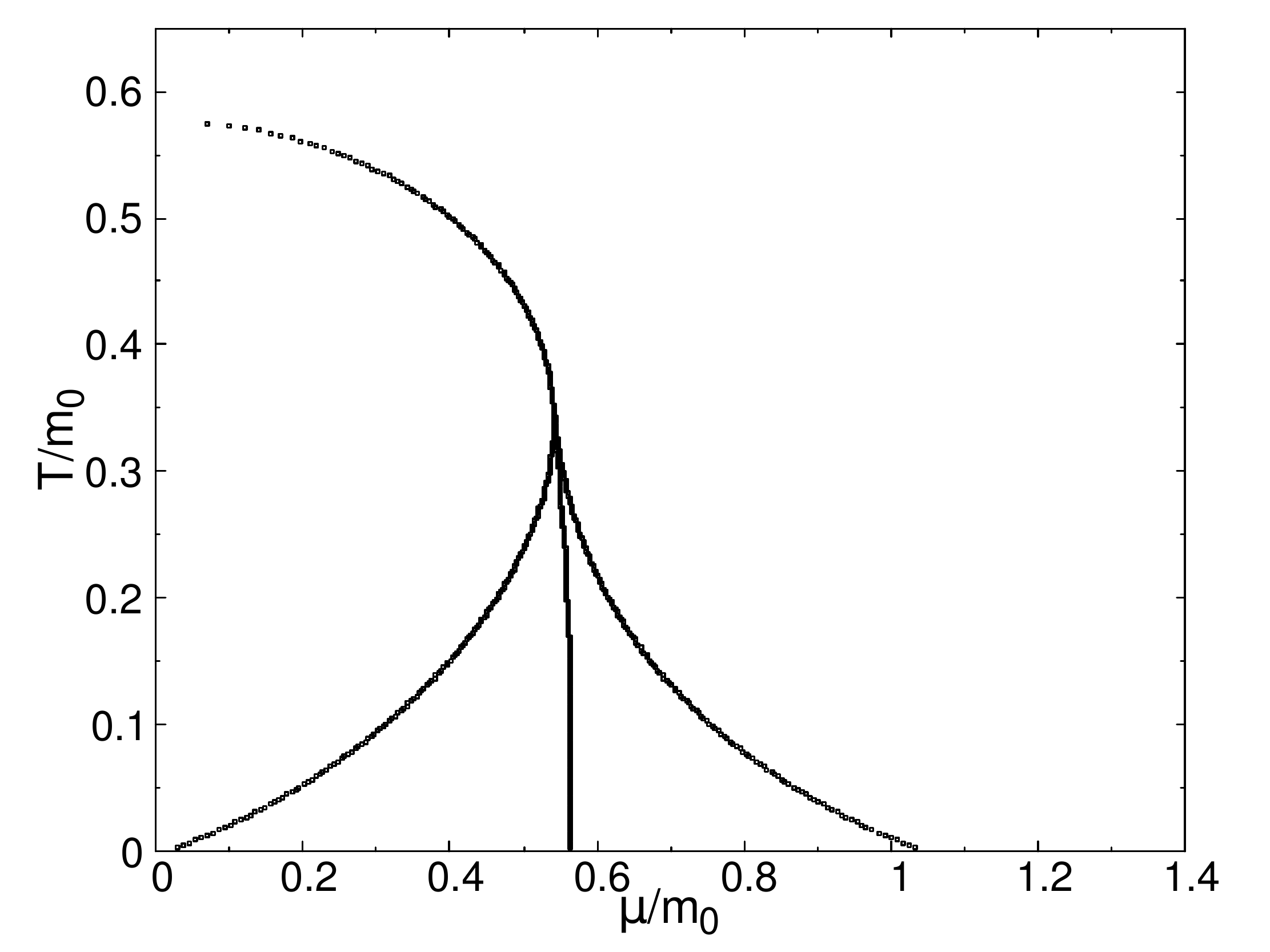}
      \put(20,40){$\Bro{2}{2}$}
      \put(36,10){$\Bro{3}{3}$}
      \put(70,40){$\Sym{1}{1}$}
      \put(50,10){$\Sym{3}{1}$}
     \end{overpic}
    \end{center}
   \end{minipage}
   &
       \begin{minipage}{0.48\hsize}
        \begin{center}
         \begin{overpic}[width=0.96\hsize,height=13.5em]
          {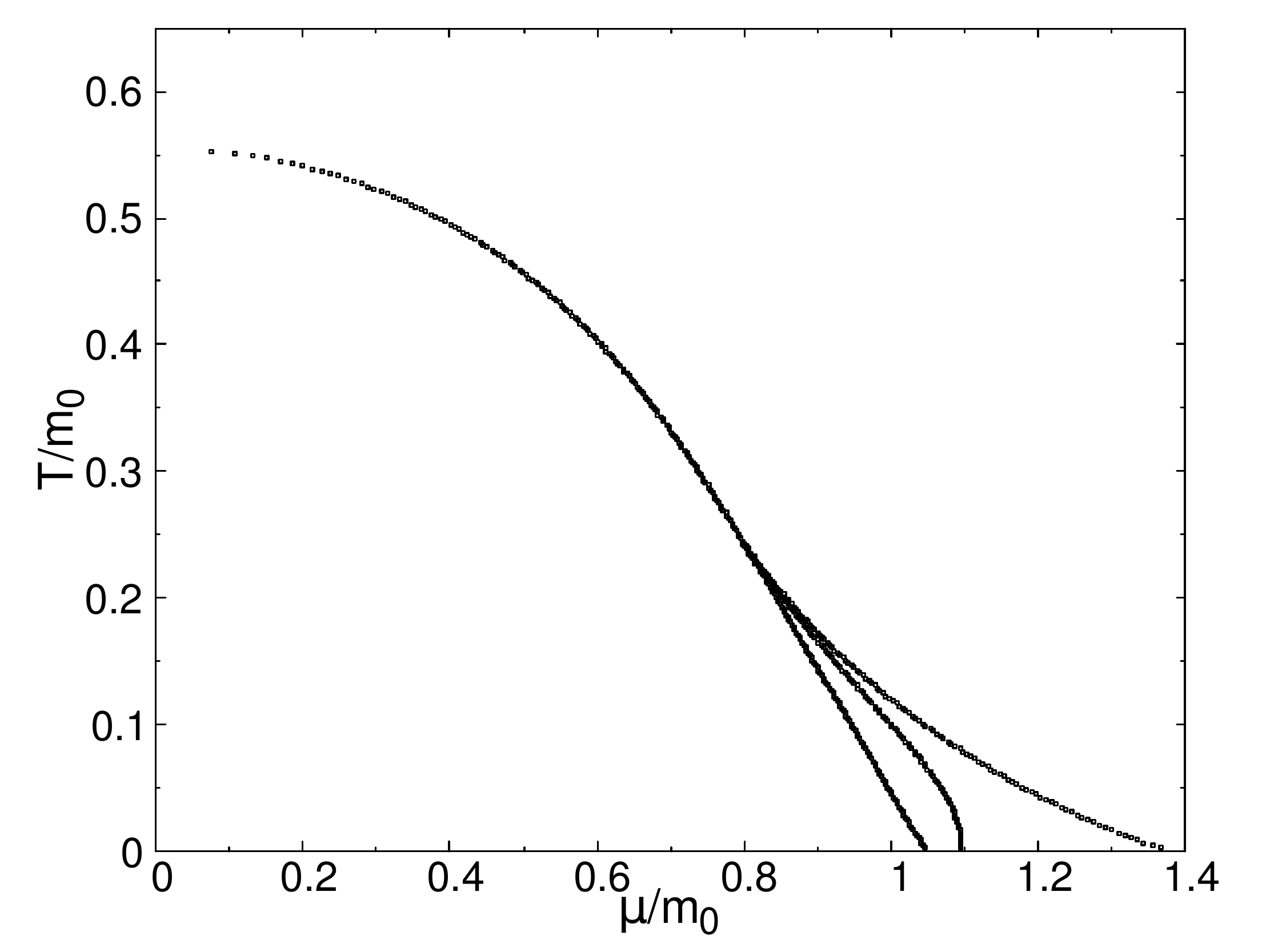}
          \put(20,40){$\Bro{2}{2}$}
          \put(56,10){$\Bro{3}{3}$ \line(1,0){10}}
          \put(80.5,17){\line(0,-1){8}}\put(78,18){$\Sym{3}{1}$}
          \put(70,40){$\Sym{1}{1}$}
         \end{overpic}
        \end{center}
       \end{minipage}
       \vspace{-0.5em}
       \\
   \begin{minipage}{0.48\hsize}
    \begin{center}
     \begin{overpic}[width=0.96\hsize,height=13.5em]
      {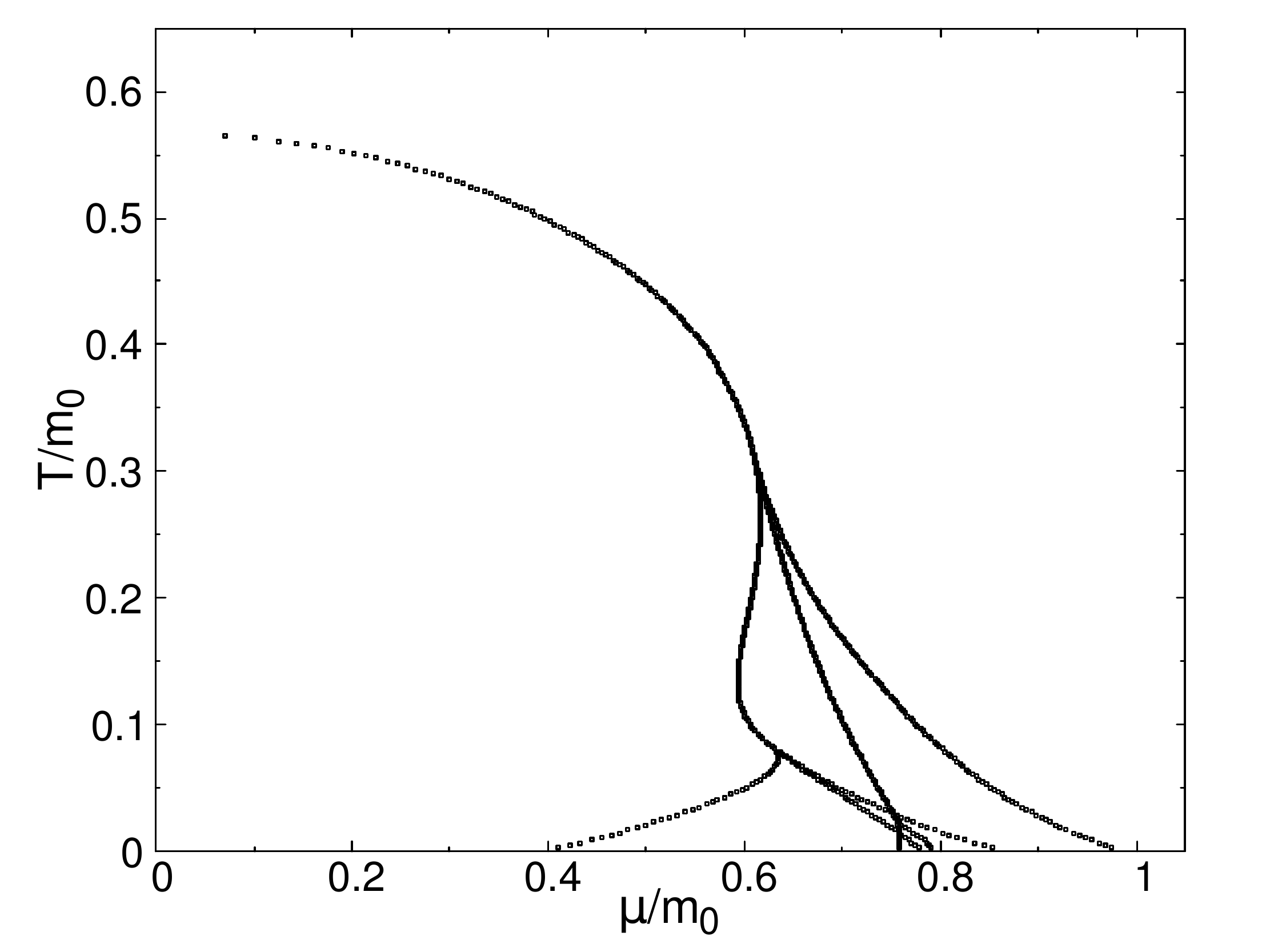}
      \put(20,40){$\Bro{2}{2}$}
      \put(48,30){$\Bro{3}{3}$ \line(1,-1){6}}
      \put(72,24.5){\line(0,-1){16.8}}\put(70,26.5){$\Bro{5}{3}$}
      \put(48,21.3){$\Bro{5}{5}$ \line(5,-4){15}}
      \put(48,12.8){$\Bro{5}{5}$ \line(3,-1){16.5}}
      \put(36,10){$\Bro{4}{4}$ \line(4,-1){10}}
      \put(70,40){$\Sym{1}{1}$}
      \put(78,18.5){\line(0,-4){9}}\put(76,20){$\Sym{3}{1}$}
      \put(86,10){\line(-5,-1){12}}\put(86,10){$\Sym{5}{1}$}
     \end{overpic}
    \end{center}
   \end{minipage}
   &
       \begin{minipage}{0.48\hsize}
        \begin{center}
         \begin{overpic}[width=0.96\hsize,height=13.5em]
          {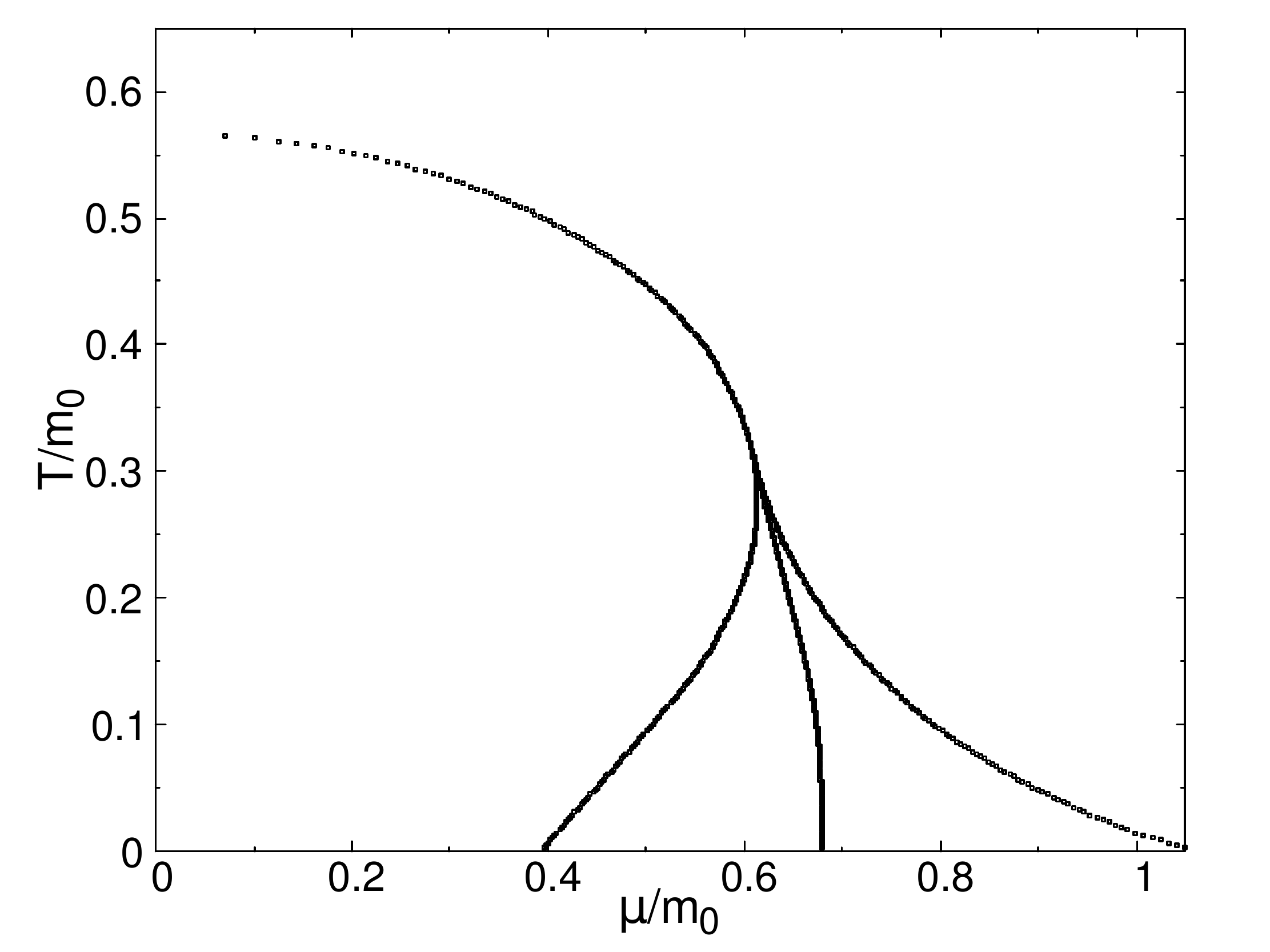}
          \put(20,40){$\Bro{2}{2}$}
          \put(56,10){$\Bro{3}{3}$}
          \put(70,40){$\Sym{1}{1}$}
          \put(70,10){$\Sym{3}{1}$}
         \end{overpic}
        \end{center}
       \end{minipage}
       \vspace{-1.5em}
  \end{tabular}
 \end{center}
 \caption{Phase diagrams in $2$ dimensions on a $\mu$-$T$ plane (above: $Lm_0=3.0$, below: $Lm_0=8.0$) for $\delta=0.0$ (lefts) and $\delta=1.0$ (rights).}
 \label{fig:phase-diagram:mu-T:2dim}
\end{figure}
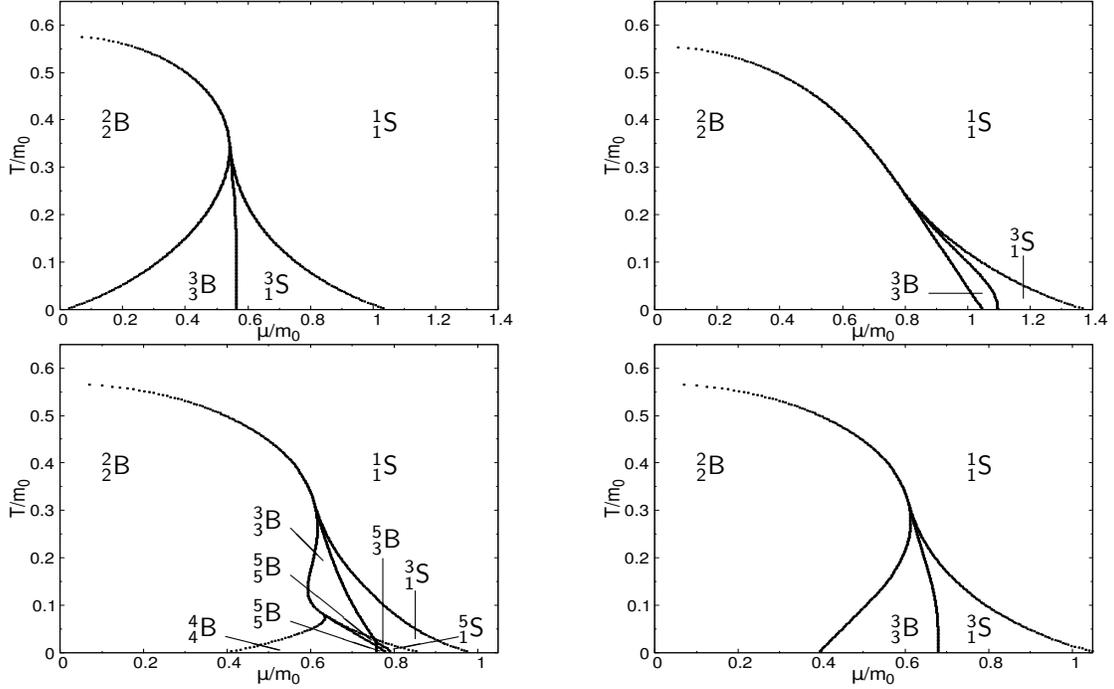

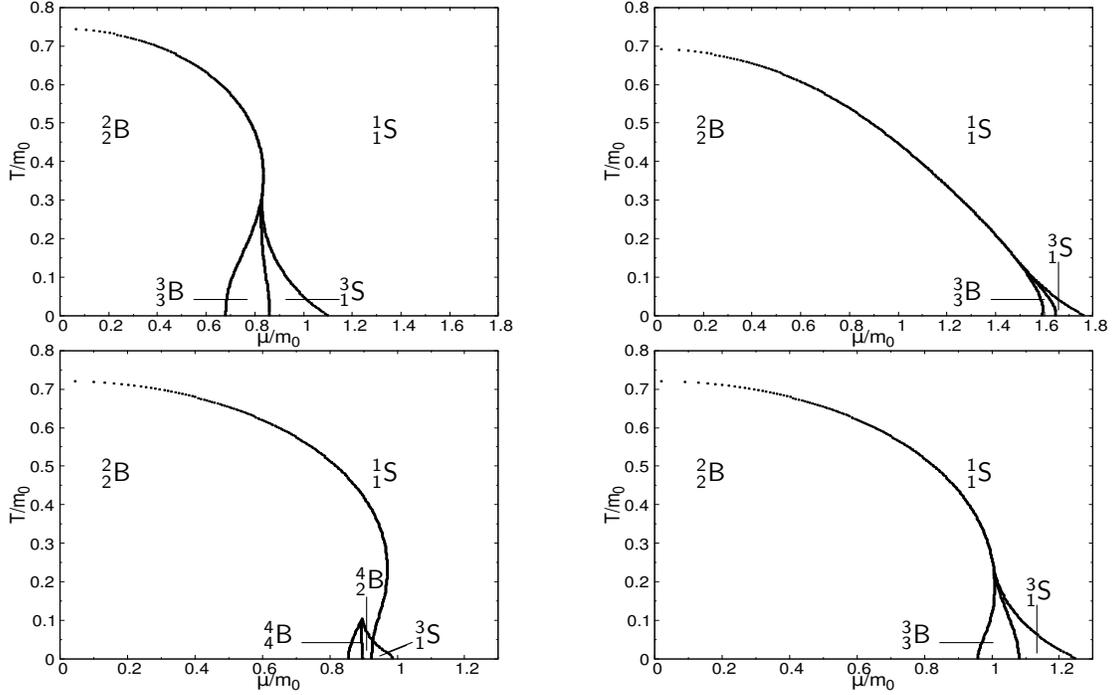
\begin{figure}[H]
 \begin{center}
  \begin{tabular}{cc}
   \begin{minipage}{0.48\hsize}
    \begin{center}
     \begin{overpic}[width=0.96\hsize,height=13.5em]
      {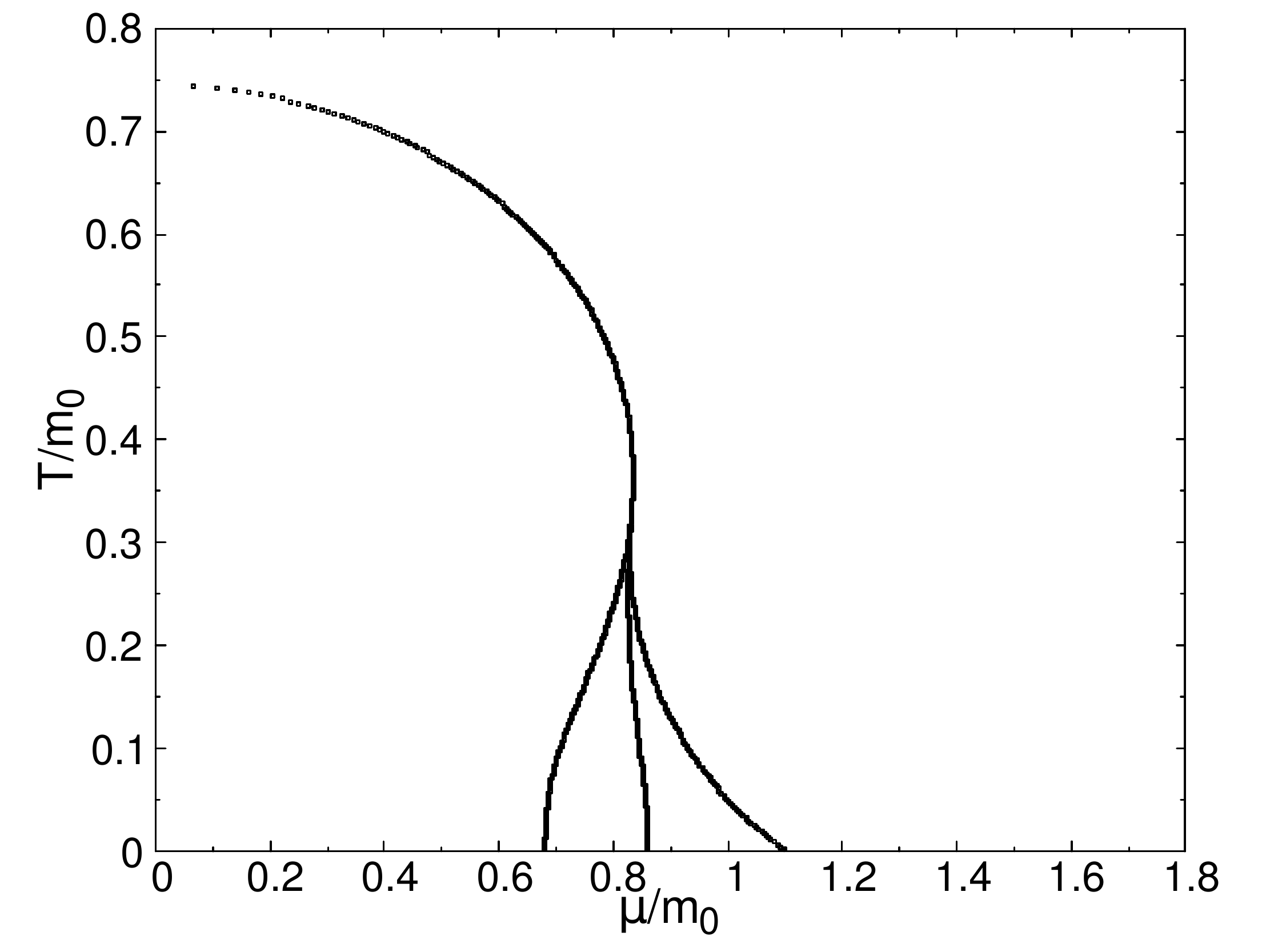}
      \put(20,40){$\Bro{2}{2}$}
      \put(30,10){$\Bro{3}{3}$ \line(1,0){10}}
      \put(70,40){$\Sym{1}{1}$}
      \put(54,10){\line(1,0){10} $\Sym{3}{1}$}
     \end{overpic}
    \end{center}
   \end{minipage}
   &
       \begin{minipage}{0.48\hsize}
        \begin{center}
         \begin{overpic}[width=0.96\hsize,height=13.5em]
          {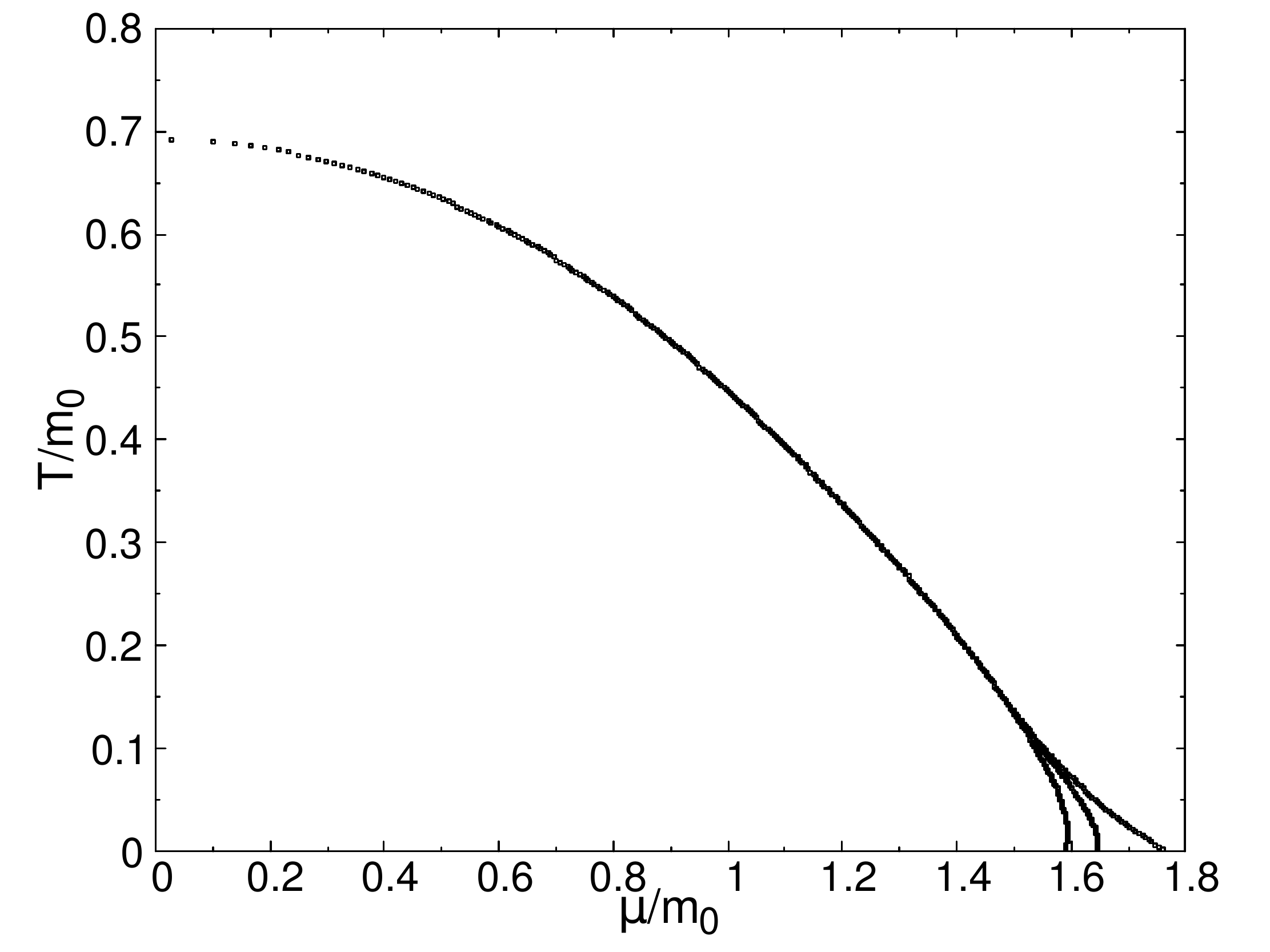}
          \put(20,40){$\Bro{2}{2}$}
          \put(67.5,10){$\Bro{3}{3}$ \line(1,0){10}}
          \put(87,17){\line(0,-1){9}}\put(85,18){$\Sym{3}{1}$}
          \put(70,40){$\Sym{1}{1}$}
         \end{overpic}
        \end{center}
       \end{minipage}
       \vspace{-0.5em}
       \\
   \begin{minipage}{0.48\hsize}
    \begin{center}
     \begin{overpic}[width=0.96\hsize,height=13.5em]
      {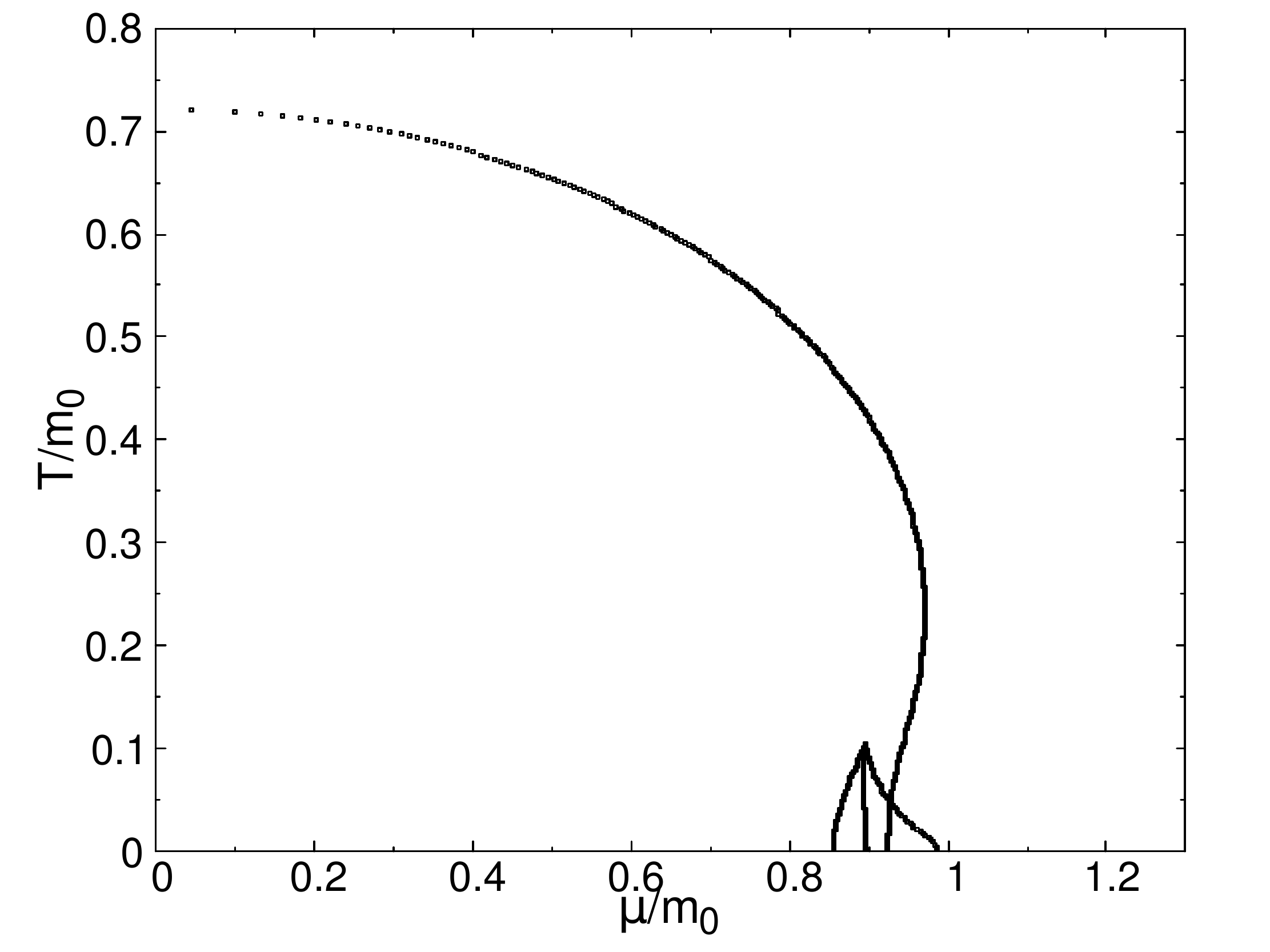}
      \put(20,40){$\Bro{2}{2}$}
      \put(69,18.5){\line(0,-1){10}}\put(67,20){$\Bro{4}{2}$}
      \put(50,10){$\Bro{4}{4}$ \line(1,0){11}}
      \put(70,40){$\Sym{1}{1}$}
      \put(77.5,10){\line(-5,-2){6}}\put(78,10){$\Sym{3}{1}$}
     \end{overpic}
    \end{center}
   \end{minipage}
   &
       \begin{minipage}{0.48\hsize}
        \begin{center}
         \begin{overpic}[width=0.96\hsize,height=13.5em]
          {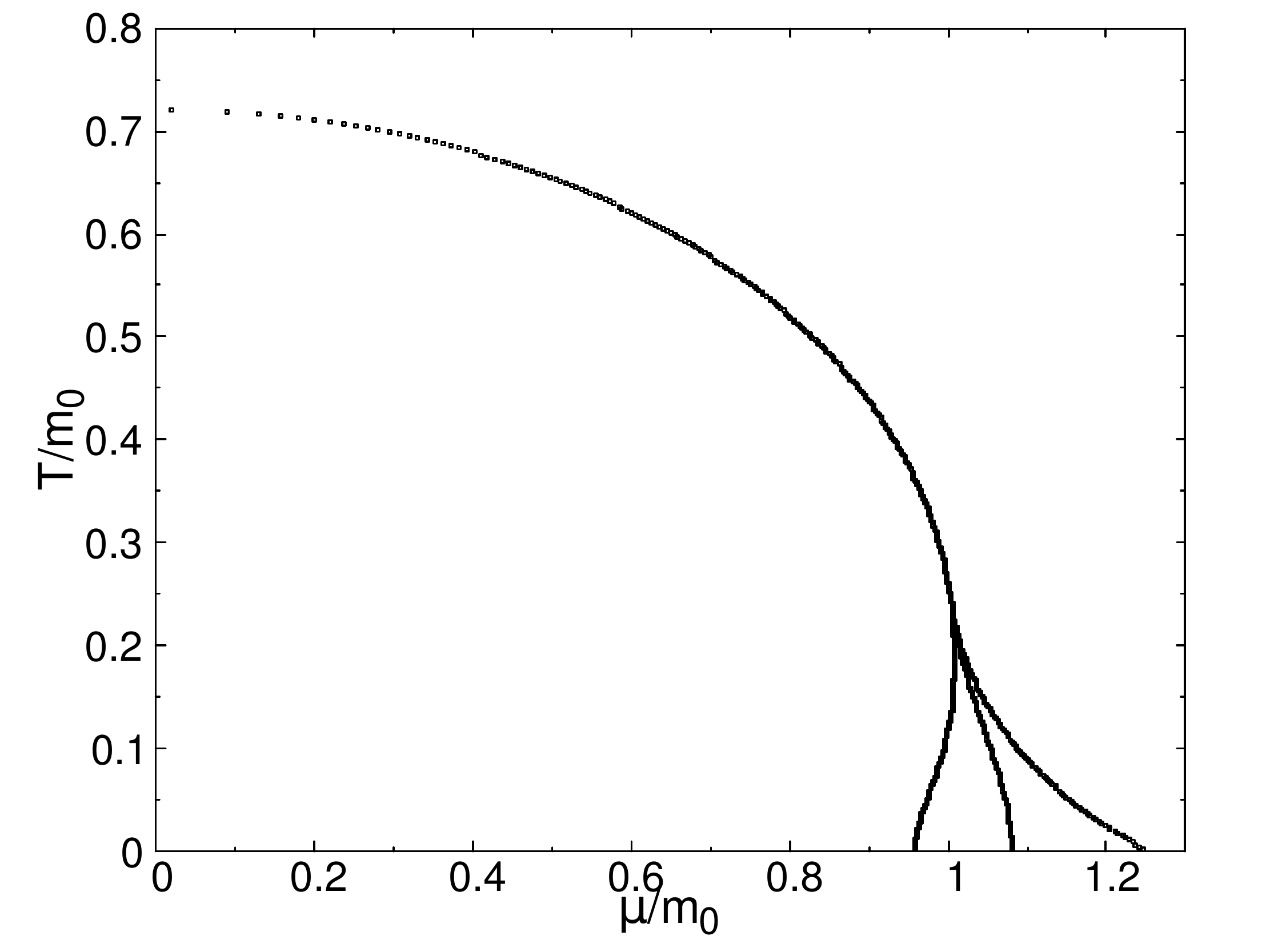}
          \put(20,40){$\Bro{2}{2}$}
          \put(58,10){$\Bro{3}{3}$ \line(1,0){10}}
          \put(83,17){\line(0,-1){9}}\put(81,18){$\Sym{3}{1}$}
          \put(70,40){$\Sym{1}{1}$}
         \end{overpic}
        \end{center}
       \end{minipage}
       \vspace{-1.5em}
  \end{tabular}
 \end{center}
 \caption{Phase diagrams in $3$ dimensions on a $\mu$-$T$ plane (above: $Lm_0=2.0$, below: $Lm_0=4.0$) for $\delta=0.0$ (lefts) and $\delta=1.0$ (rights).}
 \label{fig:phase-diagram:mu-T:3dim}
\end{figure}

We analyze the phase structure at the periodic ($\delta=0.0$) and antiperiodic ($\delta=1.0$) boundary conditions for simplicity.
Figs. \ref{fig:phase-diagram:mu-T:2dim} and \ref{fig:phase-diagram:mu-T:3dim} show the phase diagrams on the $\mu$-$T$ plane for some fixed sizes and boundary conditions.
It is known that, on non-compactified spaces, $\Rspace{}$ and $\Rspace{2}$, the critical temperatures at zero chemical potential are $e^\gamma/ \pi \simeq 0.57$ ($D=2.0$) and $1/\ln 4 \simeq 0.72$ ($D=3.0$), and the critical chemical potentials at zero temperature are $1/\sqrt{2} \simeq 0.71$ ($D=2.0$) and $1$ ($D=3.0$) with the normalization by $m_0$ \cite{Jacobs:1974ys, Wolff:1985av, Rosenstein:1990nm, Inagaki:1994ec}.
The finite-size effects primarily appear at low temperature with high chemical potential, and high temperature with low chemical potential.
The effect of the boundary condition is suppressed for large size, but at a certain size prominently appears.

For instance, in two dimensions at $Lm_0=8.0$ and $\delta=0.0$, we can observe the complex behavior of the boundaries.
As the chemical potential increases at low temperature, the domain changes  $\Bro{4}{4} \to \Bro{5}{5} \to \Bro{5}{3} \to \Sym{5}{1} \to \Sym{3}{1} \to \Sym{1}{1}$ and a mass jump appears twice at the boundaries between $\Bro{5}{5} \to \Bro{5}{3}$ and $\Bro{5}{3} \to \Sym{5}{1}$.

The basic properties are common between two and three dimensions.
Their differences come from the continuous momentum in the additional space, $\Rspace{}$.
At $Lm_0=4.0$ and $\delta=0.0$ in three dimensions, only the second-order phase transition takes place, and no critical end-point appears on the boundary between the symmetric and the broken phases.
In the broken phase, the dynamical mass discontinuously changes at the boundary between $\Bro{4}{4}$ and $\Bro{4}{2}$.

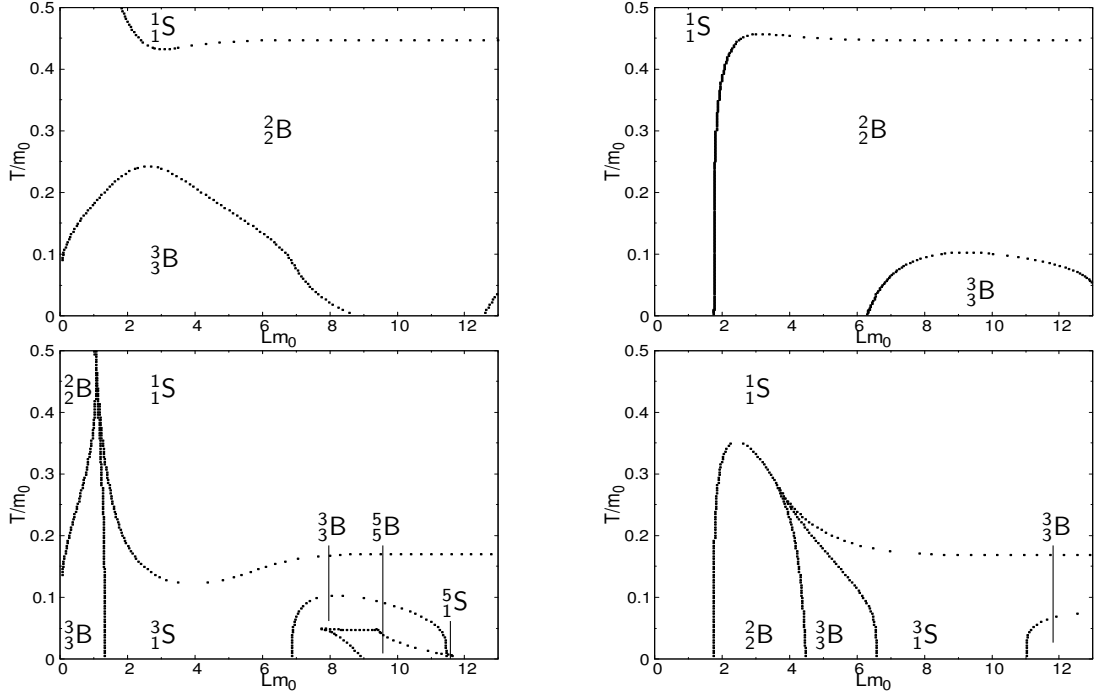
\begin{figure}[H]
 \begin{center}
  \begin{tabular}{cc}
   \begin{minipage}{0.48\hsize}
    \begin{center}
     \begin{overpic}[width=0.96\hsize,height=13.5em]
      {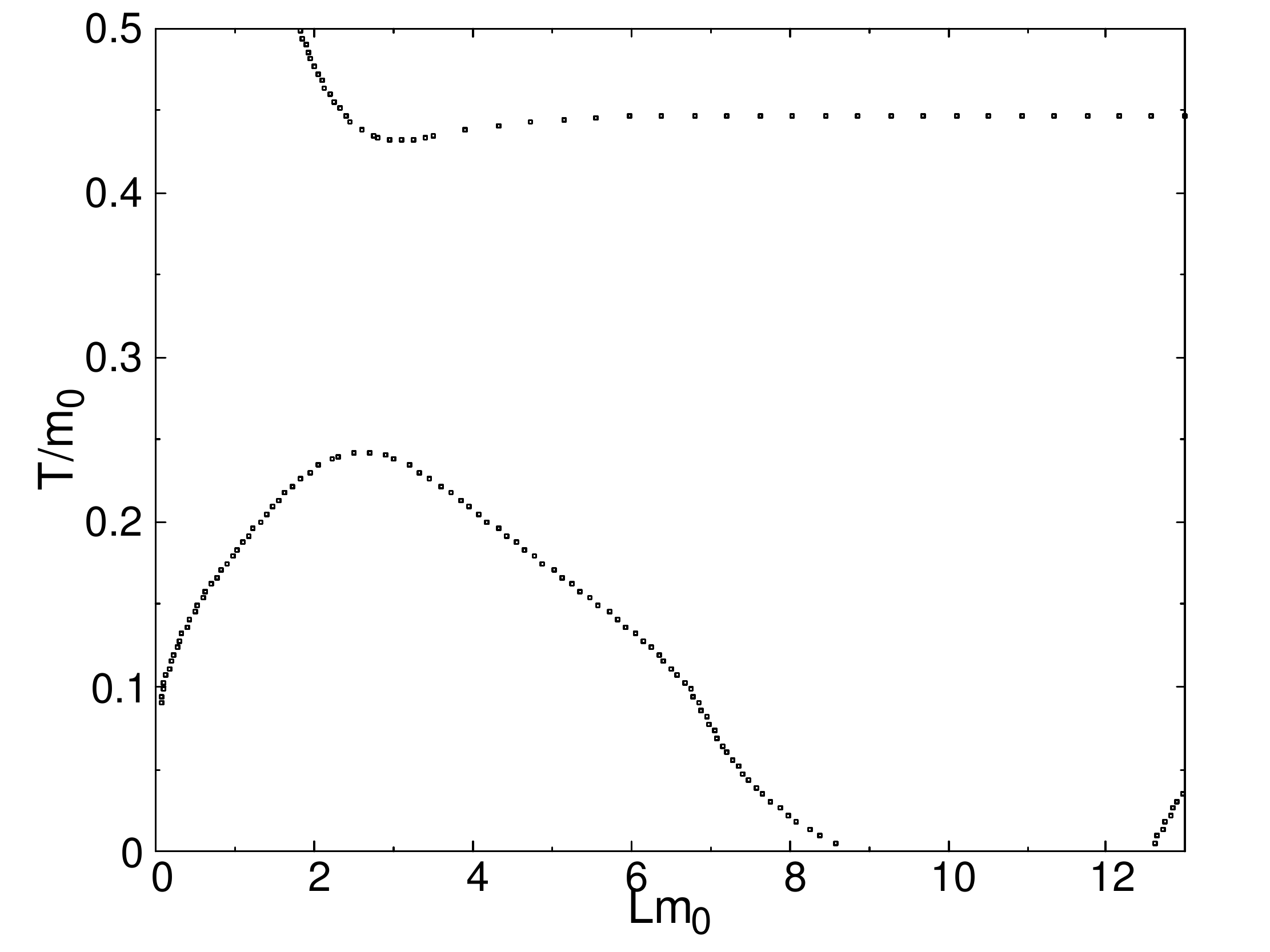}
      \put(50,40){$\Bro{2}{2}$}
      \put(29,16){$\Bro{3}{3}$}
      \put(29,59){$\Sym{1}{1}$}
     \end{overpic}
    \end{center}
   \end{minipage}
   &
       \begin{minipage}{0.48\hsize}
        \begin{center}
         \begin{overpic}[width=0.96\hsize,height=13.5em]
          {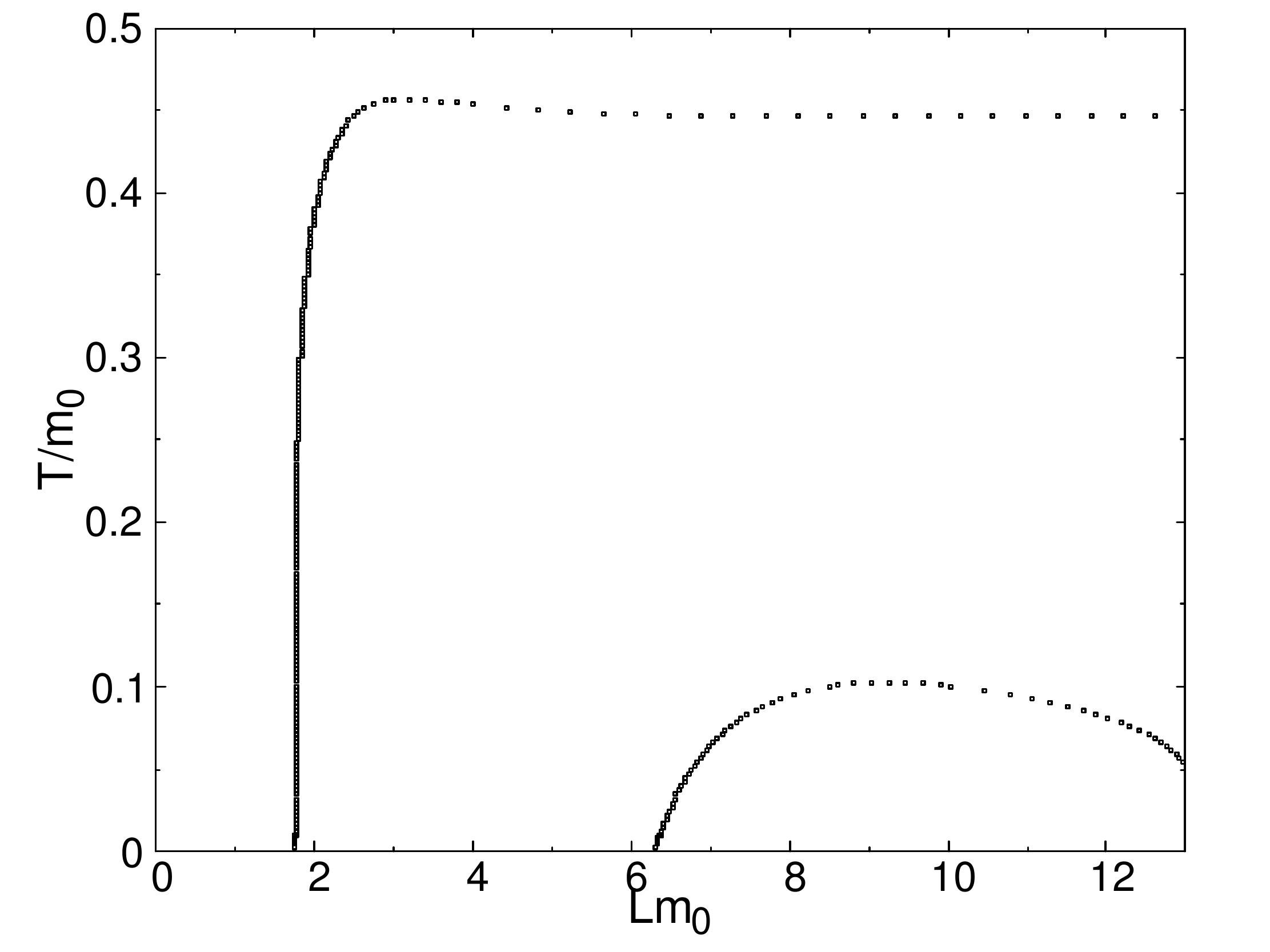}
          \put(50,40){$\Bro{2}{2}$}
          \put(70,10){$\Bro{3}{3}$}
          \put(18,59){$\Sym{1}{1}$}
         \end{overpic}
        \end{center}
       \end{minipage}
       \vspace{-0.5em}
       \\
    \begin{minipage}{0.48\hsize}
      \begin{center}
       \begin{overpic}[width=0.96\hsize,height=13.5em]
        {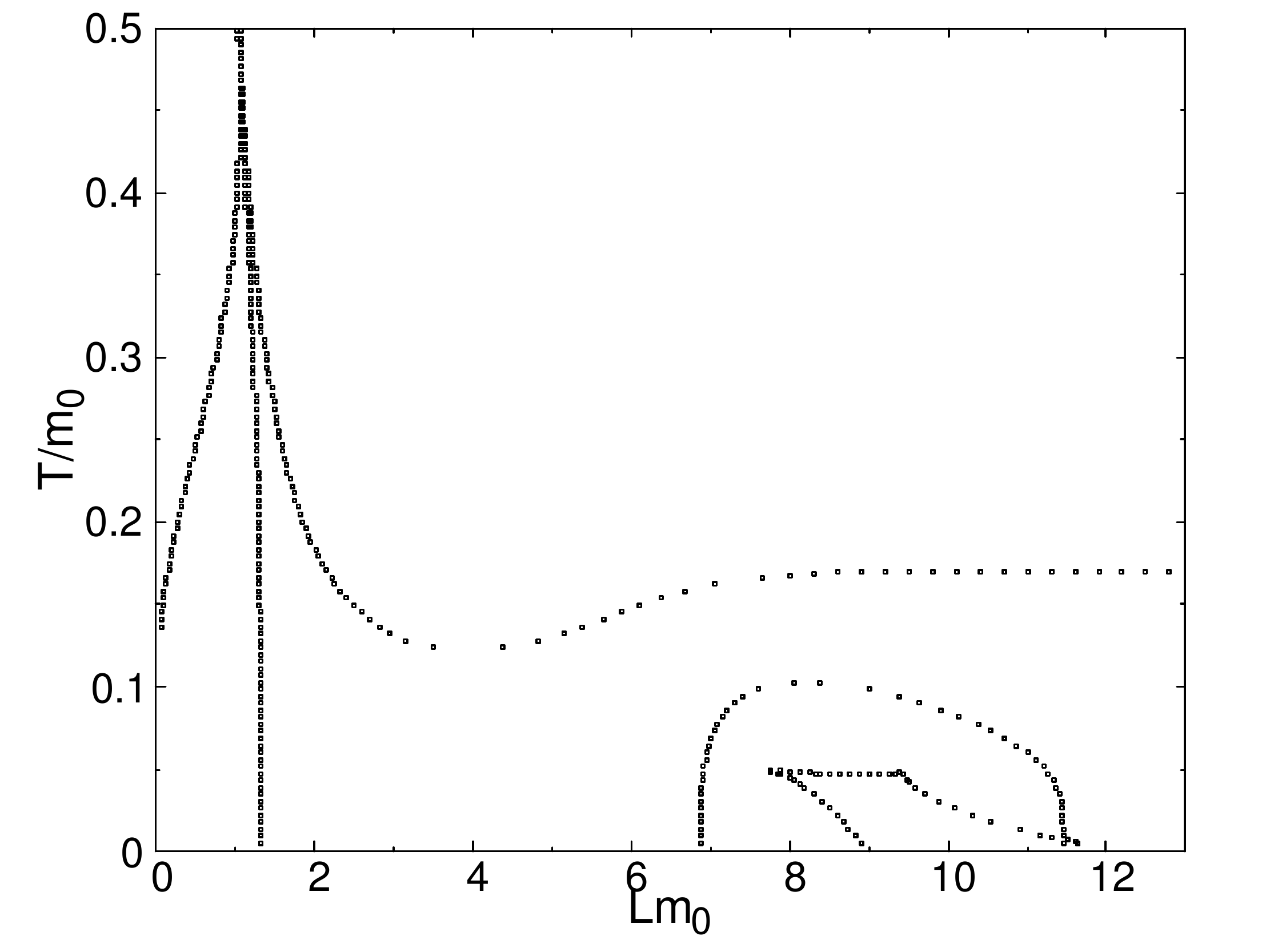}
        \put(13,55.5){$\Bro{2}{2}$}
        \put(13,10){$\Bro{3}{3}$}
        \put(62,28){\line(0,-1){14}}\put(60,30){$\Bro{3}{3}$}
        \put(72,28){\line(0,-1){20}}\put(70,30){$\Bro{5}{5}$}
        \put(29,55.5){$\Sym{1}{1}$}
        \put(84.5,14){\line(0,-1){7}}\put(83,16){$\Sym{5}{1}$}
       \put(29,10){$\Sym{3}{1}$}
       \end{overpic}
      \end{center}
    \end{minipage}
   &
       \begin{minipage}{0.48\hsize}
        \begin{center}
         \begin{overpic}[width=0.96\hsize,height=13.5em]
          {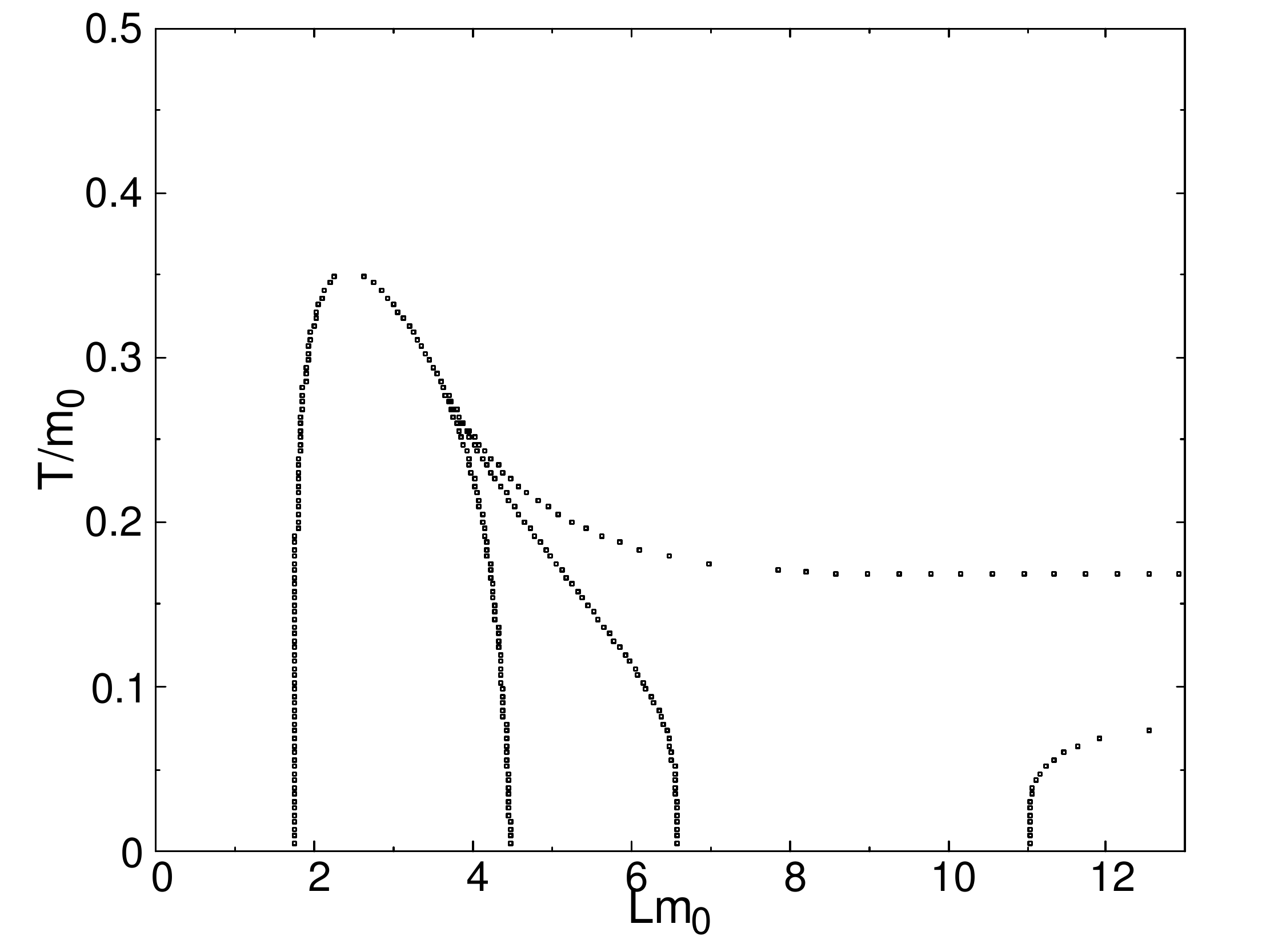}
          \put(29,10){$\Bro{2}{2}$}
          \put(42,10){$\Bro{3}{3}$}
          \put(86,28){\line(0,-1){18}}\put(84,30){$\Bro{3}{3}$}
          \put(29,55.5){$\Sym{1}{1}$}
          \put(60,10){$\Sym{3}{1}$}
         \end{overpic}
        \end{center}
       \end{minipage}
	\vspace{-1.5em}
  \end{tabular}
 \end{center}
 \caption{Phase diagrams in $2$ dimensions on a $L$-$T$ plane (above: $\mu/m_0=0.5$, below: $\mu/m_0=0.7$) for $\delta=0.0$ (lefts) and $\delta=1.0$ (rights).}
 \label{fig:phase-diagram:L-T:2dim}
\end{figure}

\begin{figure}[H]
 \begin{center}
  \begin{tabular}{cc}
   \begin{minipage}{0.48\hsize}
    \begin{center}
     \begin{overpic}[width=0.96\hsize,height=13.5em]
      {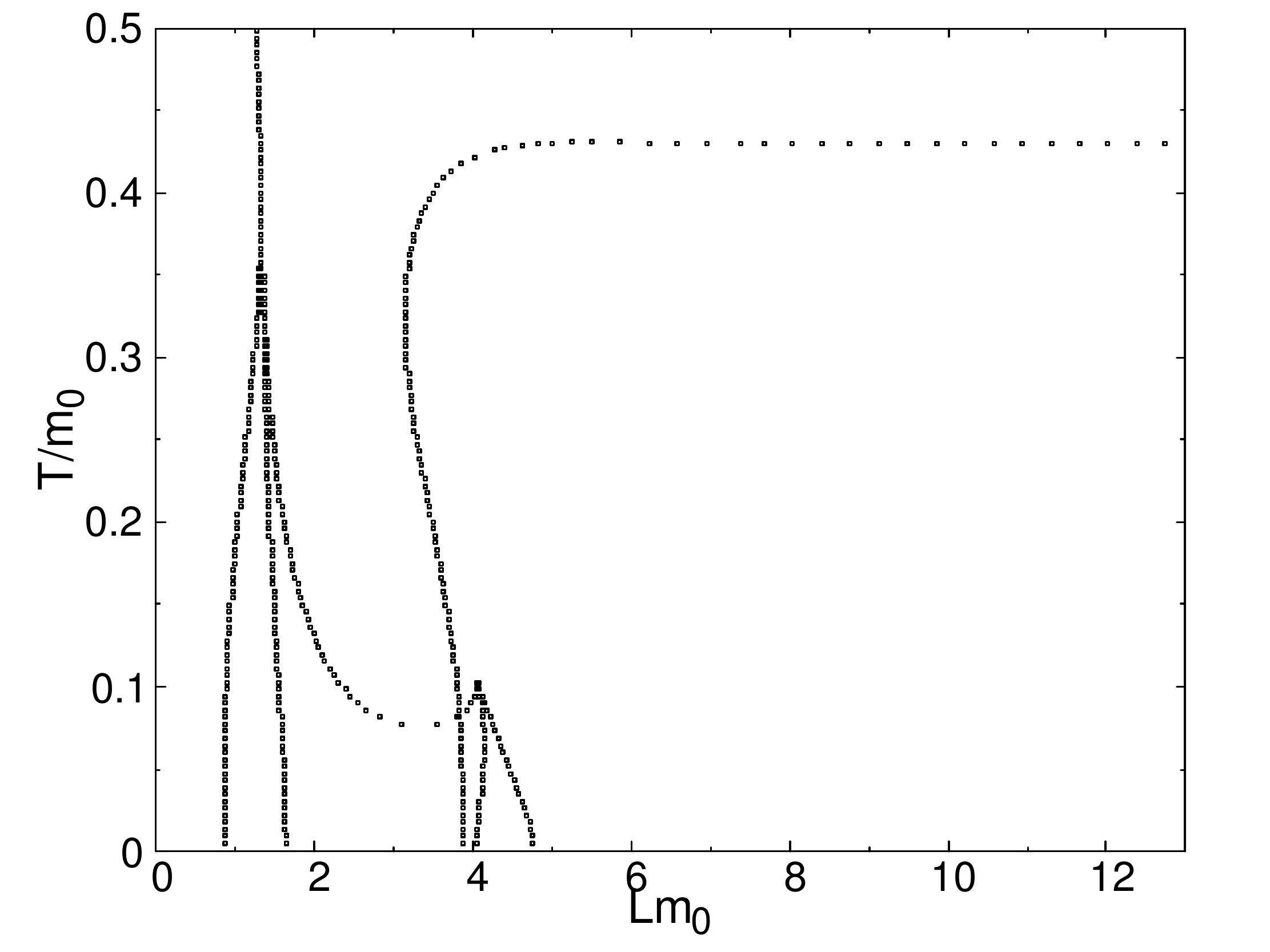}
      \put(13,55.5){$\Bro{2}{2}$}
      \put(47,40){$\Bro{2}{2}$}
      \put(37,24){\line(0,-1){10}}\put(35,26){$\Bro{4}{2}$}
      \put(46,10){\line(-1,0){7}}\put(47,10){$\Bro{4}{4}$}
      \put(18,10){$\Bro{3}{3}$}
      \put(29,55.5){$\Sym{1}{1}$}
      \put(29,10){$\Sym{3}{1}$}
     \end{overpic}
    \end{center}
   \end{minipage}
   &
       \begin{minipage}{0.48\hsize}
        \begin{center}
         \begin{overpic}[width=0.96\hsize,height=13.5em]
          {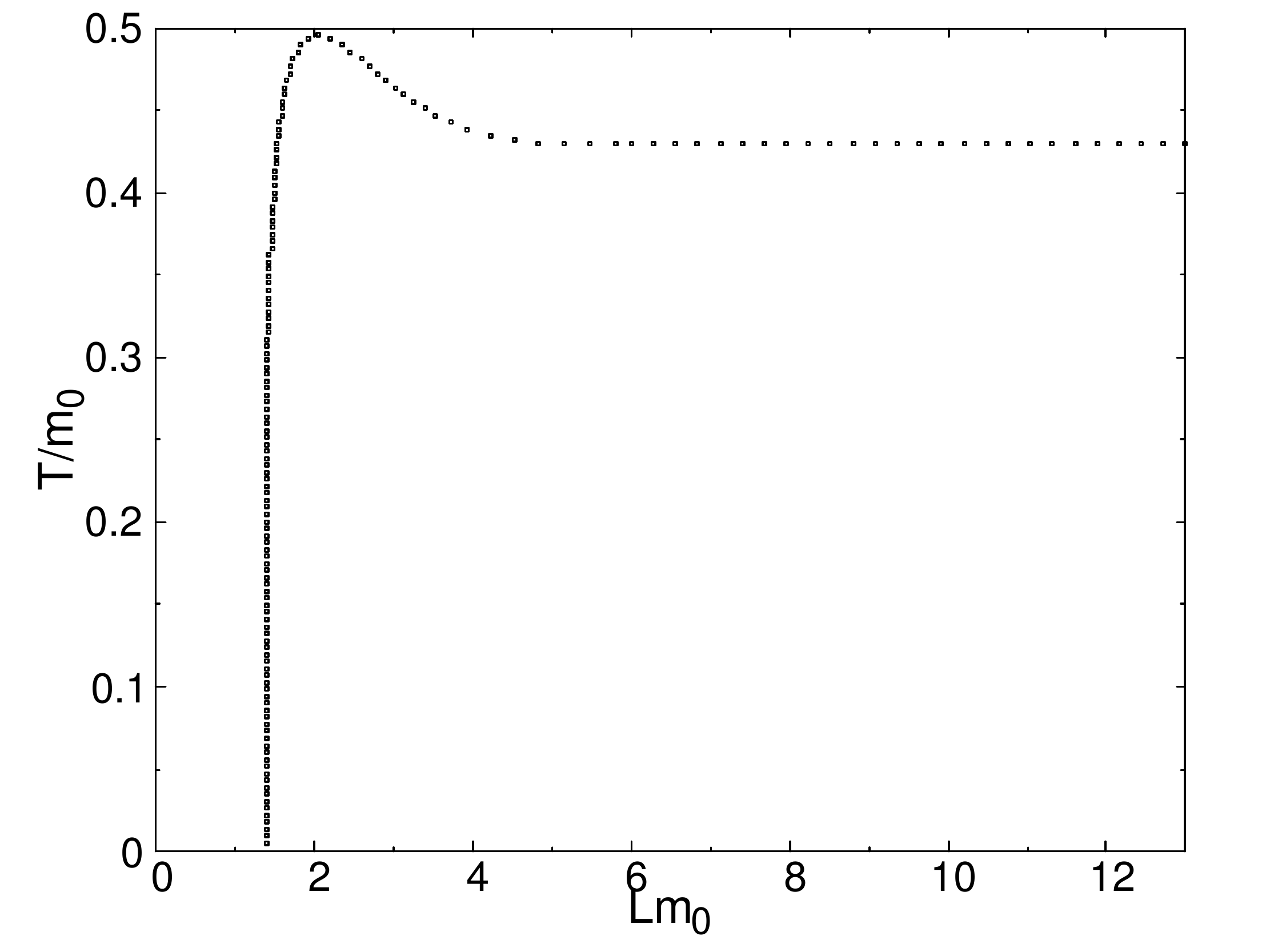}
          \put(50,40){$\Bro{2}{2}$}
          \put(13,55.5){$\Sym{1}{1}$}
         \end{overpic}
        \end{center}
       \end{minipage}
       \vspace{-0.5em}
       \\
   \begin{minipage}{0.48\hsize}
    \begin{center}
     \begin{overpic}[width=0.96\hsize,height=13.5em]
      {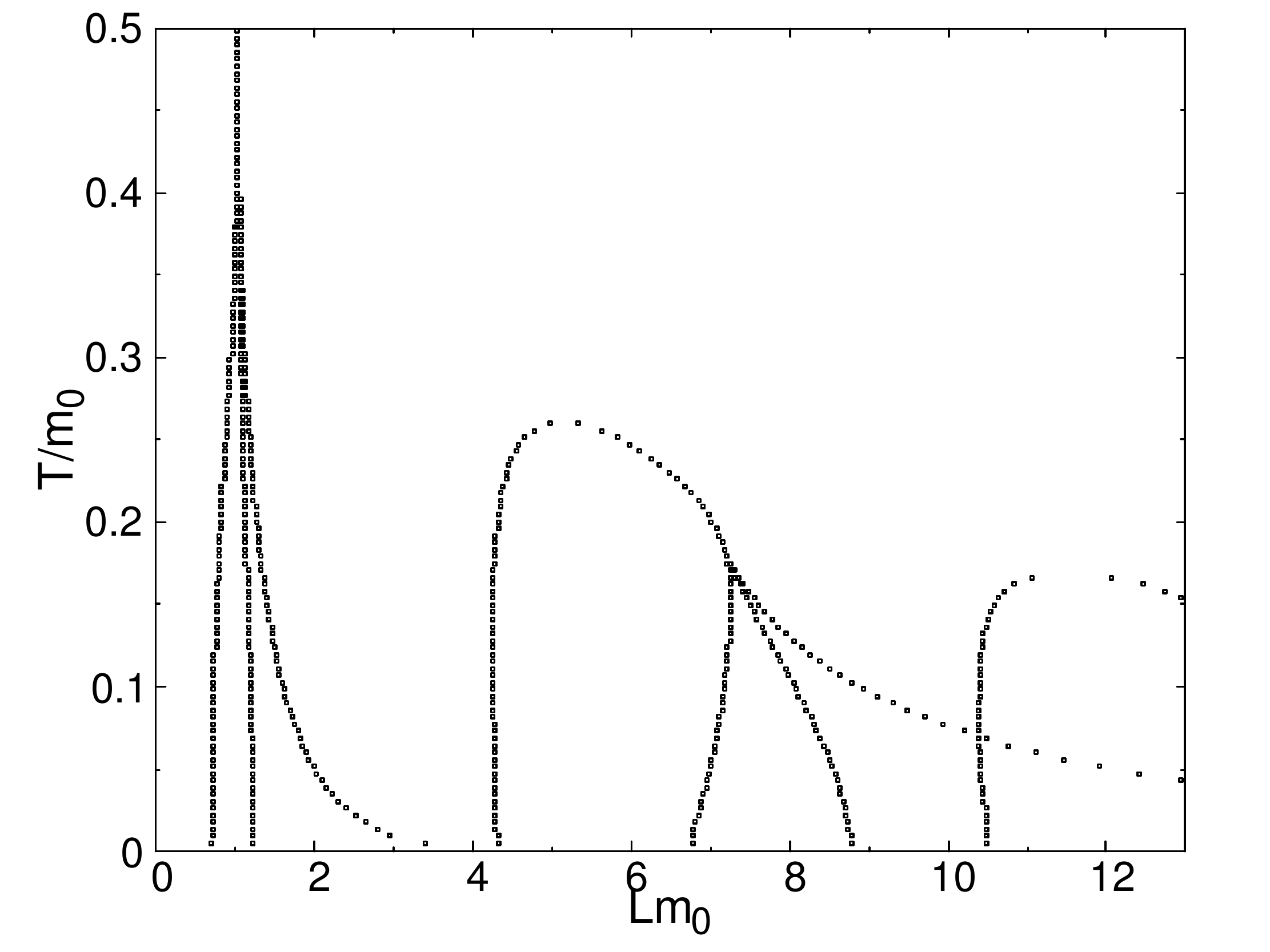}
      \put(13,55.5){$\Bro{2}{2}$}
      \put(28,18){\line(-1,0){10}}\put(29,18){$\Bro{3}{3}$}
      \put(47,18){$\Bro{2}{2}$}
      \put(58,10){$\Bro{3}{3}$}
      \put(82,18){$\Bro{2}{2}$}
      \put(82,9){$\Bro{4}{2}$}
      \put(29,55.5){$\Sym{1}{1}$}
      \put(28.5,11){\line(-1,0){6}}\put(29,11){$\Sym{3}{1}$}
      \put(68,10){$\Sym{3}{1}$}
     \end{overpic}
    \end{center}
   \end{minipage}
   &
       \begin{minipage}{0.48\hsize}
        \begin{center}
         \begin{overpic}[width=0.96\hsize,height=13.5em]
          {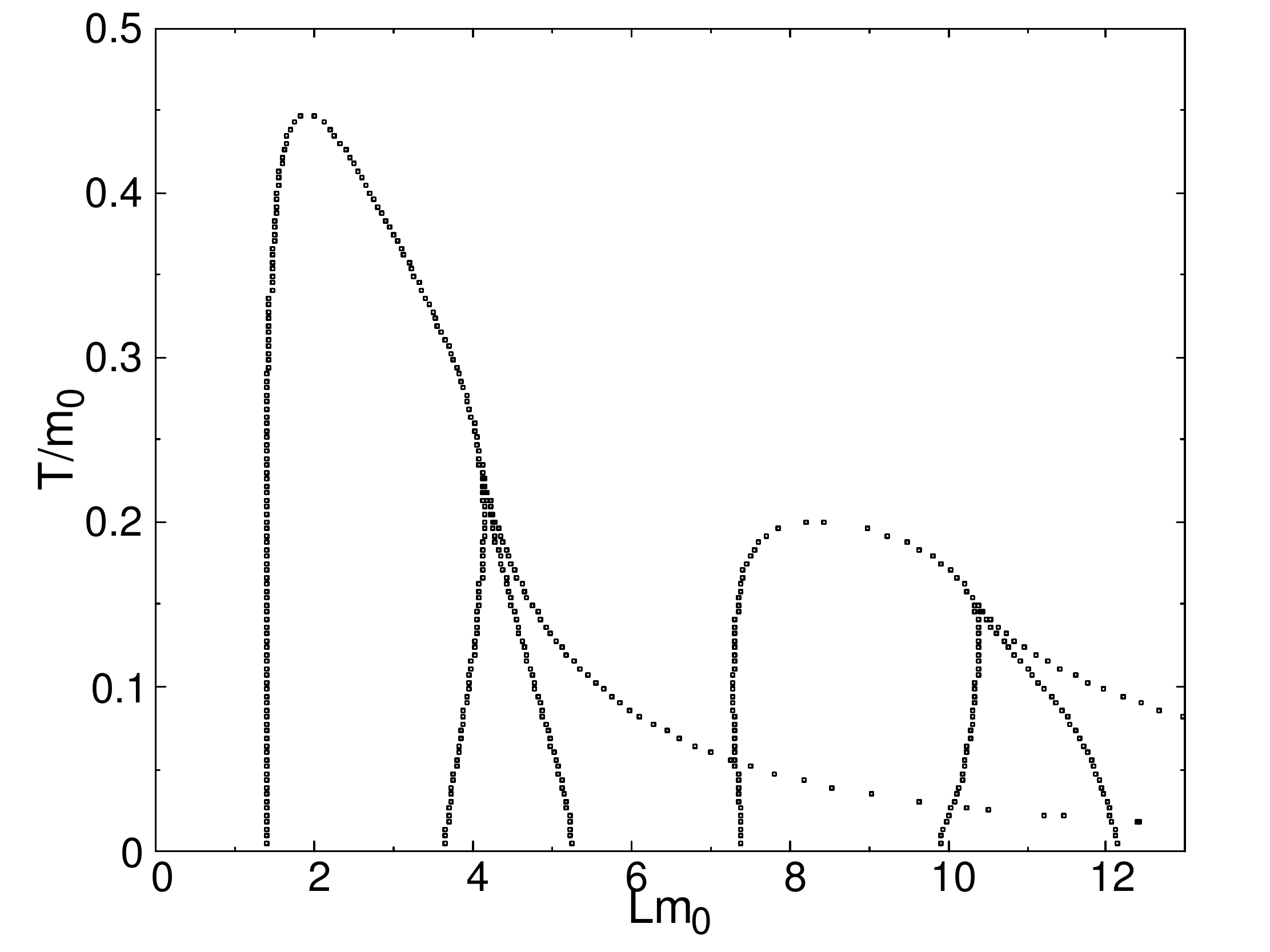}
          \put(26,18){$\Bro{2}{2}$}
          \put(37,10){$\Bro{3}{3}$}
          \put(48,20){$\Bro{4}{2}$ \line(1,-1){11}}
          \put(64,18){$\Bro{2}{2}$}
          \put(79,11){$\Bro{3}{3}$}
          \put(78.5,27){\line(0,-1){19}}\put(76.5,29){$\Bro{5}{3}$}
          \put(13,55.5){$\Sym{1}{1}$}
          \put(48,10){$\Sym{3}{1}$}
          \put(89,11){$\Sym{3}{1}$}
          \put(88.5,27){\line(0,-1){19}}\put(86.5,29){$\Sym{5}{1}$}
         \end{overpic}
        \end{center}
       \end{minipage}
	\vspace{-1.5em}
  \end{tabular}
 \end{center}
 \caption{Phase diagrams in $3$ dimensions on a $L$-$T$ plane (above: $\mu/m_0=0.5$, below: $\mu/m_0=0.7$) for $\delta=0.0$ (lefts) and $\delta=1.0$ (rights).}
 \label{fig:phase-diagram:L-T:3dim}
\end{figure}

By observing the phase diagrams on a $\mu$-$T$ plane, we can recognize the differences that emerge via the size of the system and the boundary conditions.
To determine the length dependence of the chiral symmetry, next we plot the phase structure on the $L$-$T$ plane.
We show the phase diagrams on the $L$-$T$ plane in Figs. \ref{fig:phase-diagram:L-T:2dim} and \ref{fig:phase-diagram:L-T:3dim}.
At $T=\mu=0$, the finite-size effect enhances and suppresses the chiral symmetry breaking for the periodic and the antiperiodic boundary conditions, respectively \cite{Inagaki:2019kbc}.
As is observed in Fig. \ref{fig:phase-diagram:L-T:2dim} and \ref{fig:phase-diagram:L-T:3dim}, this relation is reversed for certain intervals of size at a finite temperature and chemical potential.
For low temperatures, the broken and the symmetric phases alternate with increasing size of the system, $L$.
A first-order phase transition occurs at the boundaries between $\Bro{a}{b}$ and $\Sym{a}{1}$ ($\mathsf{b} \neq 1$).

\subsection{Dynamically generated fermion mass}
\begin{figure}[H]
 \begin{center}
  \begin{tabular}{cc}
   \hspace{-2em}
   \begin{minipage}{0.5\hsize}
    \begin{center}
     \begin{tabular}{cc}
      \begin{minipage}{0.5\hsize}
       \begin{overpic}[width=1\hsize,clip]
        {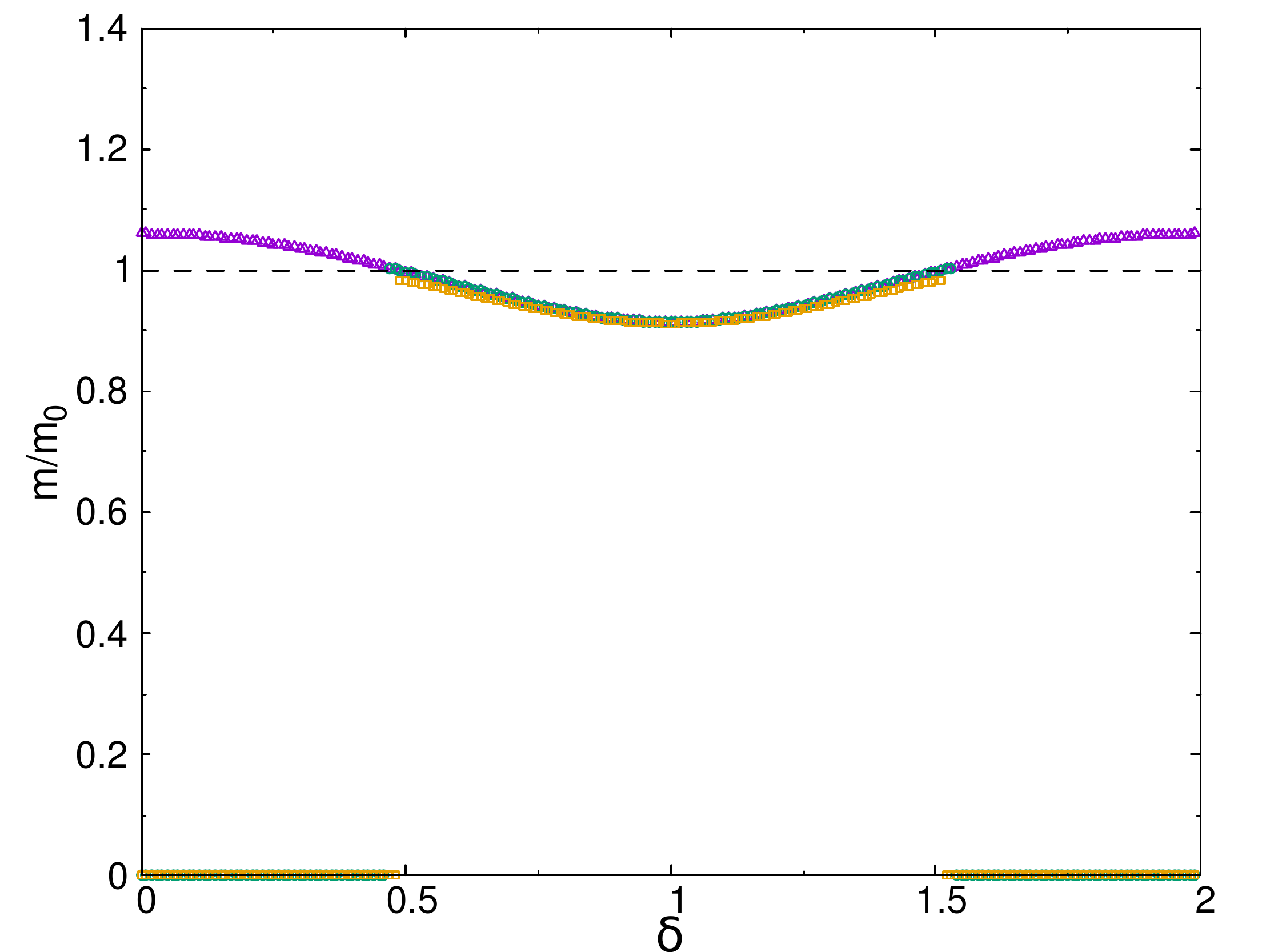}
       \end{overpic}
       \centering{\footnotesize $L m_0 = 3.0$}
      \end{minipage}
      \hspace{-2em}\vspace{-1em}
      &
      \begin{minipage}{0.5\hsize}
       \begin{overpic}[width=1\hsize,clip]
        {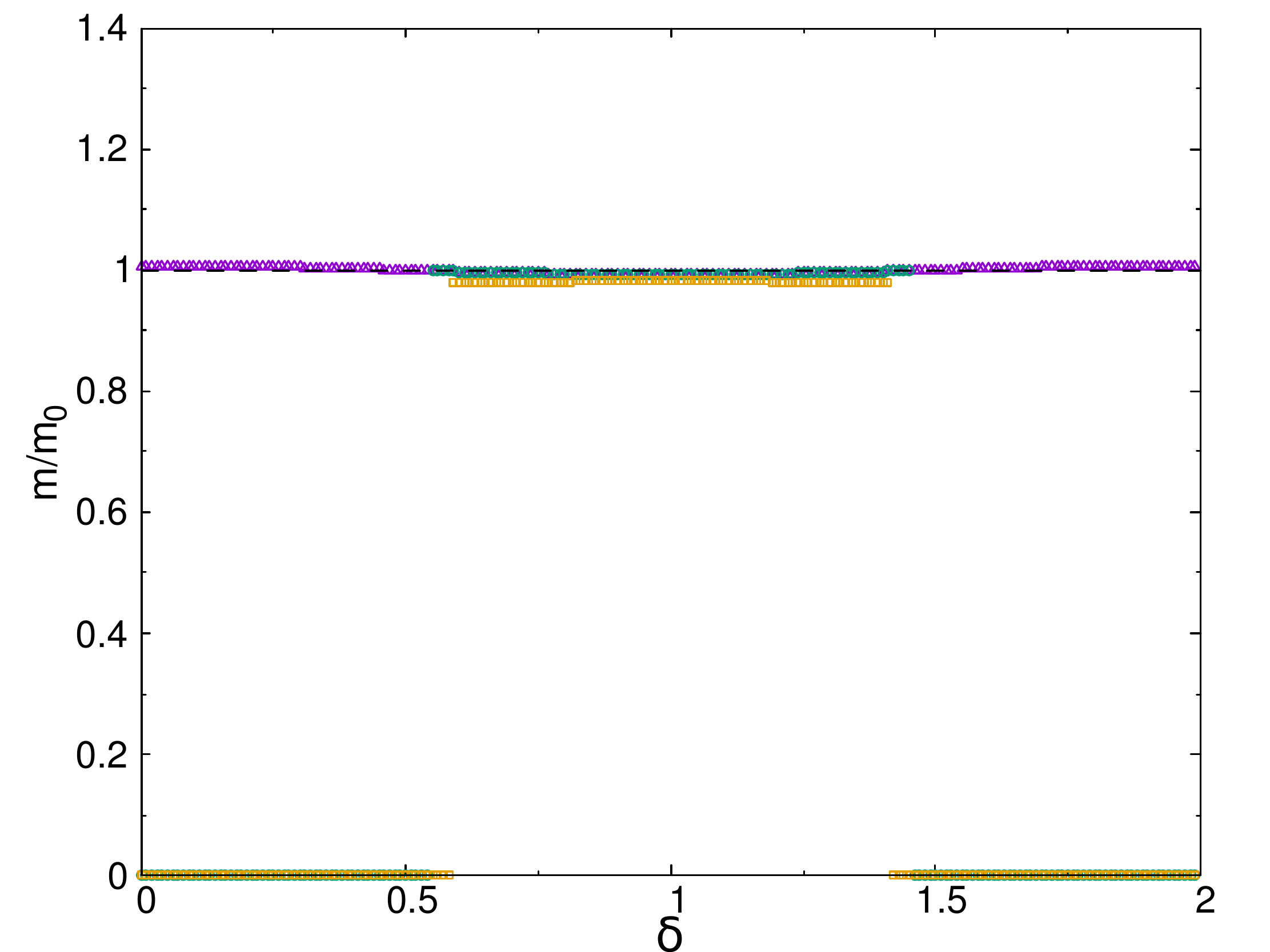}
       \end{overpic}
       \centering{\footnotesize $L m_0 = 5.0$}
      \end{minipage}
     \end{tabular}
    \end{center}
   \end{minipage}
   &
   \begin{minipage}{0.5\hsize}
    \begin{center}
     \begin{tabular}{cc}
      \begin{minipage}{0.5\hsize}
       \begin{overpic}[width=1\hsize,clip]
        {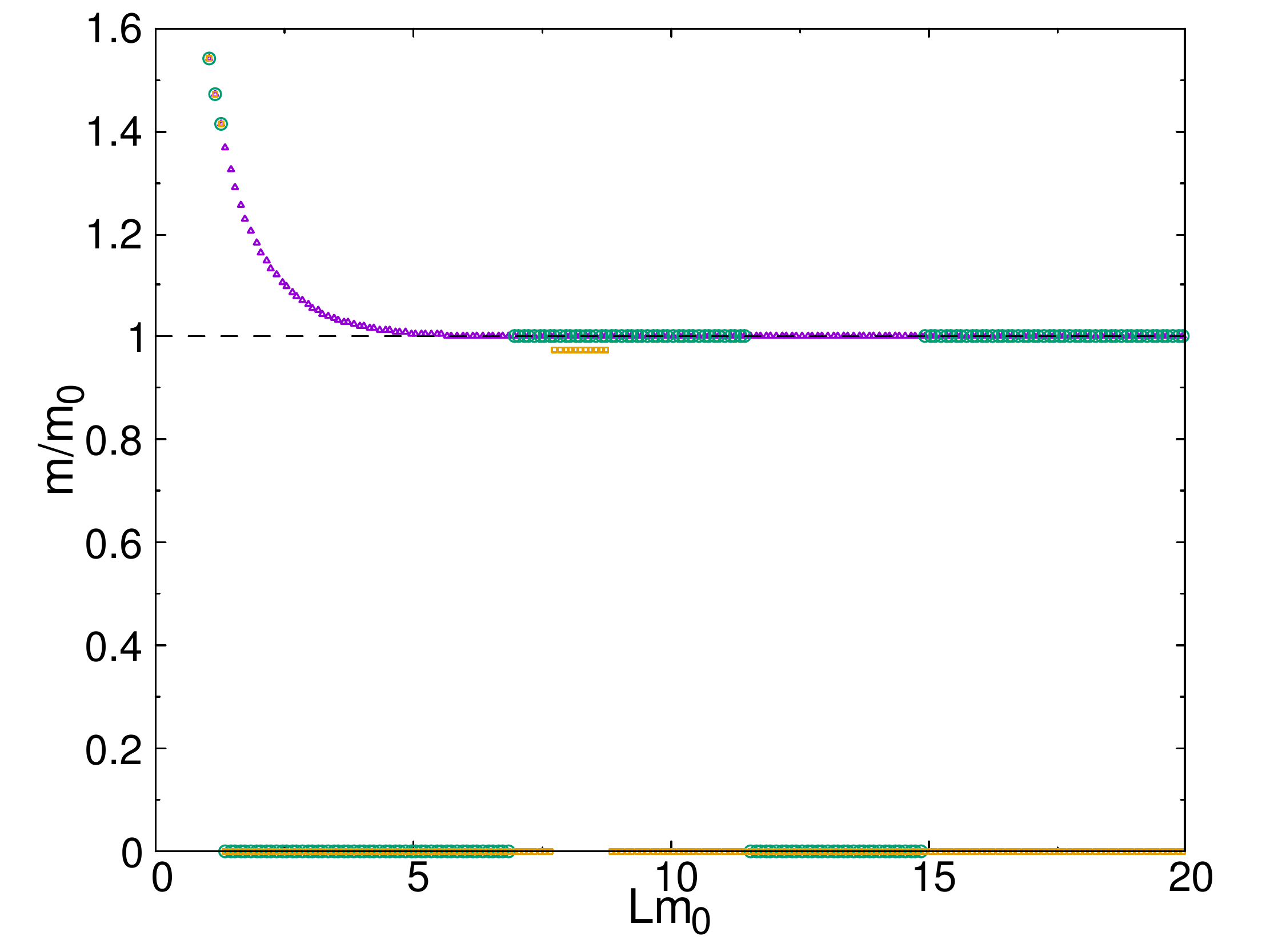}
       \end{overpic}
       \centering{\footnotesize $\delta = 0.0$}
      \end{minipage}
      \hspace{-2em}\vspace{-1em}
      &
      \begin{minipage}{0.5\hsize}
       \begin{overpic}[width=1\hsize,clip]
        {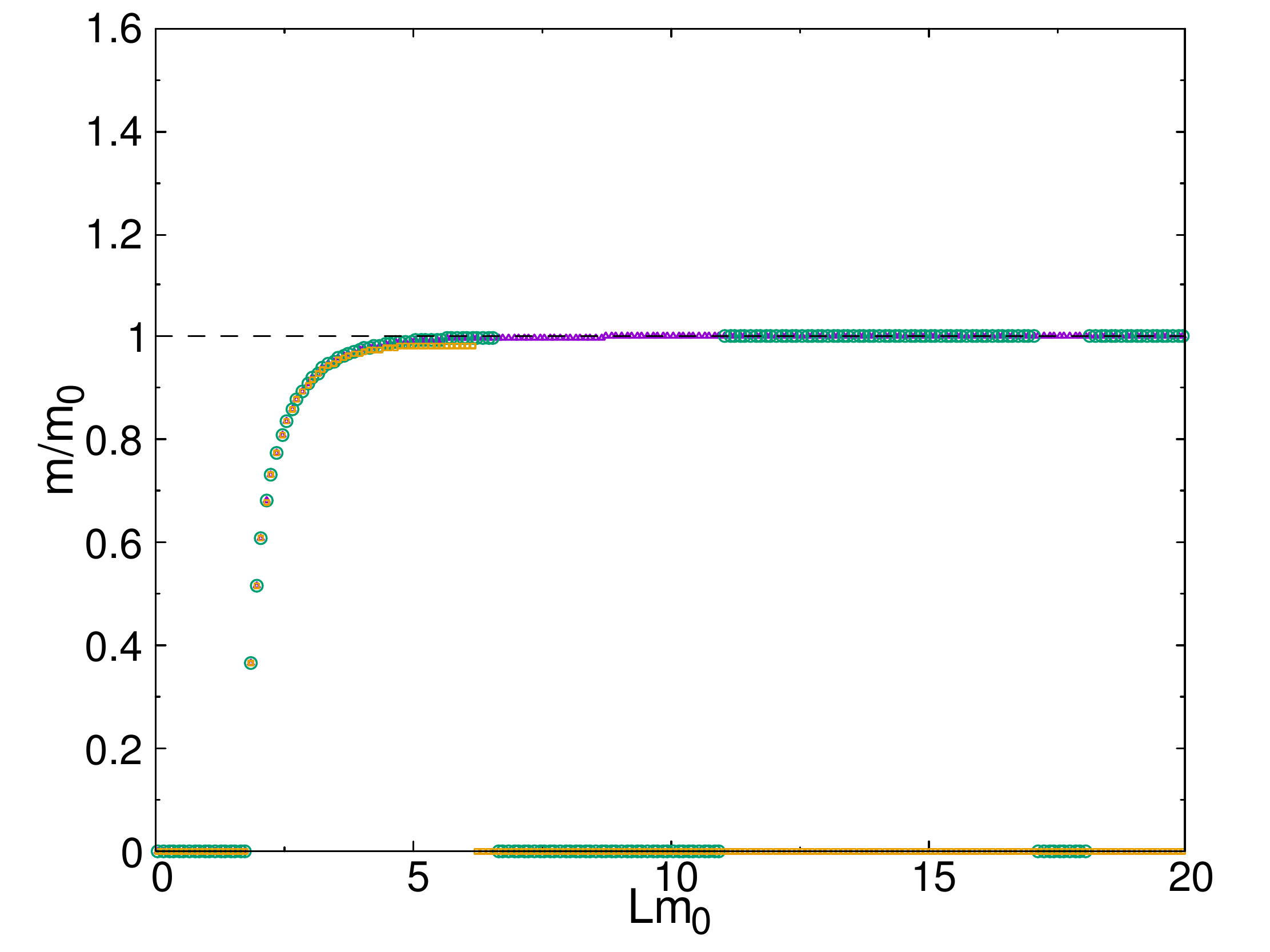}
       \end{overpic}
       \centering{\footnotesize $\delta = 1.0$}
      \end{minipage}
     \end{tabular}
    \end{center}
   \end{minipage}
  \end{tabular}
 \end{center}
 \caption{Dynamically generated fermion mass as a function of $\delta$ (left) and $L$ (right) in 2 dimensions:
 $(T/m_0, \mu/m_0; \mbox{color}) = (0.005, 0.0; \mbox{purple}), \, (0.005, 0.7; \mbox{green}), \, (0.1, 0.7; \mbox{yellow})$.}
 \label{fig:dynamically-generated-mass:2dim}
\end{figure}

\begin{figure}[H]
 \begin{center}
  \begin{tabular}{cc}
   \hspace{-2em}
   \begin{minipage}{0.5\hsize}
    \begin{center}
     \begin{tabular}{cc}
      \begin{minipage}{0.5\hsize}
       \begin{overpic}[width=1\hsize,clip]
        {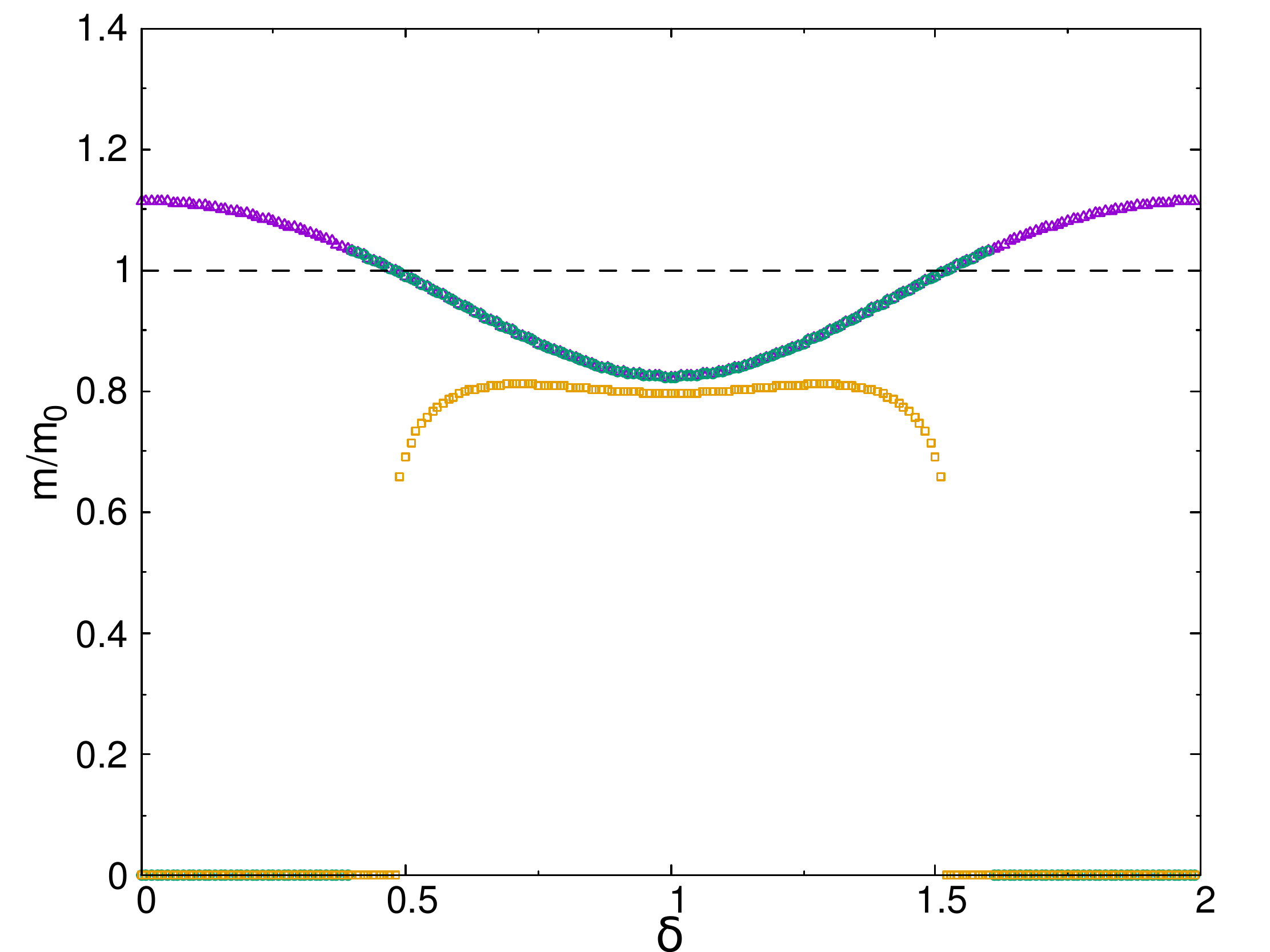}
       \end{overpic}
       \centering{\footnotesize $L m_0 = 2.0$}
      \end{minipage}
      \hspace{-2em}\vspace{-1em}
      &
      \begin{minipage}{0.5\hsize}
       \begin{overpic}[width=1\hsize,clip]
        {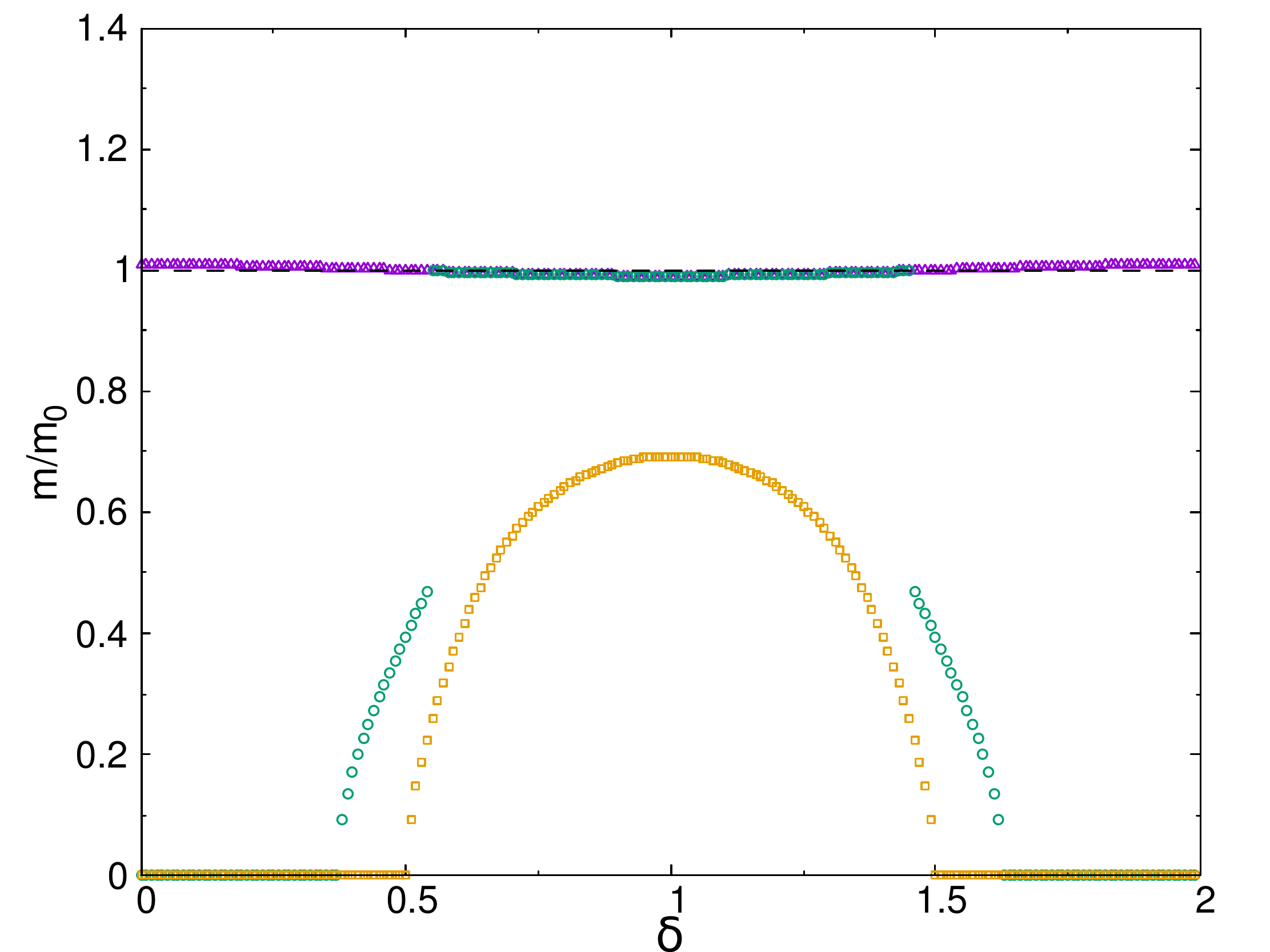}
       \end{overpic}
       \centering{\footnotesize $L m_0 = 4.0$}
      \end{minipage}
     \end{tabular}
    \end{center}
   \end{minipage}
   &
   \begin{minipage}{0.5\hsize}
    \begin{center}
     \begin{tabular}{cc}
      \begin{minipage}{0.5\hsize}
       \begin{overpic}[width=1\hsize,clip]
        {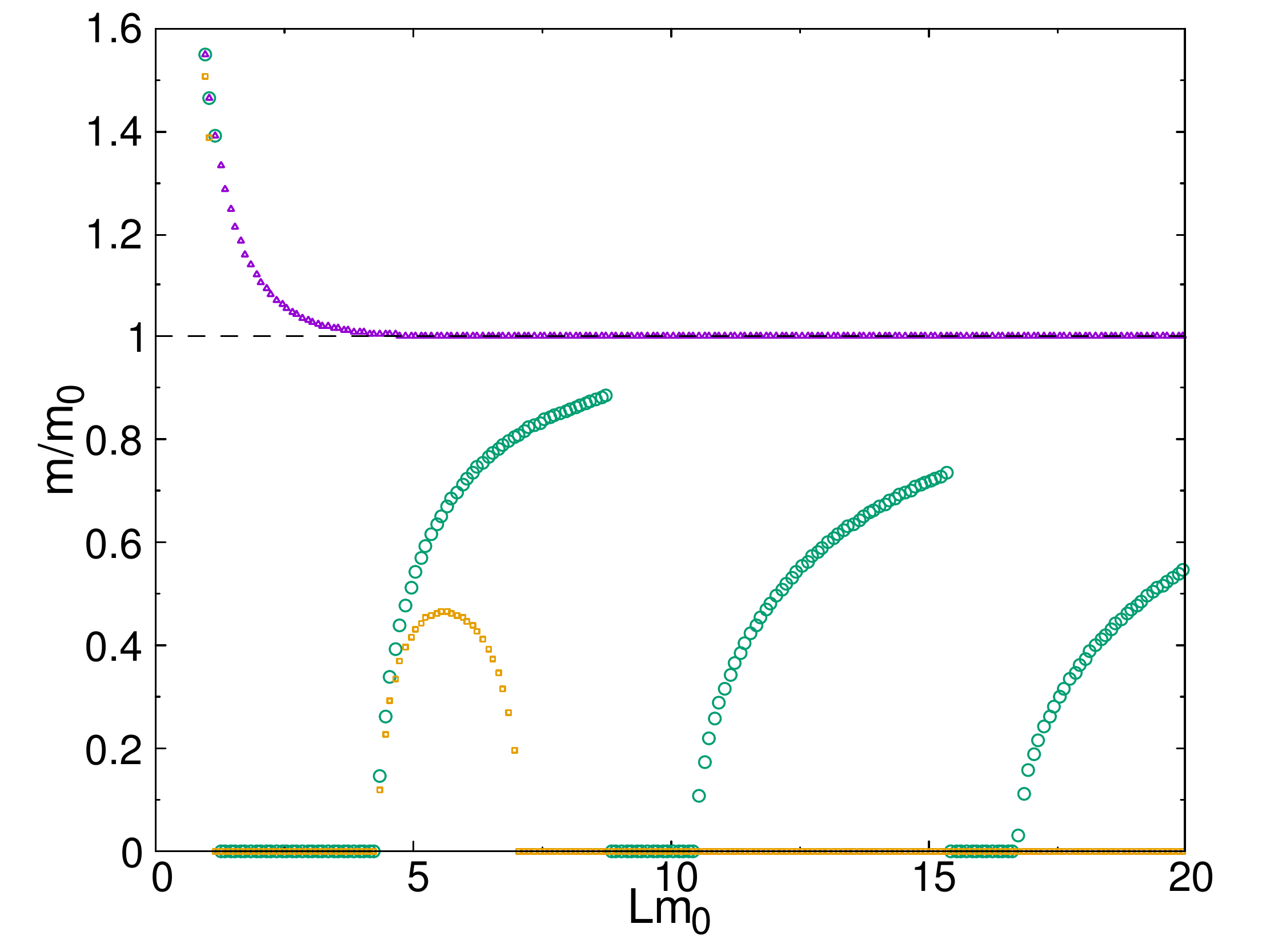}
       \end{overpic}
       \centering{\footnotesize $\delta = 0.0$}
      \end{minipage}
      \hspace{-2em}\vspace{-1em}
      &
      \begin{minipage}{0.5\hsize}
       \begin{overpic}[width=1\hsize,clip]
        {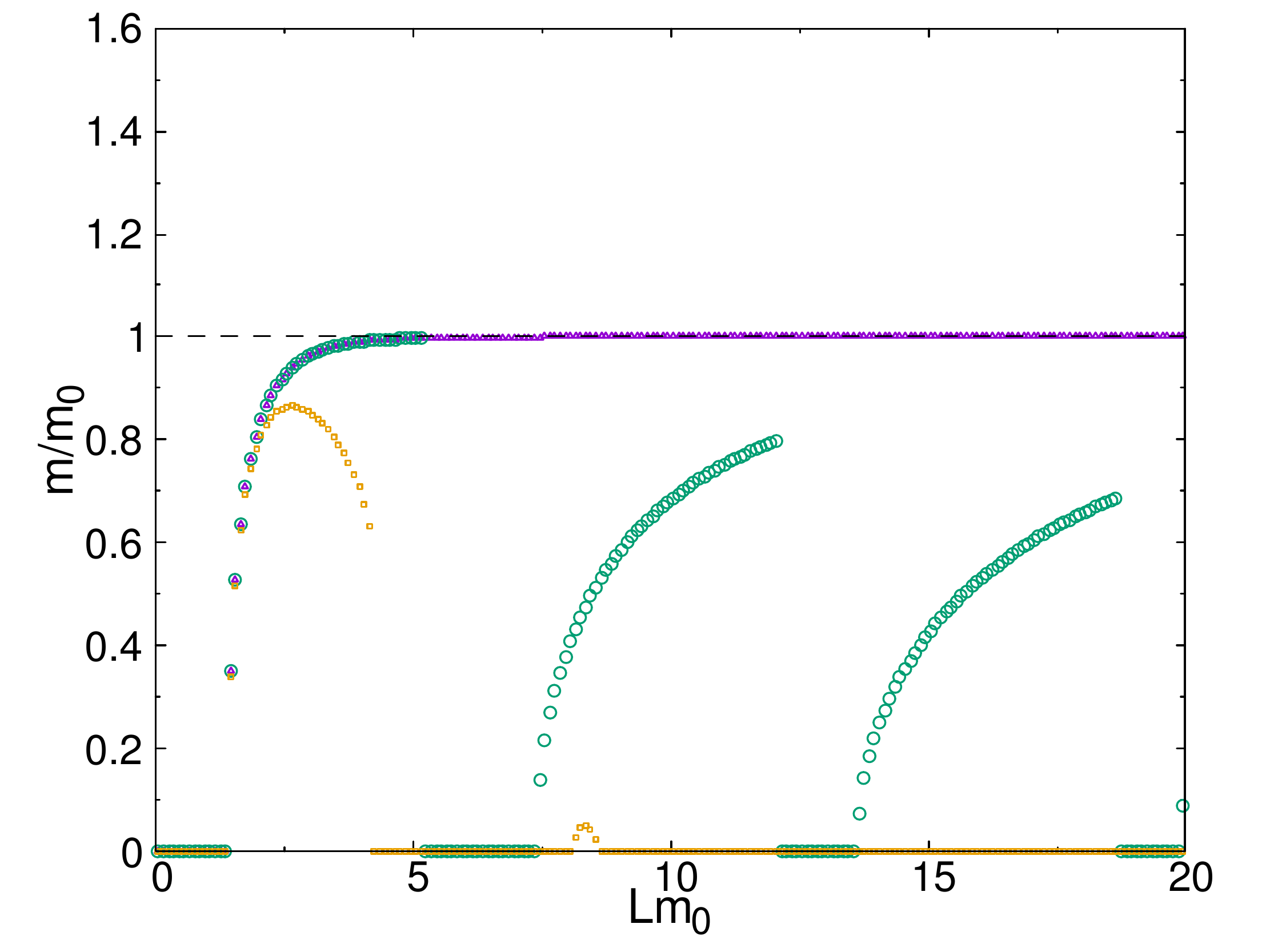}
       \end{overpic}
       \centering{\footnotesize $\delta = 1.0$}
      \end{minipage}
     \end{tabular}
    \end{center}
   \end{minipage}
  \end{tabular}
 \end{center}
 \caption{Dynamically generated fermion mass as a function of $\delta$ (left) and $L$ (right) in 3 dimensions:
 $(T/m_0, \mu/m_0; \mbox{color}) = (0.005, 0.0; \mbox{purple}), \, (0.005, 1.0; \mbox{green}), \, (0.2, 1.0; \mbox{yellow})$.}
 \label{fig:dynamically-generated-mass:3dim}
\end{figure}

The dynamically generated fermion mass, $m$, satisfies the gap equation,
\begin{align}
 \left. \frac{\partial V_D\left(\sigma; L, \delta, \beta, \mu\right)}{\partial \sigma}\right|_{\sigma=m} = 0.
 \label{eq:gap-equation}
\end{align}
We show the dynamical mass as a function of the $\Unitary{1}$ phase and the length in Figs. \ref{fig:dynamically-generated-mass:2dim} and \ref{fig:dynamically-generated-mass:3dim}.
The dashed line ($m/m_0=1$) indicates the value at $L \to \infty$, $\beta \to \infty$ and $\mu=0$.
For $\mu=0$ (purple curves) the dynamical mass at $\delta=0.0$ is heavier than that at $\delta=1.0$.
In the two graphs on the left of Figs. \ref{fig:dynamically-generated-mass:2dim} and \ref{fig:dynamically-generated-mass:3dim} the dynamical mass vanishes and the chiral symmetry is restored around the periodic boundary condition ($\delta=0.0$) slightly below the critical chemical potential.

This situation depends on the size, $L$.
Observing the dynamical mass as a function of $L$ (green and yellow curves), we see that the broken and the symmetric phases alternate with increasing size of the system, $L$.
It is consistent with the phase diagrams \ref{fig:phase-diagram:L-T:2dim} and \ref{fig:phase-diagram:L-T:3dim}.
The dynamical mass changes more smoothly in three dimensions than in two dimensions, because of the continuous momentum.


\section{Thermodynamic quantities}\label{sec:therm-quant}
In the preceding sections, we have evaluated the effective potential and shown the phase structure and the dynamically generated fermion mass.
Other thermodynamic quantities are also derived from the effective potential.
Here we discuss particle number density and pressure.

\subsection{Grand potential}\label{sec:grand-potential}
A minimum value of the effective potential can be naively regarded as the density of the grand potential, denoted by $\Omega_D \left(L, \delta, T, \mu \right)$.
We set the value of the effective potential to zero under a homogeneous and non-vanishing chiral condensate, an infinite volume, zero temperature, and a zero chemical potential.
The definition of the grand potential that we consider is given by
\begin{align}
 \Omega_D \left(L, \delta, T, \mu \right)
 = \left[ V_D \left(m; L, \delta, \beta (=1/T), \mu\right) - V_D\left(m_0\right) \right] L \mathcal{V},
 \label{eq:grand-potential:def}
\end{align}
where $L\mathcal{V}$ denotes $(D-1)$-dimensional volume.

\begin{figure}[H]
 \begin{center}
  \begin{tabular}{cc}
   \hspace{-2em}
   \begin{minipage}{0.5\hsize}
    \begin{center}
     \begin{tabular}{cc}
      \begin{minipage}{0.5\hsize}
       \begin{overpic}[width=1\hsize,clip]
        {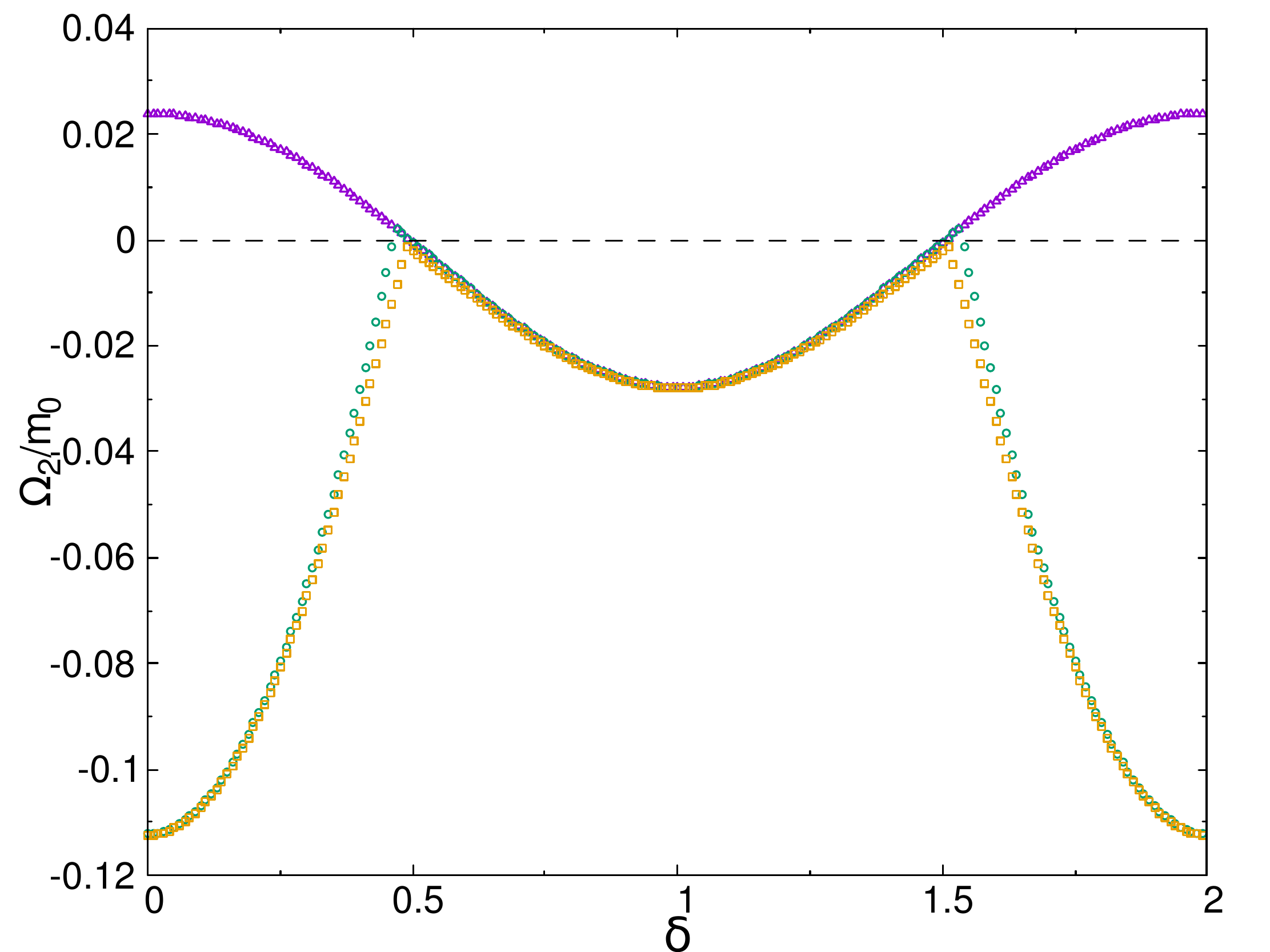}
       \end{overpic}
       \centering{\footnotesize $L m_0 = 3.0$}
      \end{minipage}
      \hspace{-2em}
      &
      \begin{minipage}{0.5\hsize}
       \begin{overpic}[width=1\hsize,clip]
        {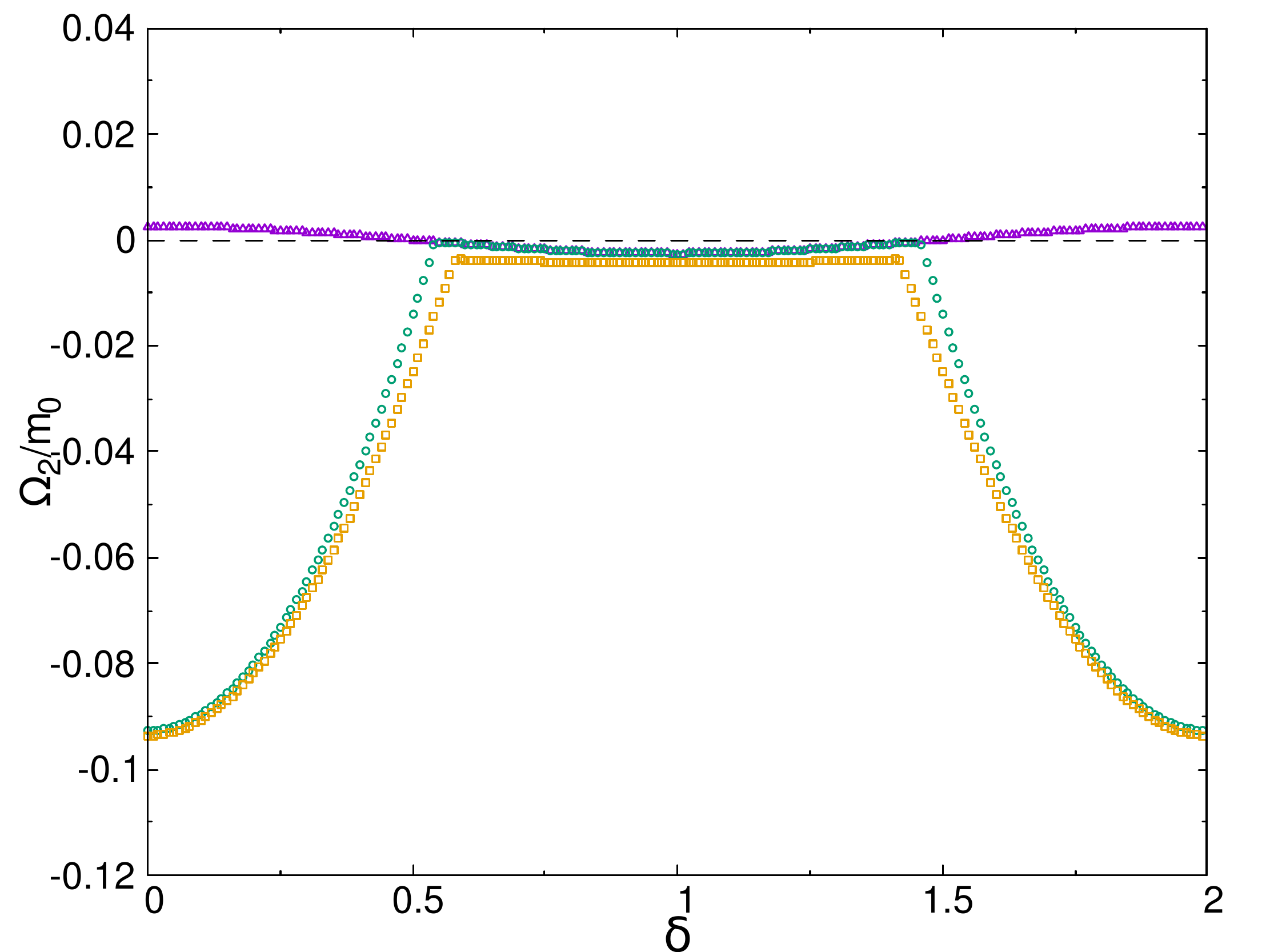}
       \end{overpic}
       \centering{\footnotesize $L m_0 = 5.0$}
      \end{minipage}
     \end{tabular}
    \end{center}
   \end{minipage}
   &
   \begin{minipage}{0.5\hsize}
    \begin{center}
     \begin{tabular}{cc}
      \begin{minipage}{0.5\hsize}
       \begin{overpic}[width=1\hsize,clip]
        {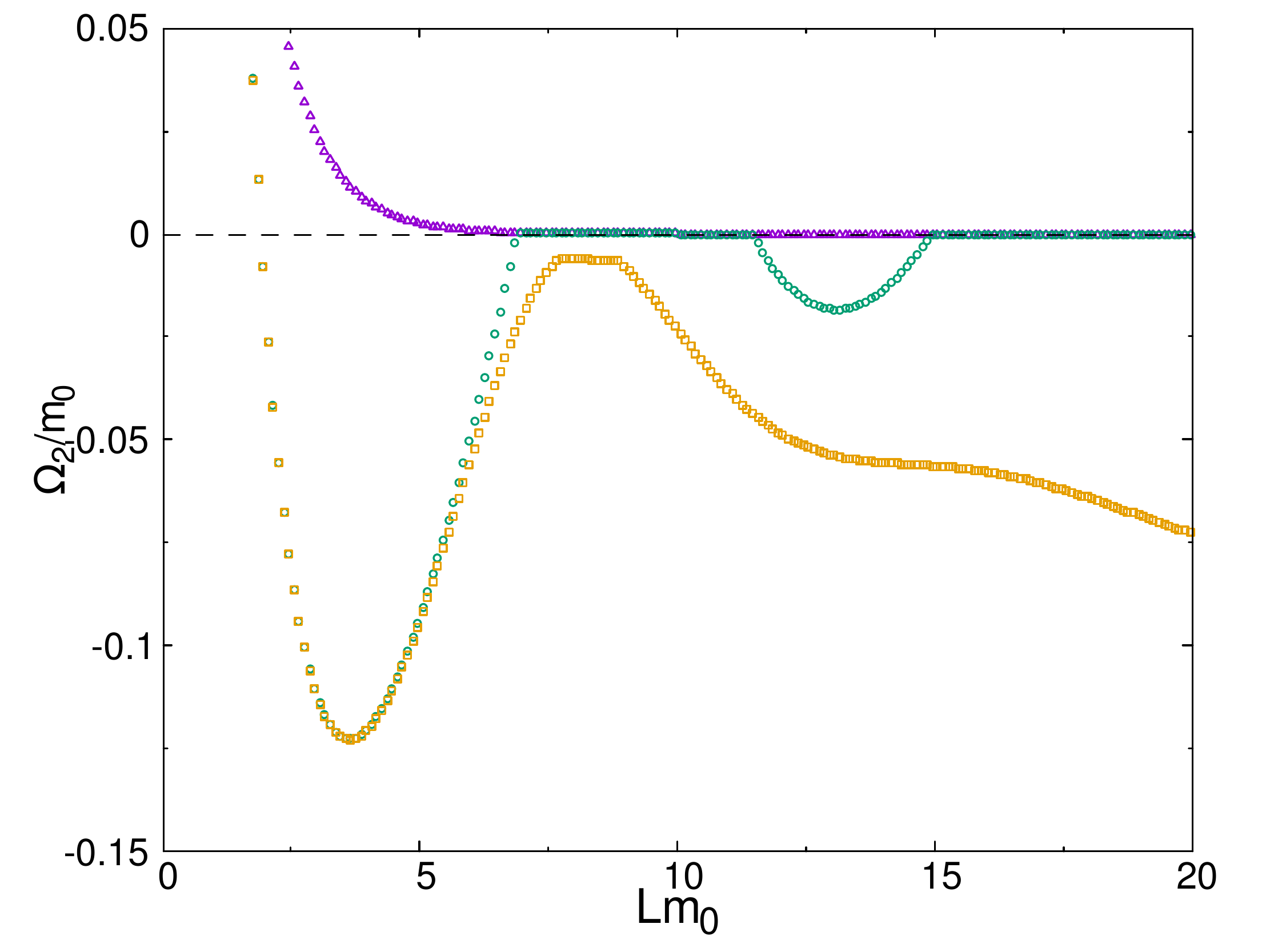}
       \end{overpic}
       \centering{\footnotesize $\delta = 0.0$}
      \end{minipage}
      \hspace{-2em}
      &
      \begin{minipage}{0.5\hsize}
       \begin{overpic}[width=1\hsize,clip]
        {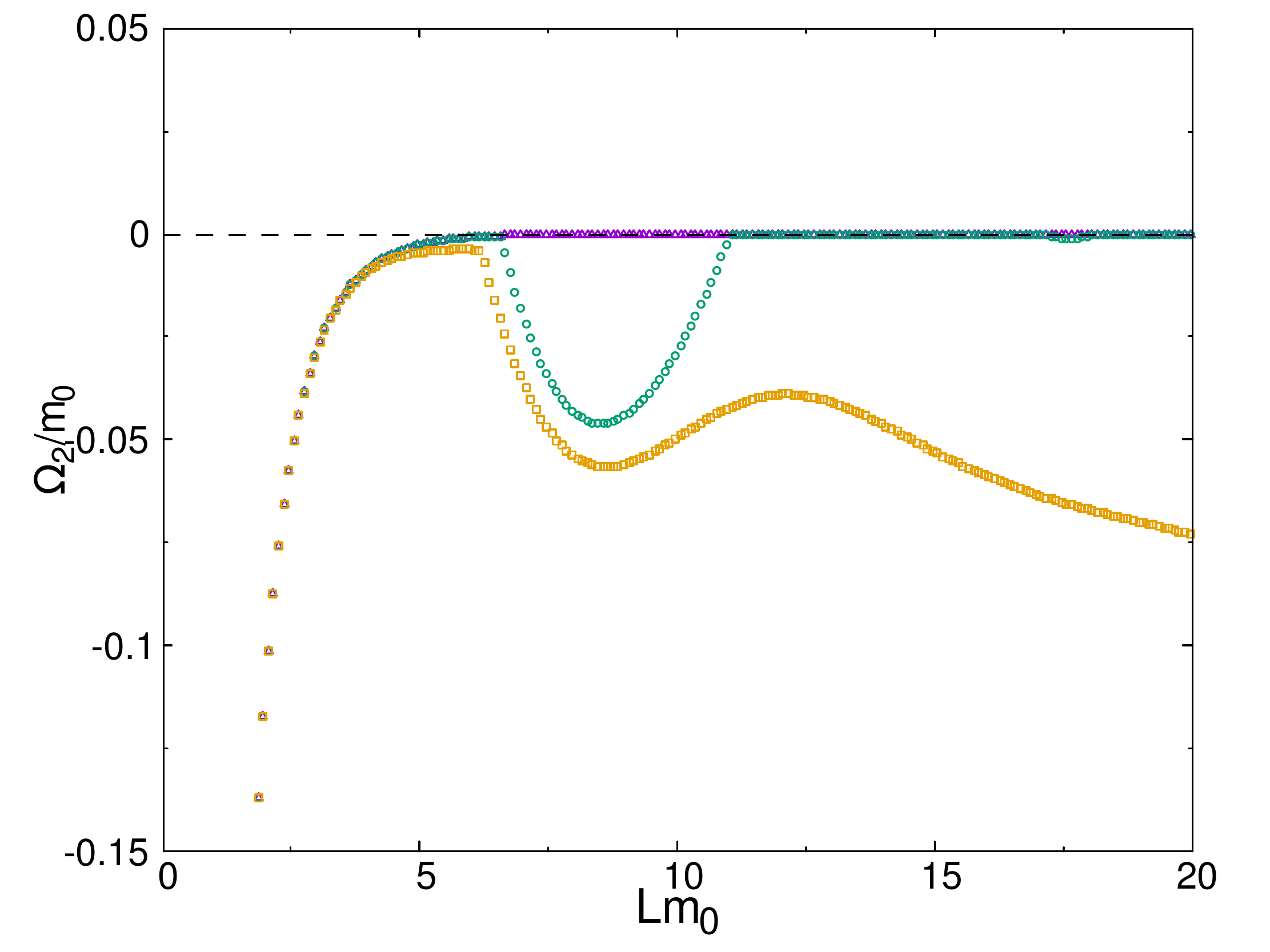}
       \end{overpic}
       \centering{\footnotesize $\delta = 1.0$}
      \end{minipage}
     \end{tabular}
    \end{center}
   \end{minipage}
  \end{tabular}
  \vspace{-2em}
 \end{center}
 \caption{Grand potential as a function of $\delta$ (left) and $L$ (right) in 2 dimensions:
 $(T/m_0, \mu/m_0; \mbox{color}) = (0.005, 0.0; \mbox{purple}), \, (0.005, 0.7; \mbox{green}), \, (0.1, 0.7; \mbox{yellow})$.}
 \label{fig:grand-potential:2d}
\end{figure}

\begin{figure}[H]
 \begin{center}
  \begin{tabular}{cc}
   \hspace{-2em}
   \begin{minipage}{0.5\hsize}
    \begin{center}
     \begin{tabular}{cc}
      \begin{minipage}{0.5\hsize}
       \begin{overpic}[width=1\hsize,clip]
        {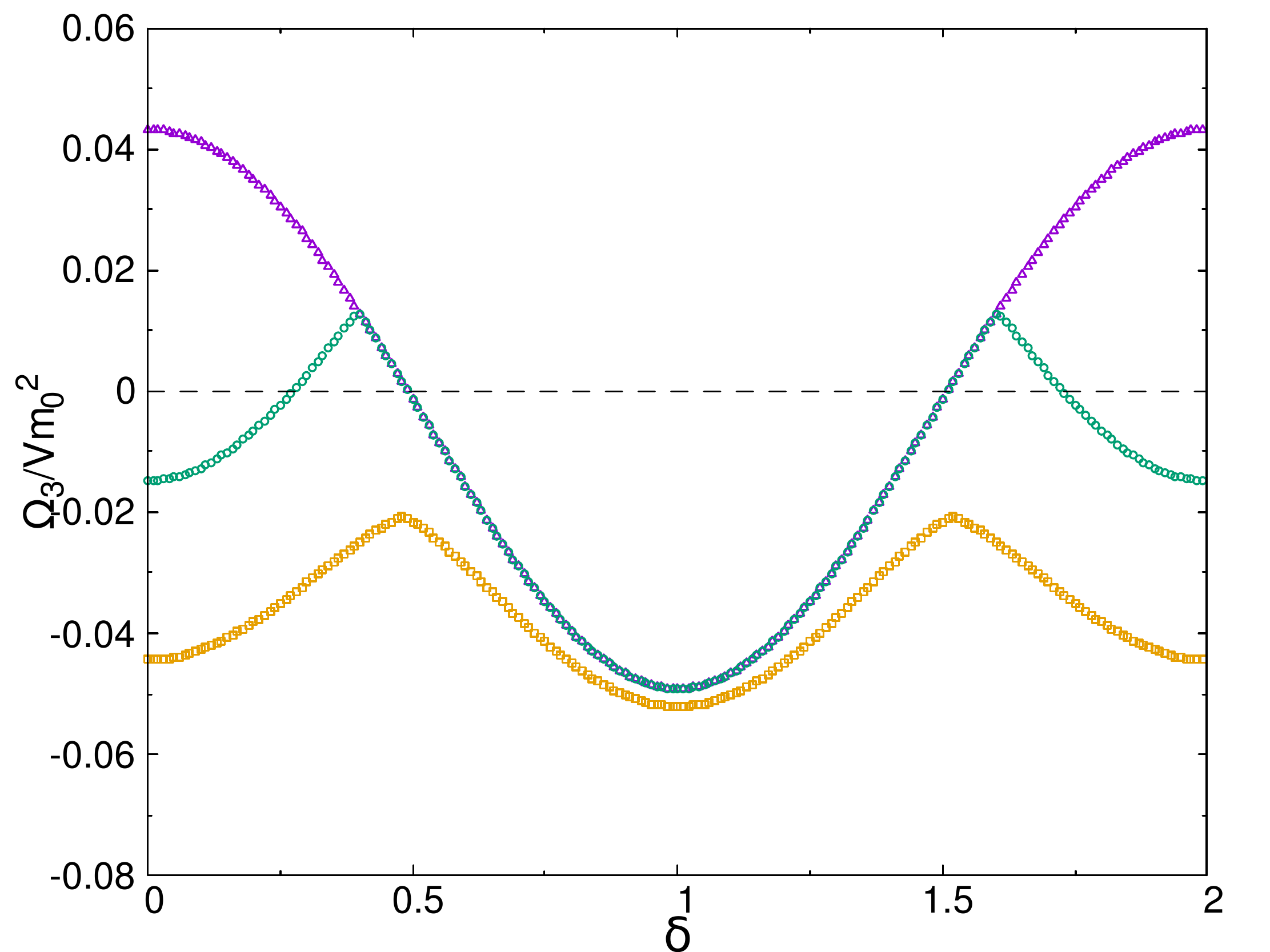}
       \end{overpic}
       \centering{\footnotesize $L m_0 = 2.0$}
      \end{minipage}
      \hspace{-2em}
      &
      \begin{minipage}{0.5\hsize}
       \begin{overpic}[width=1\hsize,clip]
        {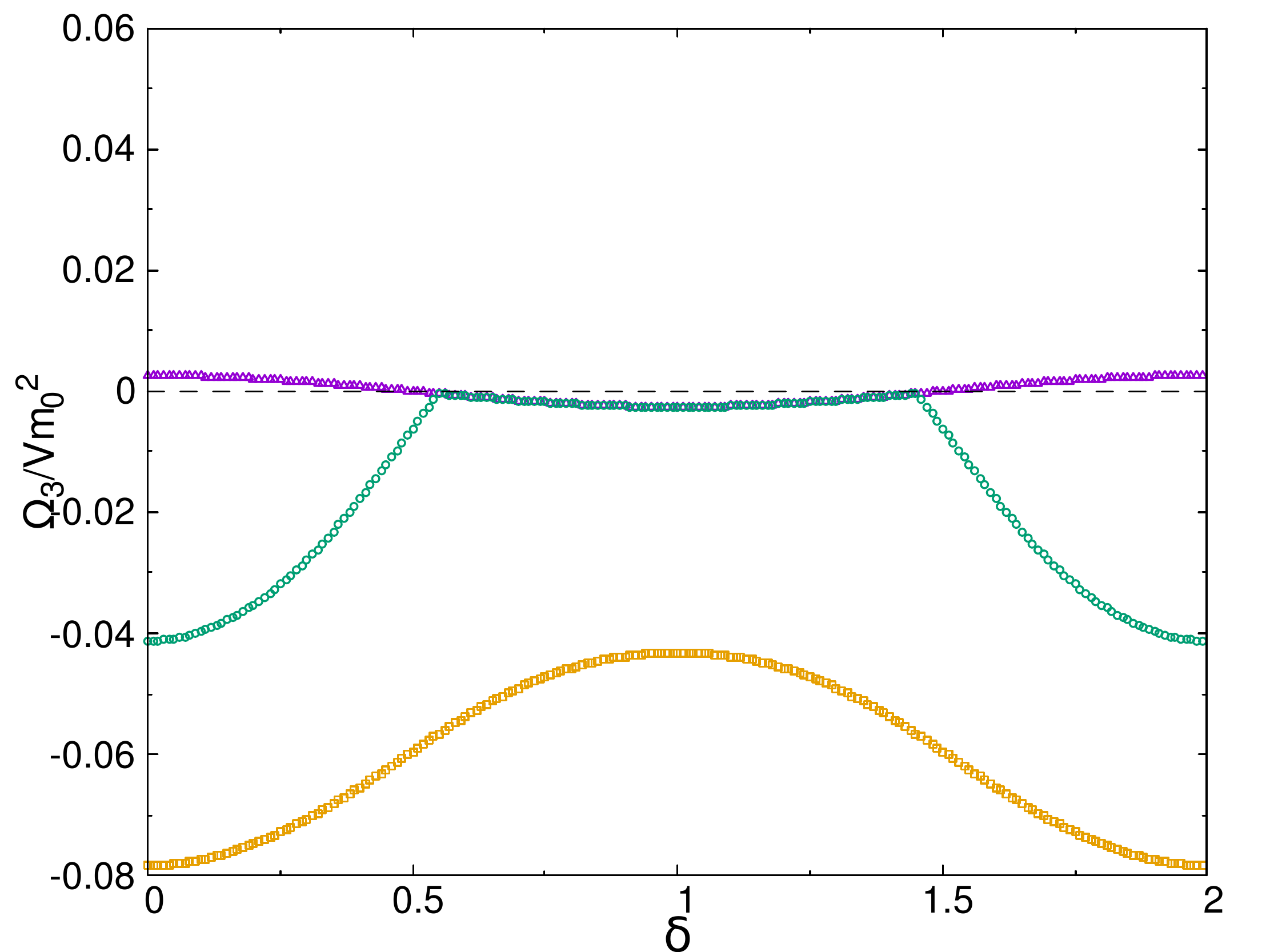}
       \end{overpic}
       \centering{\footnotesize $L m_0 = 4.0$}
      \end{minipage}
     \end{tabular}
    \end{center}
   \end{minipage}
   &
   \begin{minipage}{0.5\hsize}
    \begin{center}
     \begin{tabular}{cc}
      \begin{minipage}{0.5\hsize}
       \begin{overpic}[width=1\hsize,clip]
        {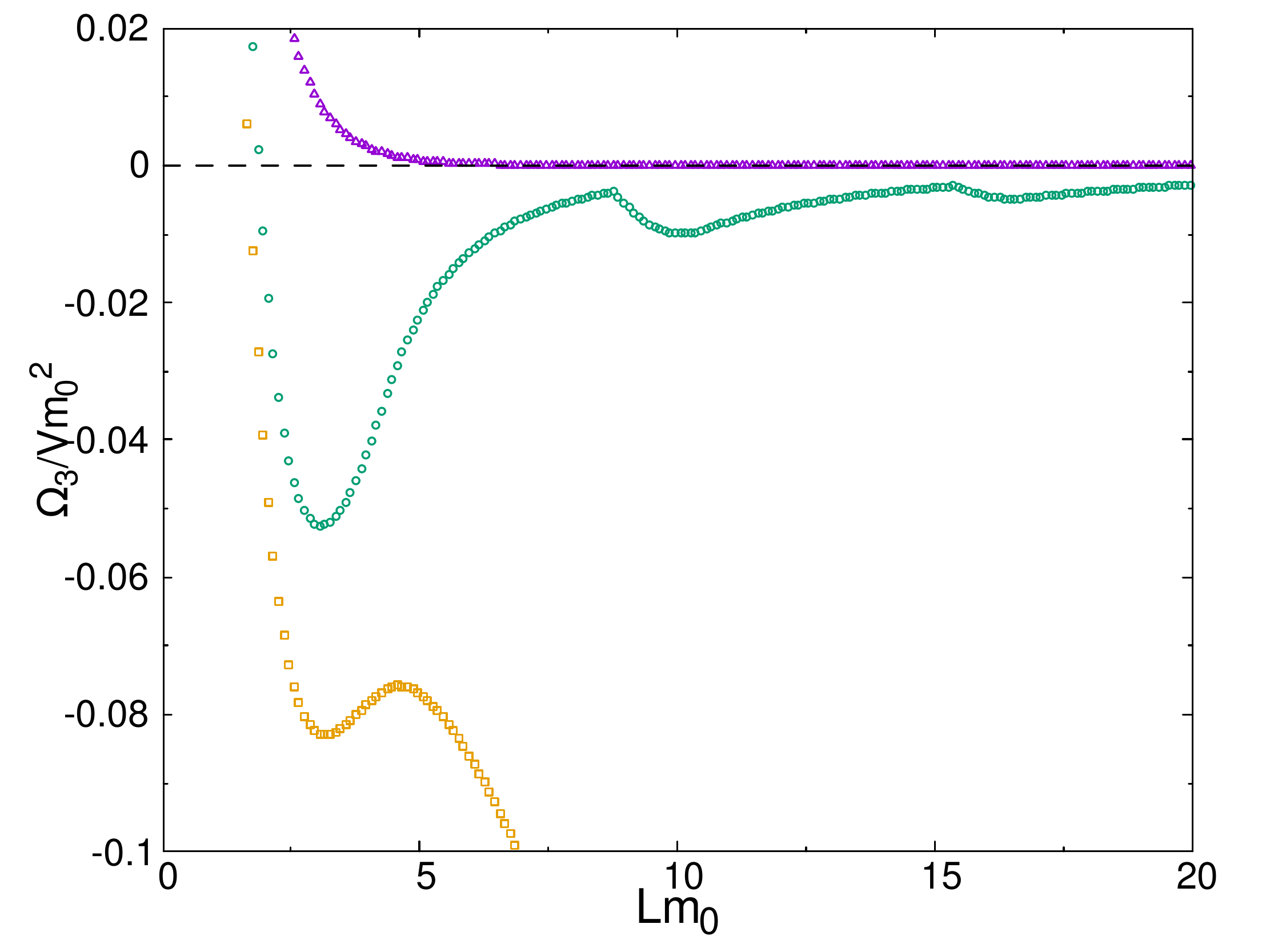}
       \end{overpic}
       \centering{\footnotesize $\delta = 0.0$}
      \end{minipage}
      \hspace{-2em}
      &
      \begin{minipage}{0.5\hsize}
       \begin{overpic}[width=1\hsize,clip]
        {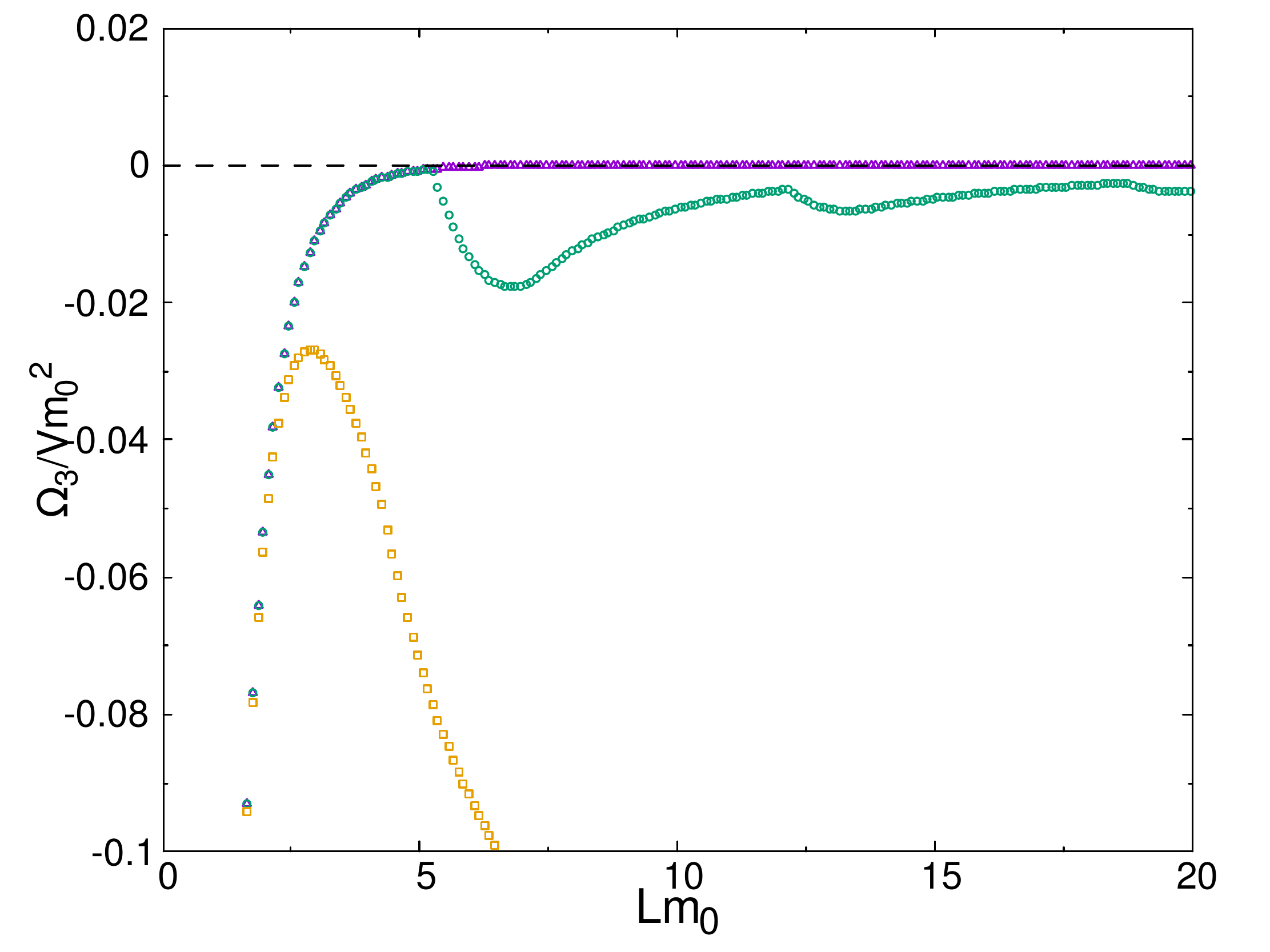}
       \end{overpic}
       \centering{\footnotesize $\delta = 1.0$}
      \end{minipage}
     \end{tabular}
    \end{center}
   \end{minipage}
  \end{tabular}
  \vspace{-2em}
 \end{center}
 \caption{Grand potential as a function of $\delta$ (left) and $L$ (right) in 3 dimensions:
 $(T/m_0, \mu/m_0; \mbox{color}) = (0.005, 0.0; \mbox{purple}), \, (0.005, 1.0; \mbox{green}), \, (0.2, 1.0; \mbox{yellow})$.}
 \label{fig:grand-potential:3d}
\end{figure}

The grand potential is shown in Figs. \ref{fig:grand-potential:2d} and \ref{fig:grand-potential:3d} as a function of the boundary condition, $\delta$, and the size, $L$.
Sharp bends are observed at the same parameters where the mass jumps appear in Figs. \ref{fig:dynamically-generated-mass:2dim} and \ref{fig:dynamically-generated-mass:3dim}.
Thus, the phase boundary is found by observing the sharp bends in Figs. \ref{fig:grand-potential:2d} and \ref{fig:grand-potential:3d}.

A stable size is a state for which the pressure is zero and becomes negative (positive) for a larger (smaller) size.
We find the existence of a stable size for a finite chemical potential at low temperature.
For a finite chemical potential (green curve), it is observed that the grand potential at $\delta=0.0$ is minimized at $Lm_0 \sim 4.0$.
This size is realized in the chirally symmetric phase.
This stable state disappears at $\mu=0$, because the grand potential (purple curves) is a monotonic function of $L$ and divergent at the small $L$ limit.

\subsection{Particle number density}\label{sec:part-numb-dens}
\begin{figure}[H]
 \begin{center}
  \begin{tabular}{cc}
   \hspace{-2em}
   \begin{minipage}{0.5\hsize}
    \begin{center}
     \begin{tabular}{cc}
      \begin{minipage}{0.5\hsize}
       \begin{overpic}[width=1\hsize,clip]
        {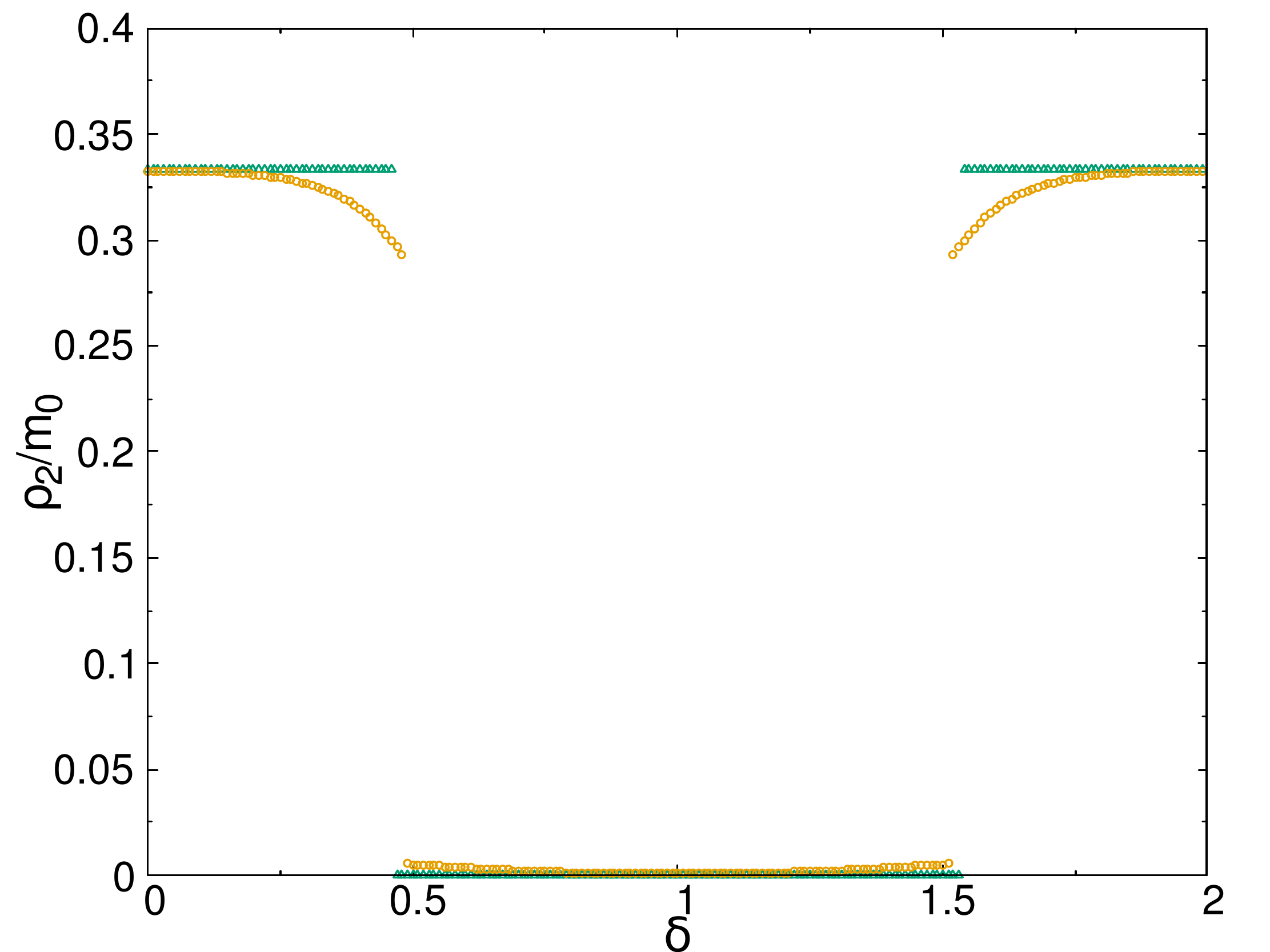}
       \end{overpic}
       \centering{\footnotesize $L m_0 = 3.0$}
      \end{minipage}
      \hspace{-2em}
      &
      \begin{minipage}{0.5\hsize}
       \begin{overpic}[width=1\hsize,clip]
        {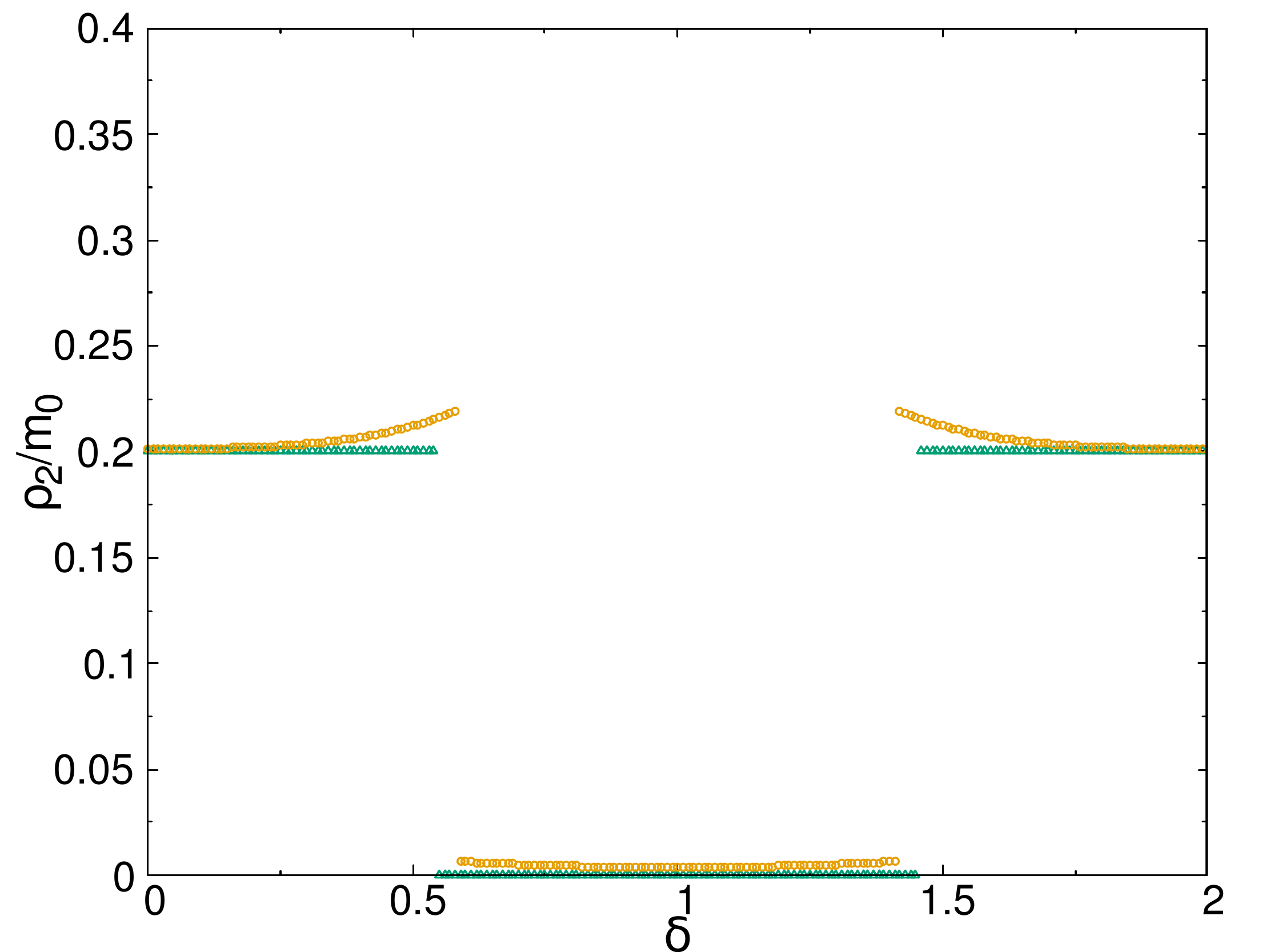}
       \end{overpic}
       \centering{\footnotesize $L m_0 = 5.0$}
      \end{minipage}
     \end{tabular}
    \end{center}
   \end{minipage}
   &
   \begin{minipage}{0.5\hsize}
    \begin{center}
     \begin{tabular}{cc}
      \begin{minipage}{0.5\hsize}
       \begin{overpic}[width=1\hsize,clip]
        {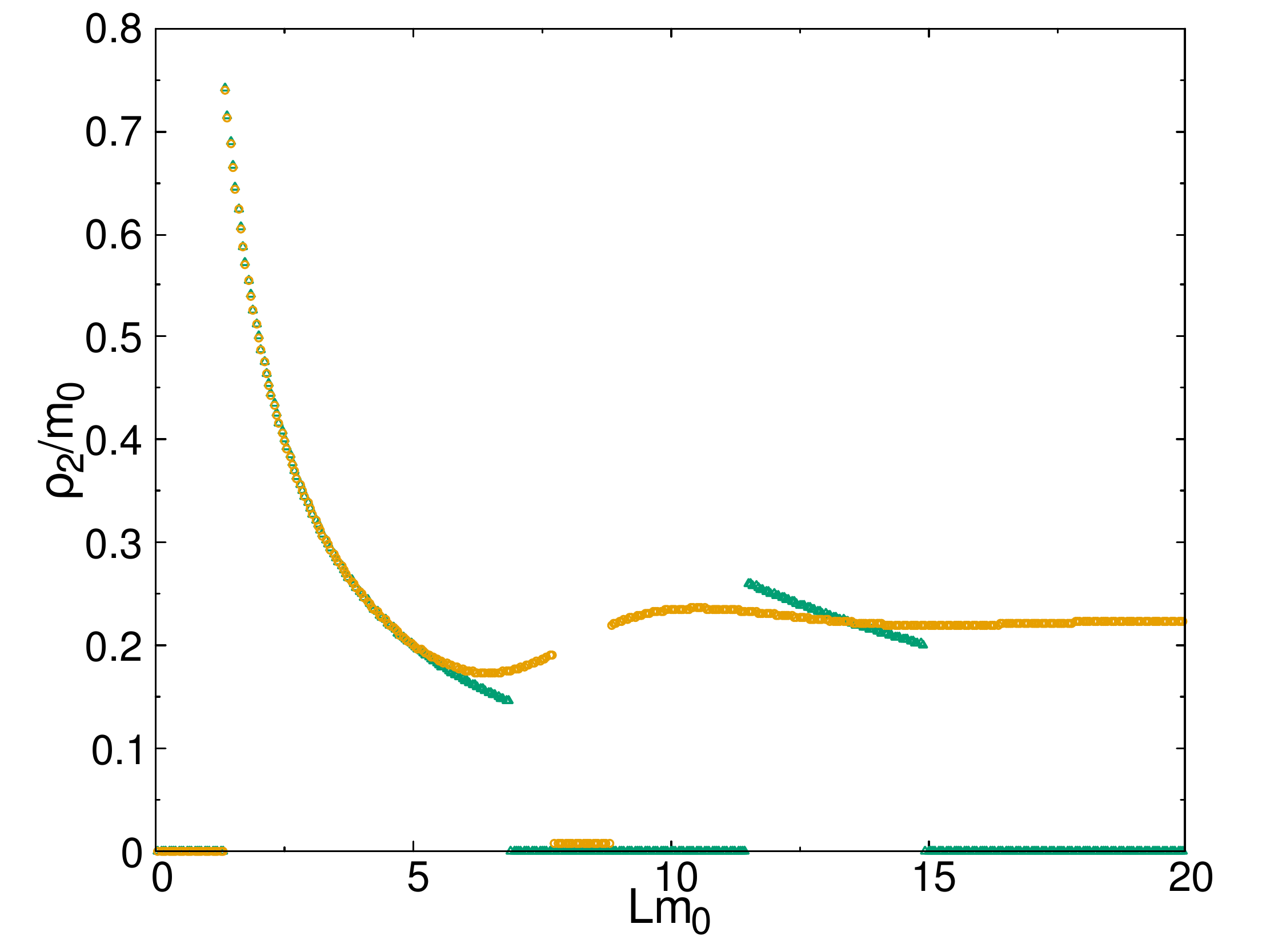}
       \end{overpic}
       \centering{\footnotesize $\delta = 0.0$}
      \end{minipage}
      \hspace{-2em}
      &
      \begin{minipage}{0.5\hsize}
       \begin{overpic}[width=1\hsize,clip]
        {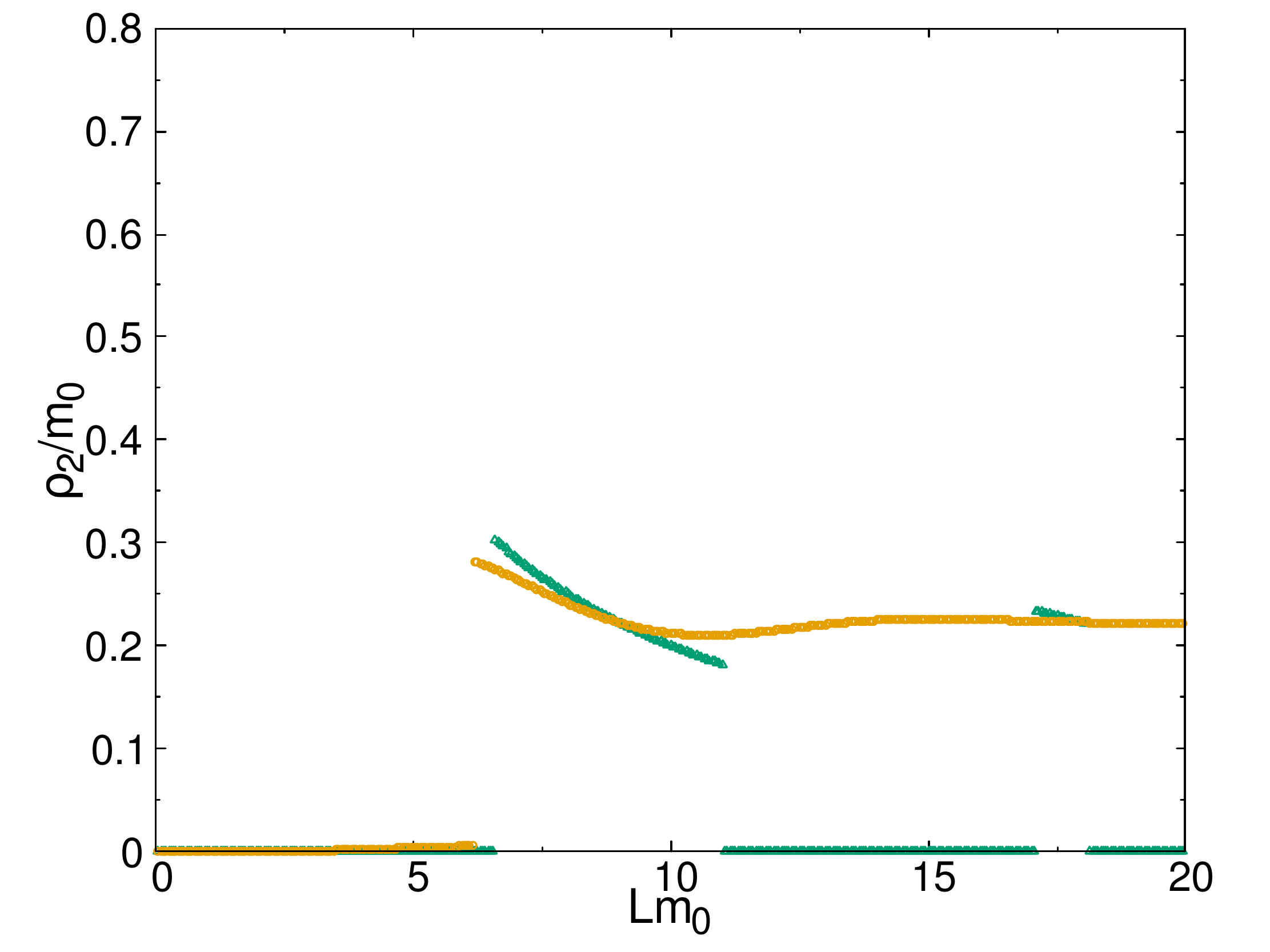}
       \end{overpic}
       \centering{\footnotesize $\delta = 1.0$}
      \end{minipage}
     \end{tabular}
    \end{center}
   \end{minipage}
  \end{tabular}
  \vspace{-2em}
 \end{center}
 \caption{Particle number density fixed to $\mu=0.7$ as a function of $\delta$ (left) and $L$ (right) in 2 dimensions.
 The green and yellow curve denotes $T/m_0 = 0.005$ and $T/m_0 = 0.1$ respectively.}
 \label{fig:particle-number-density:2d}
\end{figure}

To investigate the origin of the complex behavior of the phase diagrams, we study the contribution of the chemical potential.
The particle number density is defined by the derivative of the grand potential with respect to the chemical potential,
\begin{align}
 \rho_D \left(L, \delta, T, \mu \right)
 = - \frac{1}{\mathcal{V} L} \frac{ \partial \Omega_D \left(L, \delta, T, \mu \right)}{\partial \mu}.
 \label{eq:pnd:def}
\end{align}
We normalize the particle number density by $m_0$ and obtain
\begin{align}
 &\frac{\rho_D \left(L, \delta, T, \mu \right)}{m_0^{D-1}} \notag\\
 = &\frac{ \tr I}{C(D-1)}\frac{1}{L m_0}
 \sum_{n=-\infty}^{\infty} \int_0^\infty \frac{\dmeasure{}{q}}{m_0} \left( \frac{q}{m_0} \right)^{D-3} \hspace{-0.5em}\frac{\sinh \mu /T}{\cosh \sqrt{ q^2 + k_{\delta, n}^2 + m^2}/T + \cosh \mu/T}.
 \label{eq:pnd:d}
\end{align}
In two dimensions the expression reads
\begin{align}
 \frac{\rho_2 \left(L, \delta, T, \mu \right)}{m_0}
 & = \frac{1}{L m_0} \sum_{n=-\infty}^{\infty} \frac{\sinh \mu/T }{\cosh \sqrt{ k_{\delta, n}^2 + m^2}/T + \cosh \mu/T}.
 \label{eq:pdn:2d}
\end{align}
The numerical results are shown in Fig. \ref{fig:particle-number-density:2d} as a function of $\delta$ and $L$.
At the limit $T \to 0$,
\begin{equation}
 \frac{\sinh \mu/T }{\cosh \sqrt{ k_{\delta, n}^2 + m^2}/T + \cosh \mu/T}
  \longrightarrow
  \sgn(\mu) \theta\left(\left|\mu\right| - \sqrt{k_{\delta, n}^2 + m^2} \right),
\end{equation}
a non-zero lower bound appears for $k_{\delta, n}^2$, except for the periodic boundary condition.
This implies that the particle number, $\rho_2(L, \delta, T, \mu)L$, becomes an integer at $T \to 0$.
Because of the degeneracy of the states, the possible values are restricted to $0, 1, 3, 5, \dots$ and $0, 2, 4, \dots$ for $\delta= 0.0$ and $1.0$ respectively.

\begin{figure}[H]
 \begin{center}
  \begin{tabular}{cc}
   \hspace{-2em}
   \begin{minipage}{0.5\hsize}
    \begin{center}
     \begin{tabular}{cc}
      \begin{minipage}{0.5\hsize}
       \begin{overpic}[width=1\hsize,clip]
        {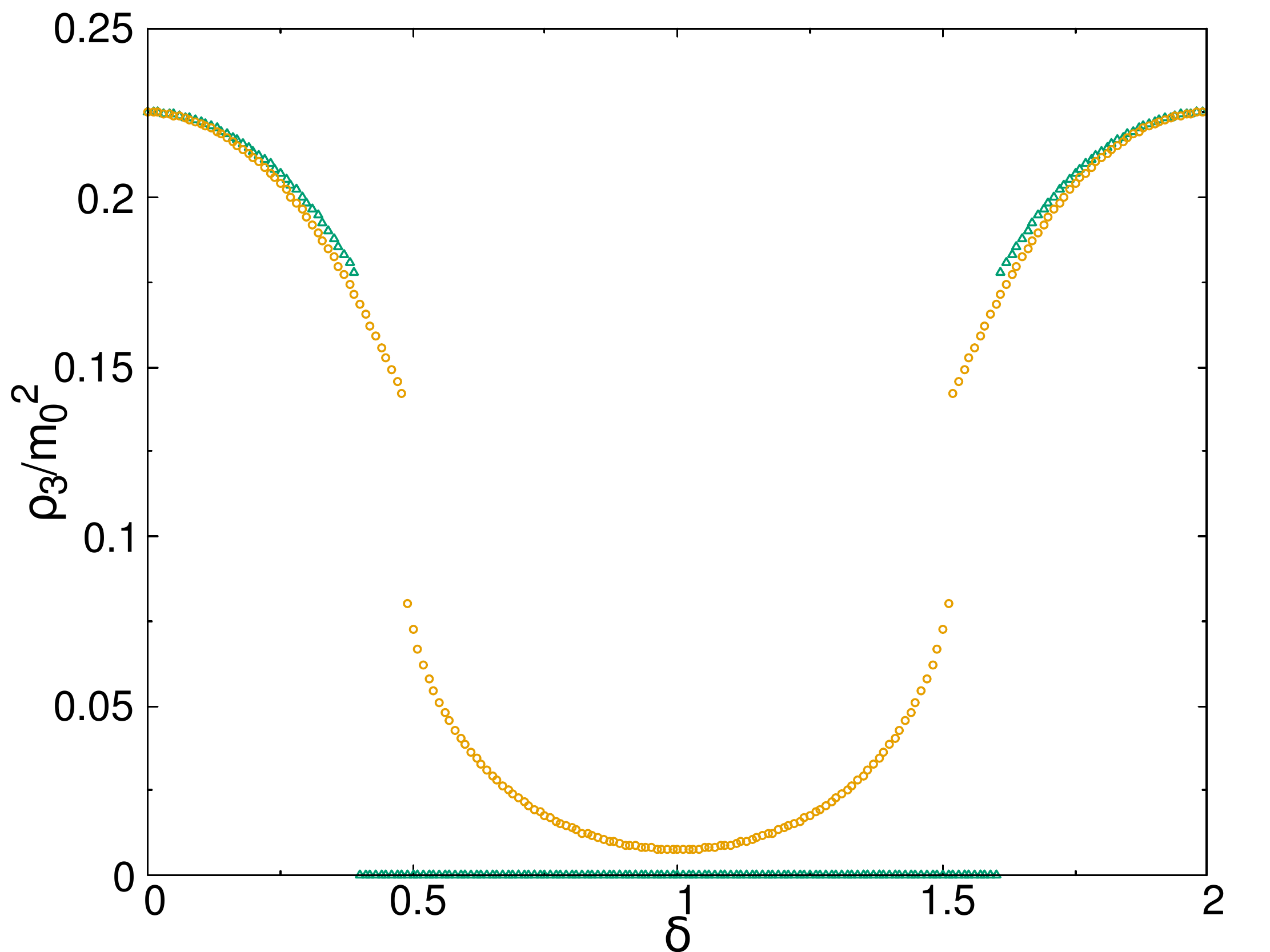}
       \end{overpic}
       \centering{\footnotesize $L m_0 = 2.0$}
      \end{minipage}
      \hspace{-2em}
      &
      \begin{minipage}{0.5\hsize}
       \begin{overpic}[width=1\hsize,clip]
        {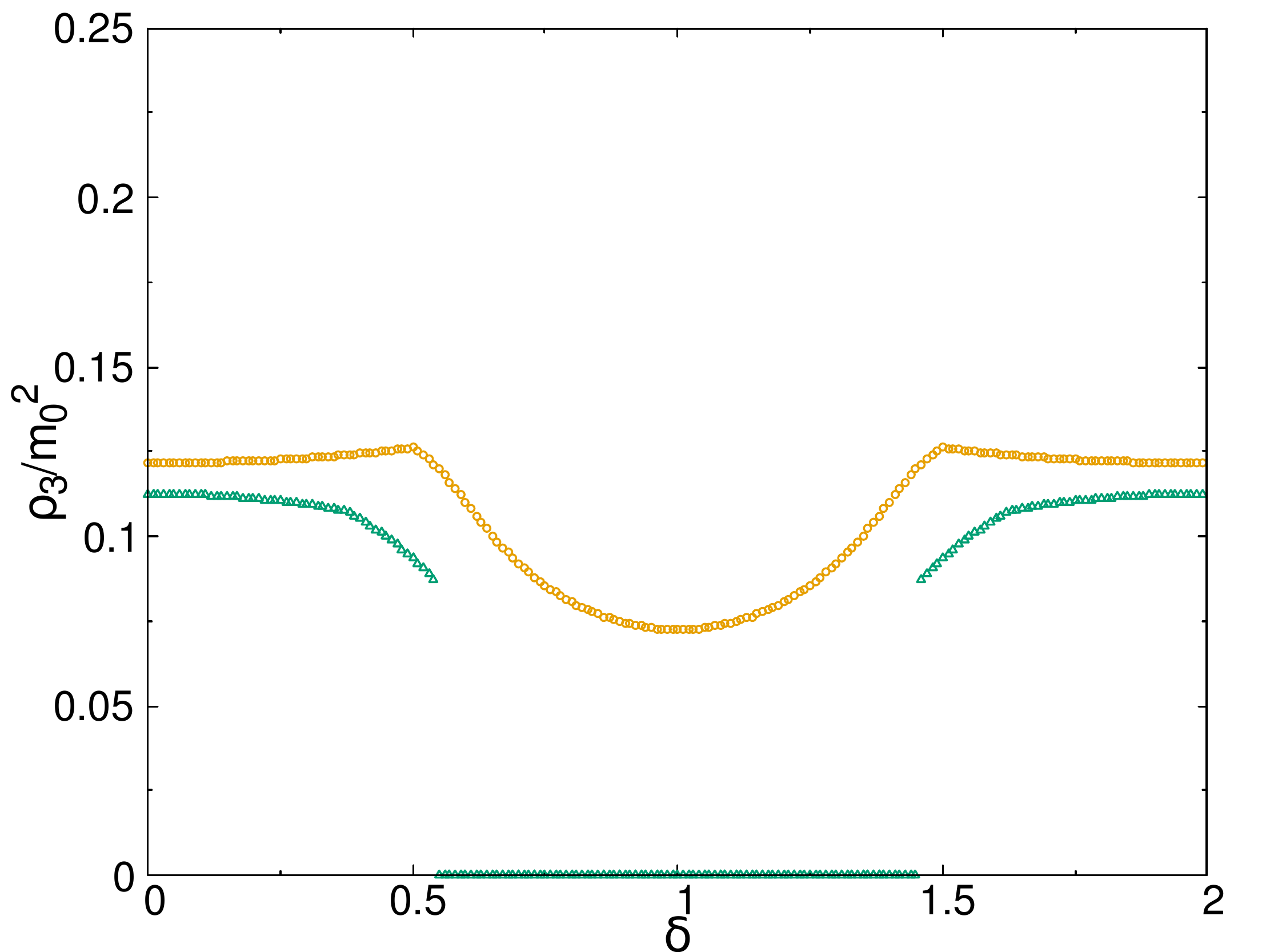}
       \end{overpic}
       \centering{\footnotesize $L m_0 = 4.0$}
      \end{minipage}
     \end{tabular}
    \end{center}
   \end{minipage}
   &
   \begin{minipage}{0.5\hsize}
    \begin{center}
     \begin{tabular}{cc}
      \begin{minipage}{0.5\hsize}
       \begin{overpic}[width=1\hsize,clip]
        {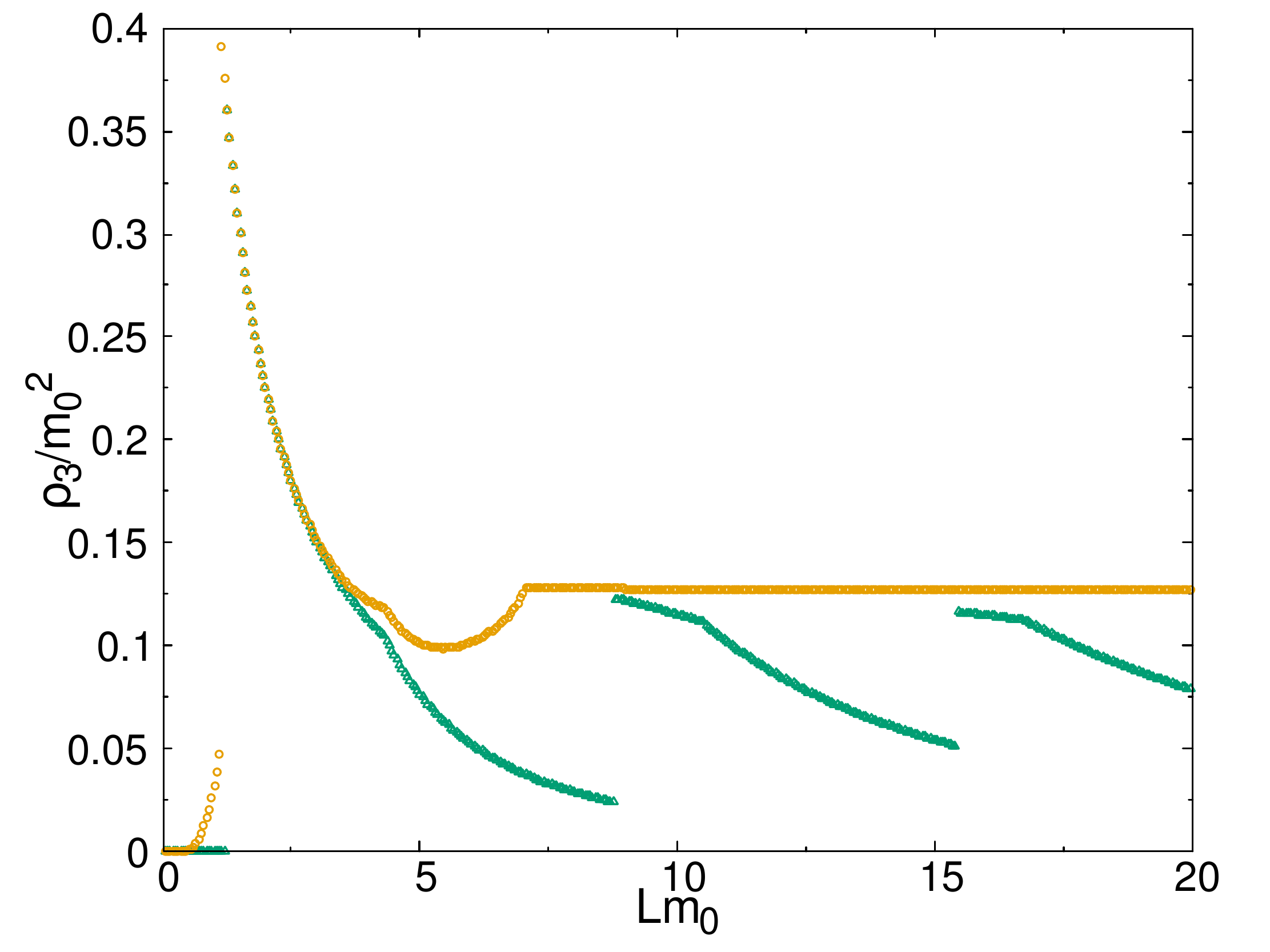}
       \end{overpic}
       \centering{\footnotesize $\delta = 0.0$}
      \end{minipage}
      \hspace{-2em}
      &
      \begin{minipage}{0.5\hsize}
       \begin{overpic}[width=1\hsize,clip]
        {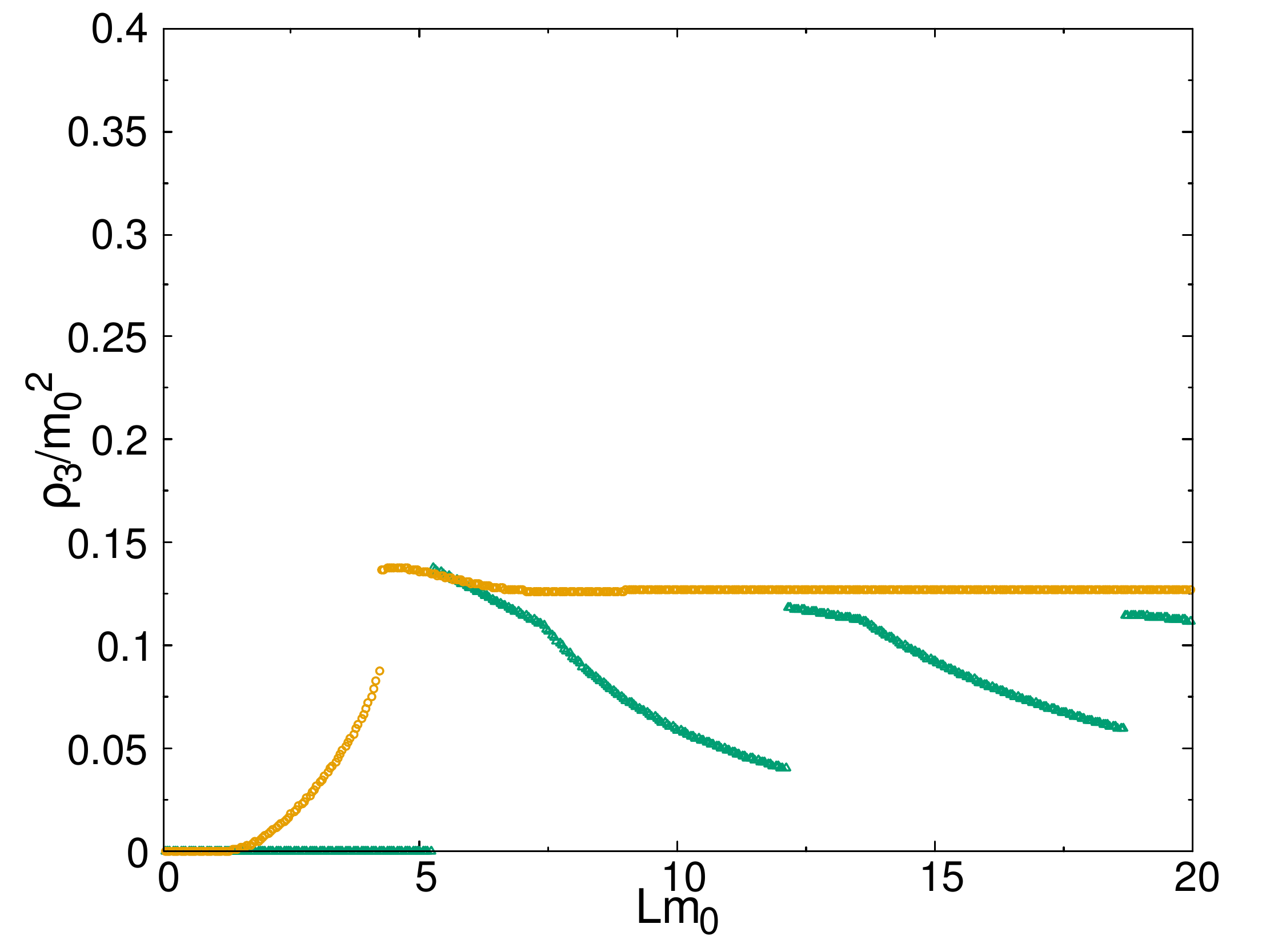}
       \end{overpic}
       \centering{\footnotesize $\delta = 1.0$}
      \end{minipage}
     \end{tabular}
    \end{center}
   \end{minipage}
  \end{tabular}
  \vspace{-2em}
 \end{center}
 \caption{Particle number density fixed to $\mu=1.0$ as a function of $\delta$ (left) and $L$ (right) in 3 dimensions.
 The green and yellow curve denotes $T/m_0 = 0.005$ and $T/m_0 = 0.2$ respectively.}
 \label{fig:particle-number-density:3d}
\end{figure}

We find the correspondences between Figs. \ref{fig:dynamically-generated-mass:2dim} and \ref{fig:particle-number-density:2d}.
The particle number density vanishes in the broken phase for $m > 0$.
In the symmetric phase, the particle number density vanishes because the non-zero lower bound for $k_{\delta=1.0, n}^2$ extremely suppresses the summation in Eq. \eqref{eq:pdn:2d} for $Lm_0 \lesssim 1.7$ at $\delta=1.0$.
At $T/m_0=0.1$ we observe pre-transitional phenomena because of the finite temperature effect.
As is shown in Fig. \ref{fig:particle-number-density:3d} at $D=3.0$, the continuous momentum for the additional dimension enhances the pre-transitional phenomena at a finite temperature and induces a finite number density near the phase boundary.
Through the analysis of the particle number density and the comparison with the dynamical mass, the complex behavior in the phase diagrams is caused by a balance between the particle production and the mass generation.

\subsection{Pressure}\label{sec:pressure}
In the preceding section, we have analyzed the particle number density and mentioned the correspondence between the particle number density and the dynamical mass.

Here we evaluate pressure to find a stable size.
The pressure is given by the derivative of the grand potential with respect to $L$,
\begin{equation}
 P_D\left(L, \delta, T, \mu\right)
 = - \frac{1}{\mathcal{V}} \frac{\partial \Omega_D\left(L, \delta, T, \mu\right)}{\partial L},
 \label{eq:pressure:def}
\end{equation}
and expressed as
\begin{align}
 \frac{P_D\left(L, \delta, T, \mu\right)}{m_0^D}
 &= - \frac{\tr I \cdot \Gammaf{1-\frac{D}{2}}}{\left(4 \pi\right)^{\frac{D}{2}}} \left[ \frac{1}{2} \left( \left(\frac{m}{m_0}\right)^2 -1 \right) - \frac{1}{D} \left( \left(\frac{m}{m_0}\right)^{2 \cdot \frac{D}{2}} -1 \right) \right]\notag\\
 & \hspace{1em} - \frac{\tr I}{C(D+1)}
 \int^\infty_0 \frac{\dmeasure{}{q}}{m_0} \left( \frac{q}{m_0}\right)^{D-2} \frac{\sqrt{q^2 + m^2}}{m_0} \frac{ \exp \left(-L\sqrt{q^2 + m^2}\right) - \cos \pi \delta }{\cosh L\sqrt{q^2 + m^2} - \cos \pi \delta} \notag\\
 & \hspace{1em} + \frac{\tr I}{C(D)} \frac{1}{L m_0}\sum_{n=-\infty}^\infty \int^\infty_0 \frac{\dmeasure{}{q}}{m_0} \left( \frac{q}{m_0}\right)^{D-3} \notag\\
 & \hspace{6em} \times \frac{k_{\delta, n}^2}{ m_0 \sqrt{k_{\delta, n}^2 + q^2 + m^2}} \frac{ \exp \left(-\sqrt{k_{\delta, n}^2 + q^2 + m^2}/T\right) + \cosh \mu/T }{\cosh \sqrt{k_{\delta, n}^2 + q^2 + m^2}/T + \cosh \mu/T}.
 \label{eq:pressure:d}
\end{align}
In two dimensions it reads
\begin{align}
 \frac{P_2\left(L, \delta, T, \mu\right)}{m_0^2}
 =& -\frac{1}{4\pi} \left[ 1 - \left( 1 - \ln\left(\frac{m}{m_0}\right)^2 \right) \left(\frac{m}{m_0}\right)^2 \right] \notag\\
 &- \frac{1}{\pi} \int^\infty_0 \frac{\dmeasure{}{q}}{m_0} \frac{\sqrt{q^2 + m^2}}{m_0} \frac{ \exp \left(-L\sqrt{q^2 + m^2}\right) - \cos \pi \delta }{\cosh L\sqrt{q^2 + m^2} - \cos \pi \delta} \notag\\
 &+ \frac{1}{L m_0} \sum_{n=-\infty}^\infty \frac{k_{\delta, n}^2}{ m_0 \sqrt{k_{\delta, n}^2 + m^2}} \frac{ \exp \left(-\sqrt{k_{\delta, n}^2 + m^2}/T\right) + \cosh \mu/T }{\cosh \sqrt{k_{\delta, n}^2 + m^2}/T + \cosh \mu/T}.
 \label{eq:pressure:2}
\end{align}

\begin{figure}[H]
 \begin{center}
  \begin{tabular}{cc}
   \hspace{-2em}
   \begin{minipage}{0.5\hsize}
    \begin{center}
     \begin{tabular}{cc}
      \begin{minipage}{0.5\hsize}
       \begin{overpic}[width=1\hsize,clip]
        {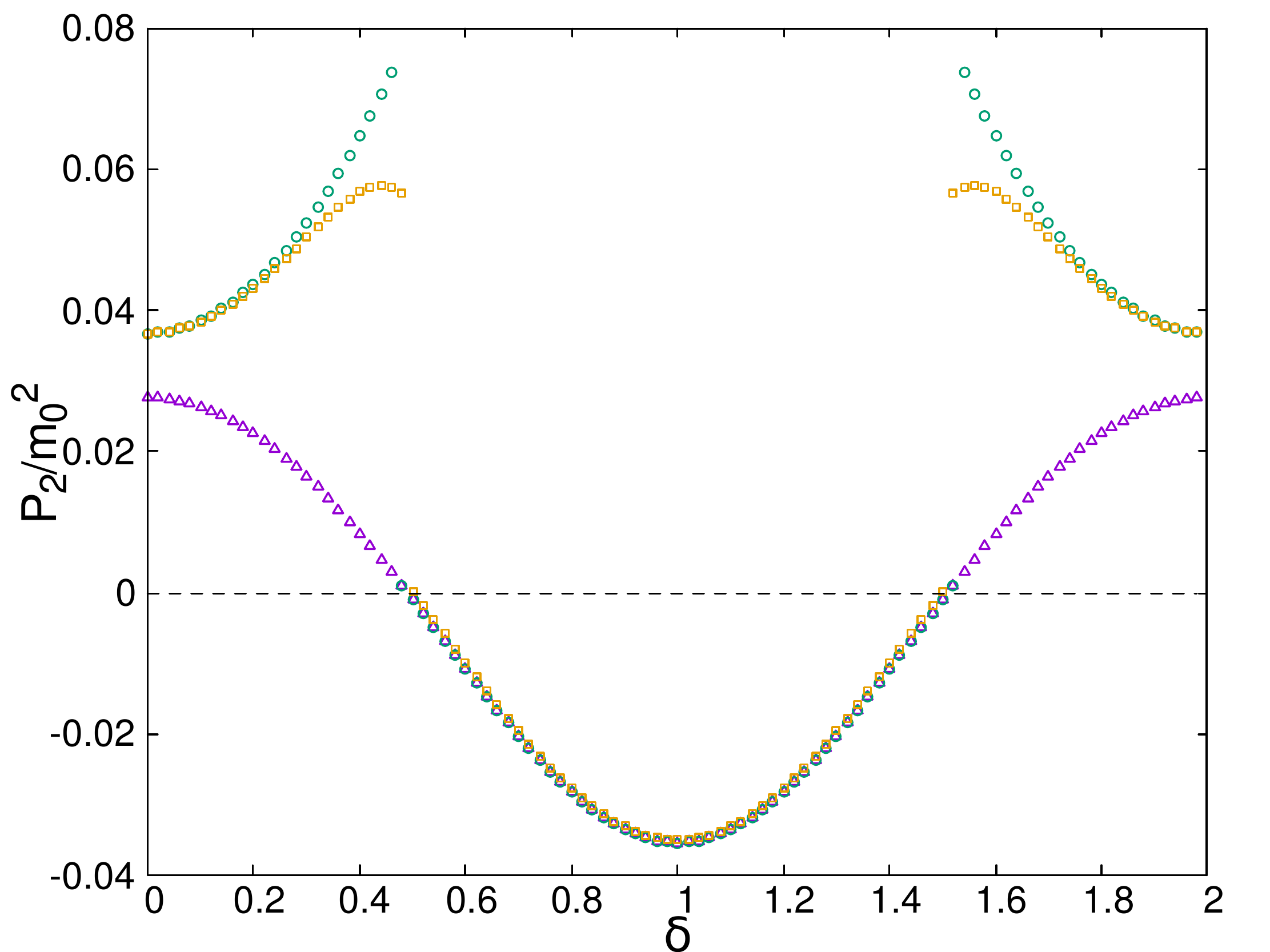}
       \end{overpic}
       \centering{\footnotesize $L m_0 = 3.0$}
      \end{minipage}
      \hspace{-2em}
      &
      \begin{minipage}{0.5\hsize}
       \begin{overpic}[width=1\hsize,clip]
        {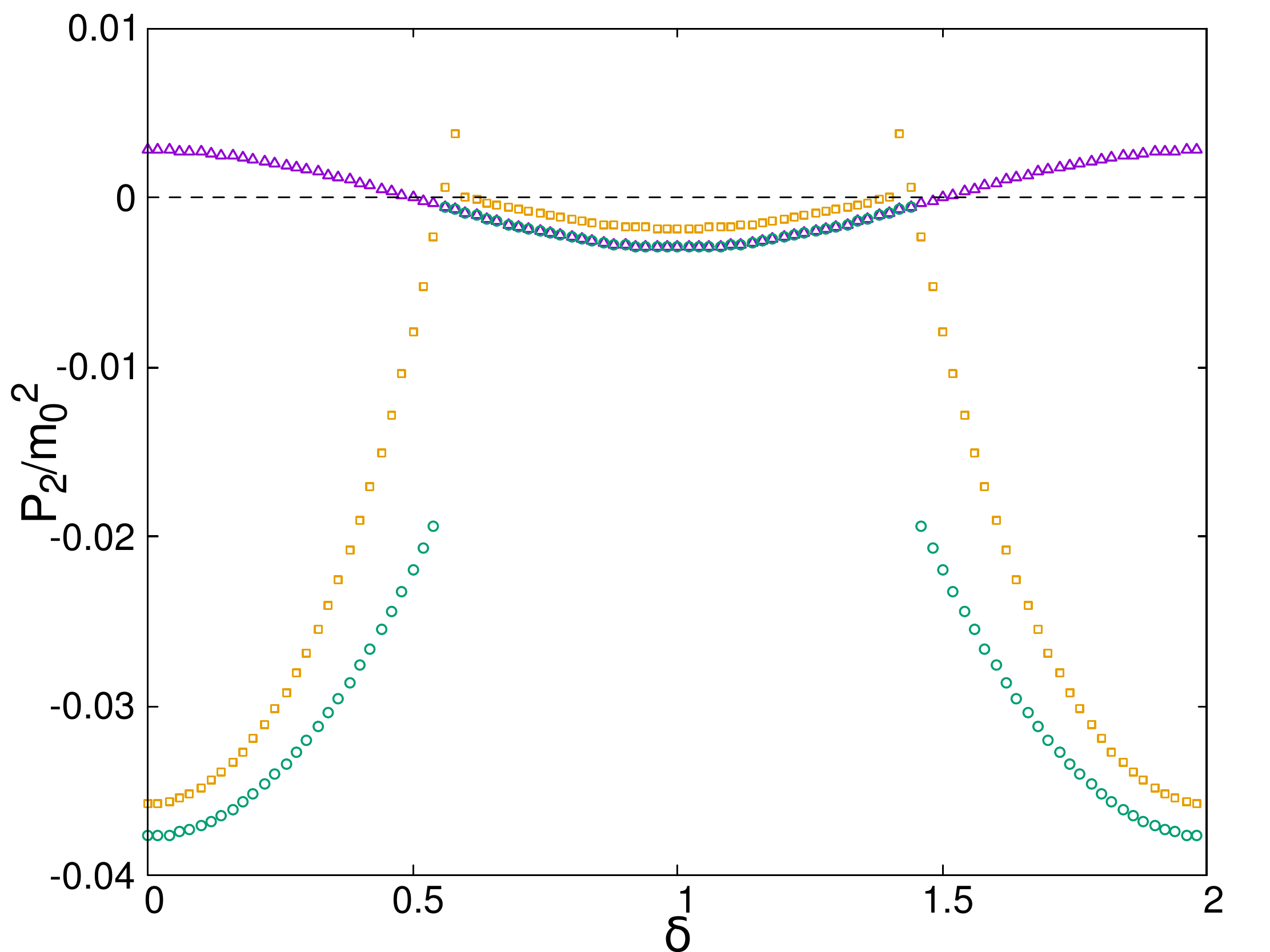}
       \end{overpic}
       \centering{\footnotesize $L m_0 = 5.0$}
      \end{minipage}
     \end{tabular}
    \end{center}
   \end{minipage}
   &
   \begin{minipage}{0.5\hsize}
    \begin{center}
     \begin{tabular}{cc}
      \begin{minipage}{0.5\hsize}
       \begin{overpic}[width=1\hsize,clip]
        {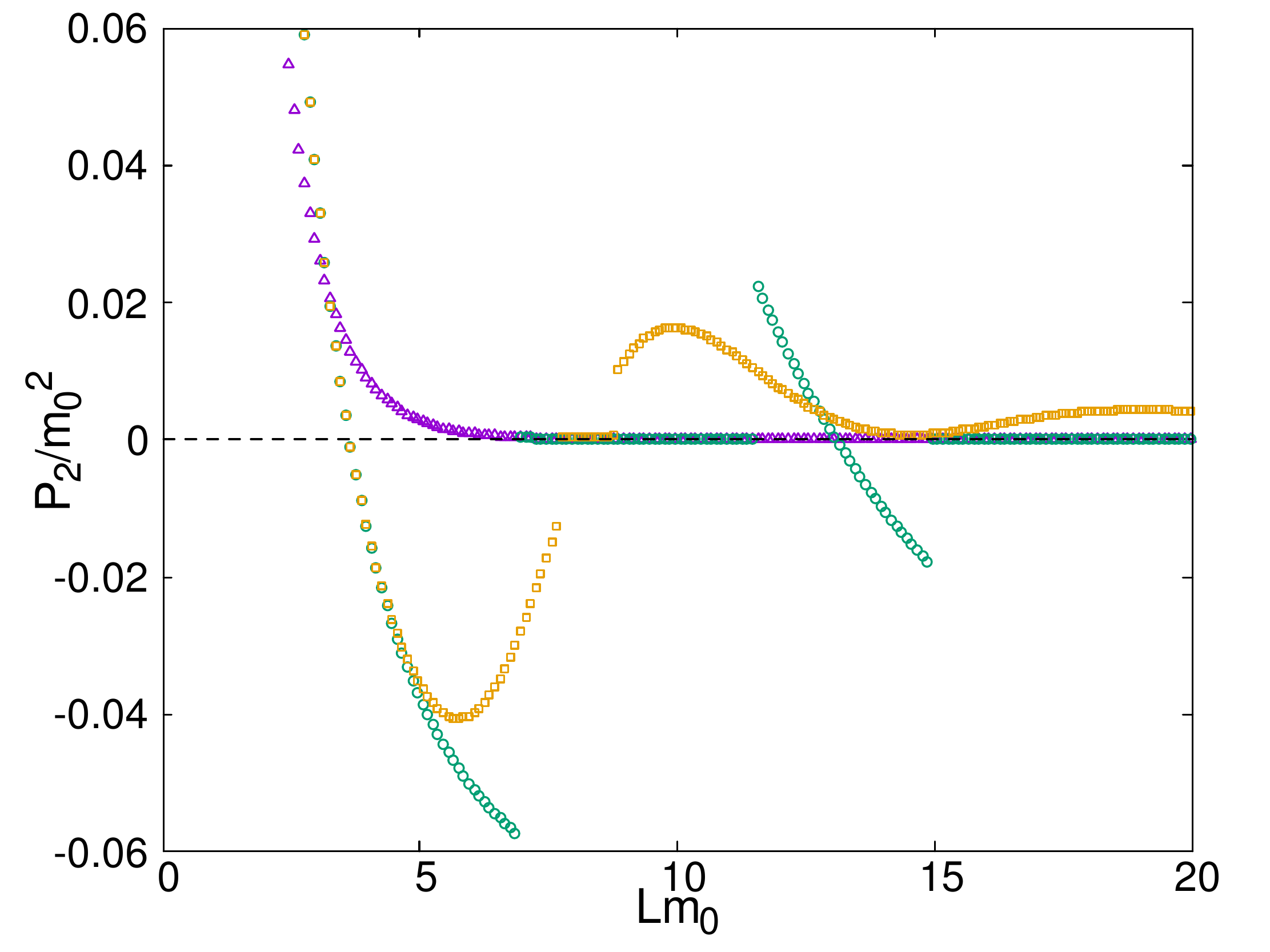}
       \end{overpic}
       \centering{\footnotesize $\delta = 0.0$}
      \end{minipage}
      \hspace{-2em}
      &
      \begin{minipage}{0.5\hsize}
       \begin{overpic}[width=1\hsize,clip]
        {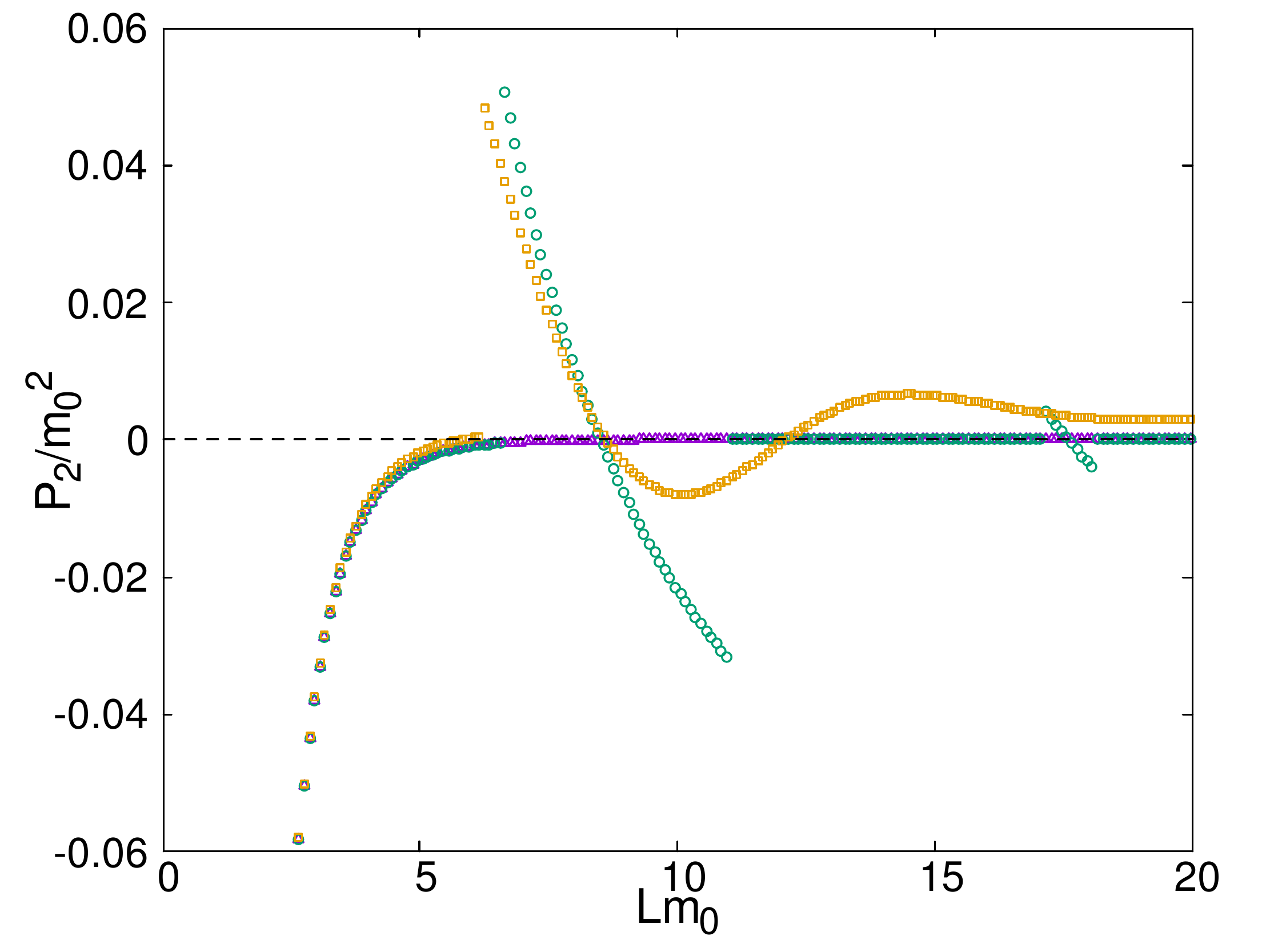}
       \end{overpic}
       \centering{\footnotesize $\delta = 1.0$}
      \end{minipage}
     \end{tabular}
    \end{center}
   \end{minipage}
  \end{tabular}
  \vspace{-2em}
 \end{center}
 \caption{Pressure as a function of $\delta$ (left) and $L$ (right) in 2 dimensions.
 $(T/m_0, \mu/m_0; \mbox{color}) = (0.005, 0.0; \mbox{purple}), \, (0.005, 0.7; \mbox{green}), \, (0.1, 0.7; \mbox{yellow})$.}
 \label{fig:pressure:2d}
\end{figure}

\begin{figure}[H]
 \begin{center}
  \begin{tabular}{cc}
   \hspace{-2em}
   \begin{minipage}{0.5\hsize}
    \begin{center}
     \begin{tabular}{cc}
      \begin{minipage}{0.5\hsize}
       \begin{overpic}[width=1\hsize,clip]
        {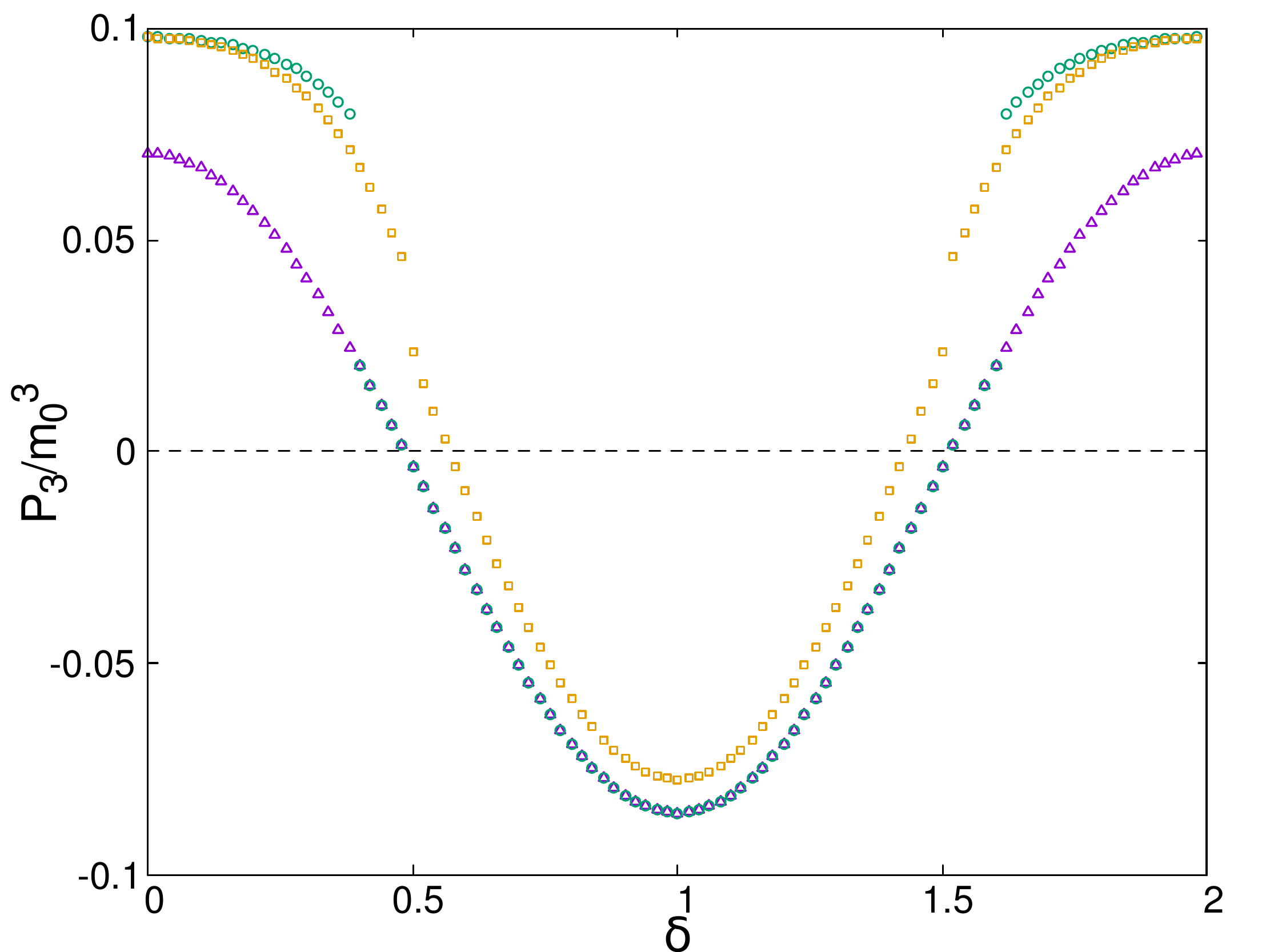}
       \end{overpic}
       \centering{\footnotesize $L m_0 = 2.0$}
      \end{minipage}
      \hspace{-2em}
      &
      \begin{minipage}{0.5\hsize}
       \begin{overpic}[width=1\hsize,clip]
        {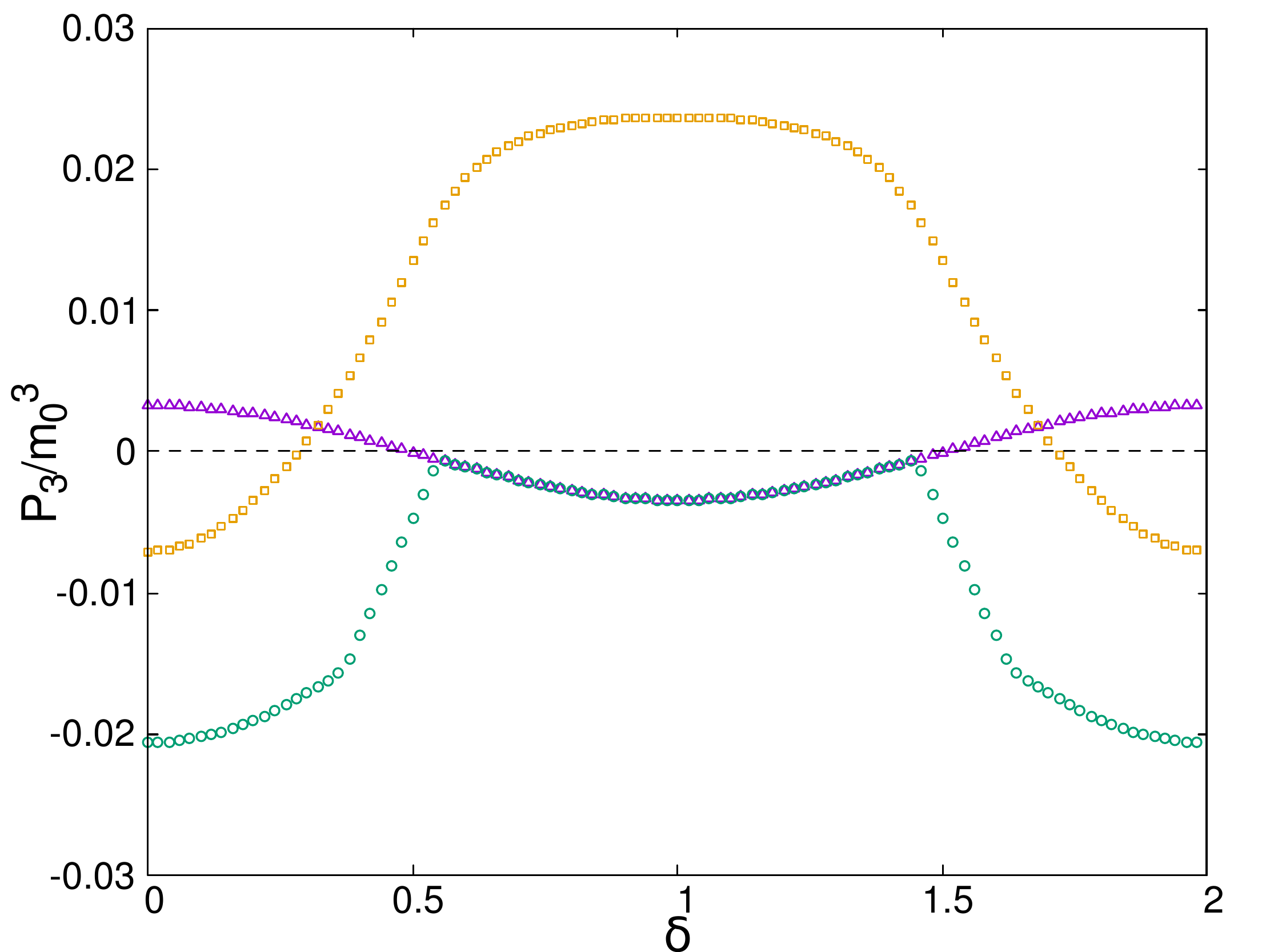}
       \end{overpic}
       \centering{\footnotesize $L m_0 = 4.0$}
      \end{minipage}
     \end{tabular}
    \end{center}
   \end{minipage}
   &
   \begin{minipage}{0.5\hsize}
    \begin{center}
     \begin{tabular}{cc}
      \begin{minipage}{0.5\hsize}
       \begin{overpic}[width=1\hsize,clip]
        {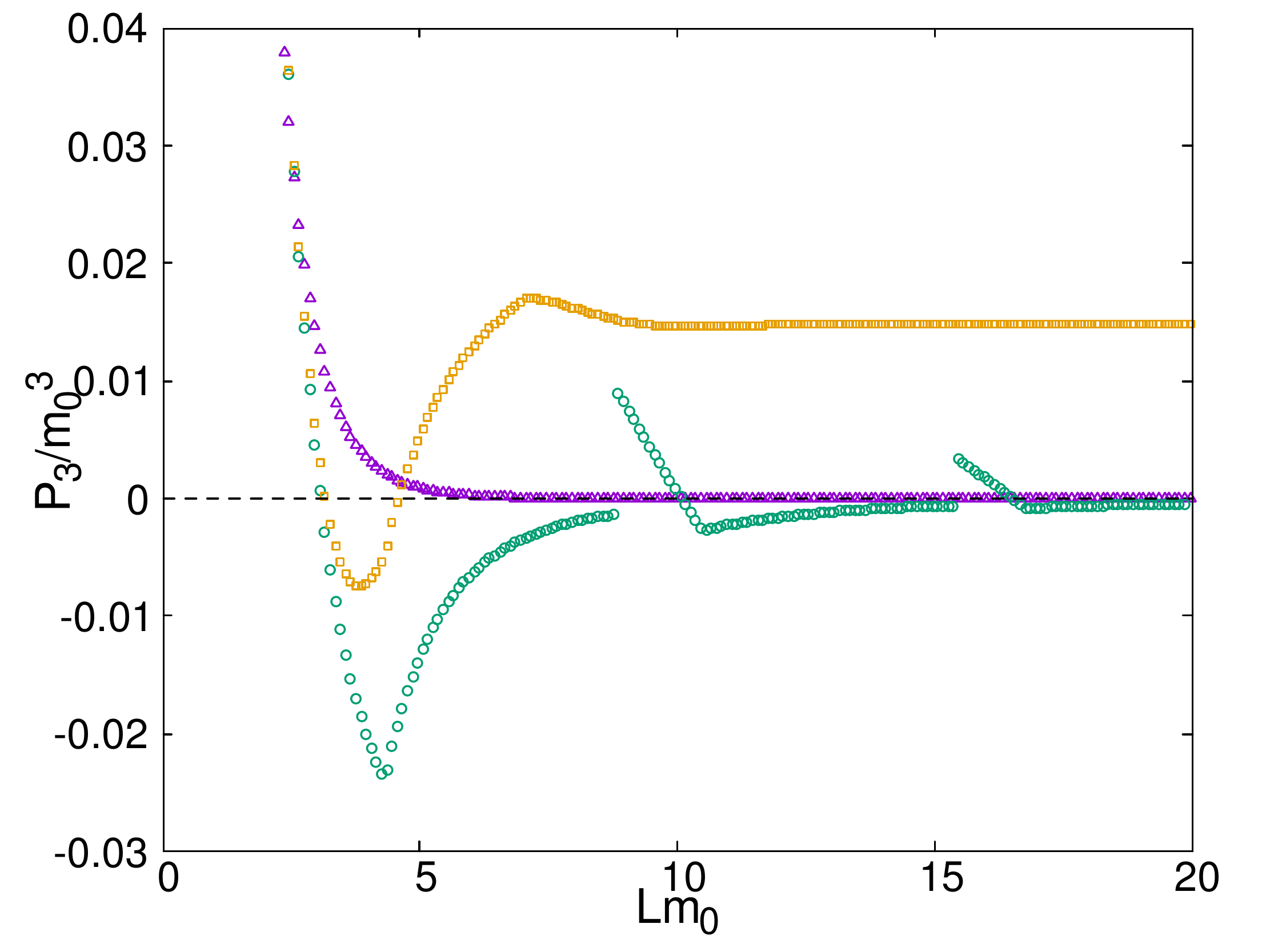}
       \end{overpic}
       \centering{\footnotesize $\delta = 0.0$}
      \end{minipage}
      \hspace{-2em}
      &
      \begin{minipage}{0.5\hsize}
       \begin{overpic}[width=1\hsize,clip]
        {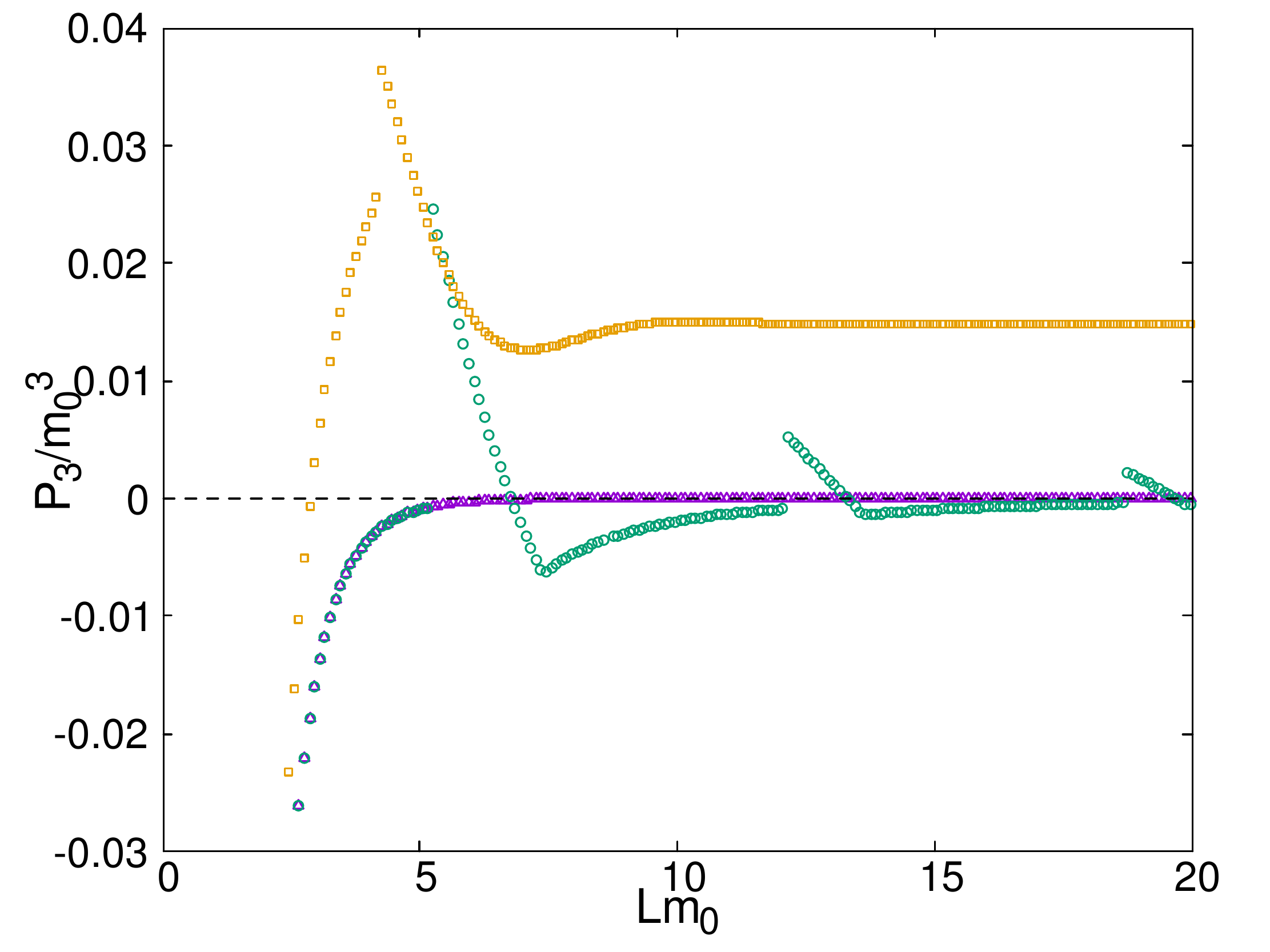}
       \end{overpic}
       \centering{\footnotesize $\delta= 1.0$}
      \end{minipage}
     \end{tabular}
    \end{center}
   \end{minipage}
  \end{tabular}
  \vspace{-2em}
 \end{center}
  \caption{Pressure as a function of $\delta$ (left) and $L$ (right) in 3 dimensions.
 $(T/m_0, \mu/m_0; \mbox{color}) = (0.005, 0.0; \mbox{purple}), \, (0.005, 1.0; \mbox{green}), \, (0.2, 1.0; \mbox{yellow})$.}
 \label{fig:pressure:3d}
\end{figure}

In Figs. \ref{fig:pressure:2d} and \ref{fig:pressure:3d} the behavior of the pressure is shown as a function of $\delta$ and $L$.
For a finite chemical potential, the discontinuous changes of the pressure are observed at the same points where the dynamical mass and the particle number density change discontinuously.
The critical $L$ and $\delta$ for a second-order phase transition are found by observing the bends of the pressure curve.

\begin{figure}[H]
 \begin{center}
  \begin{tabular}{cc}
   \begin{minipage}{0.48\hsize}
    \begin{center}
     \begin{overpic}[width=1\hsize,clip]
      {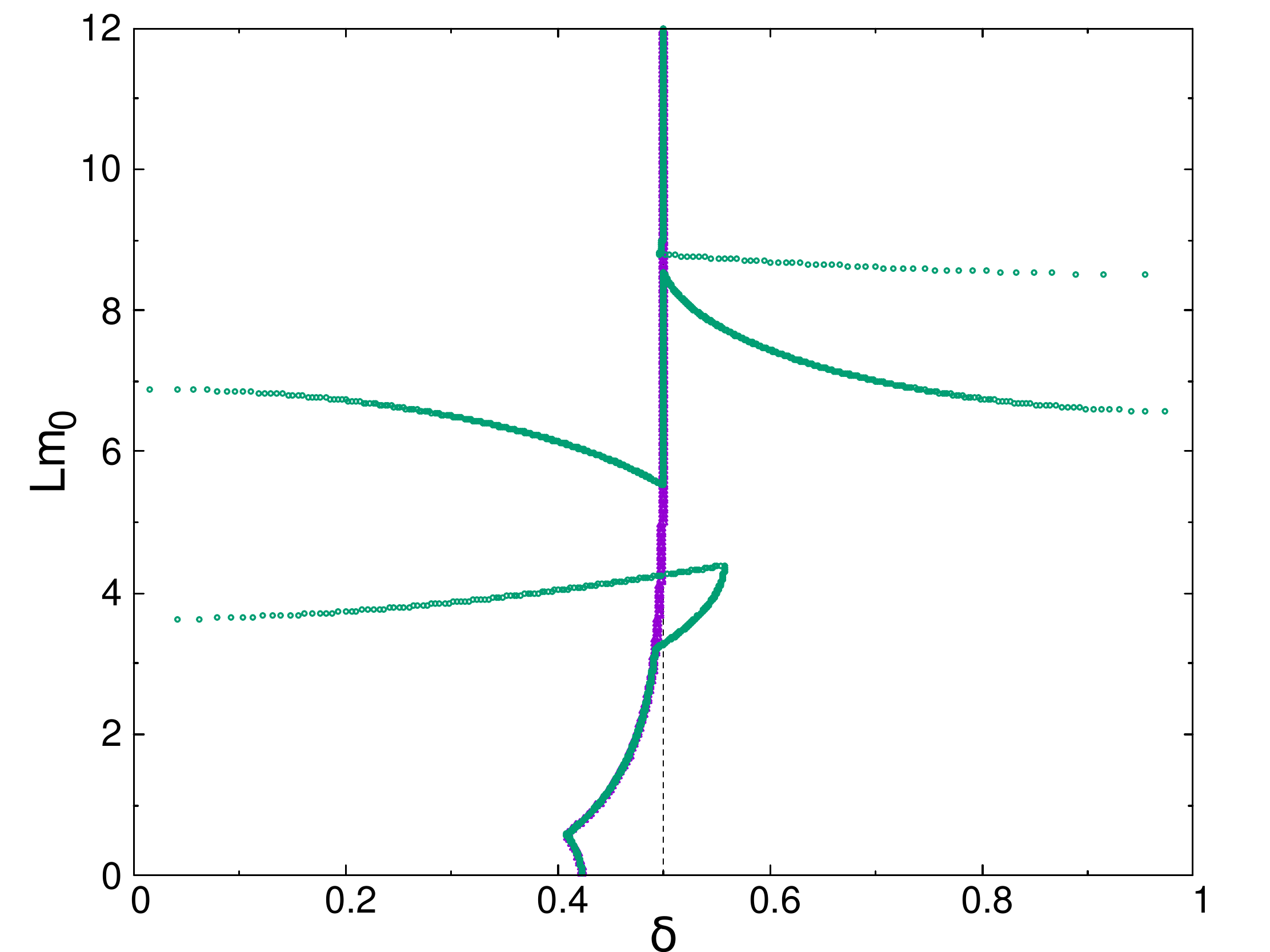}
      \put(20,16){\scriptsize \sfbtcpg{Rep.}{Rep.}}
      \put(70,16){\scriptsize \sfbtcpg{Att.}{Att.}}
      \put(20,34){\scriptsize \sfbtcpg{Rep.}{Att.}}
       \put(70,48){\scriptsize \sfbtcpg{Att.}{Rep.}}
       \put(20,62){\scriptsize \sfbtcpg{Rep.}{Rep.}}
      \put(70,62){\scriptsize \sfbtcpg{Att.}{Att.}}
     \end{overpic}
     {\footnotesize $T/m_0 = 0.005$}
    \end{center}
   \end{minipage}
   &
   \begin{minipage}{0.48\hsize}
    \begin{center}
     \begin{overpic}[width=1\hsize,clip]
      {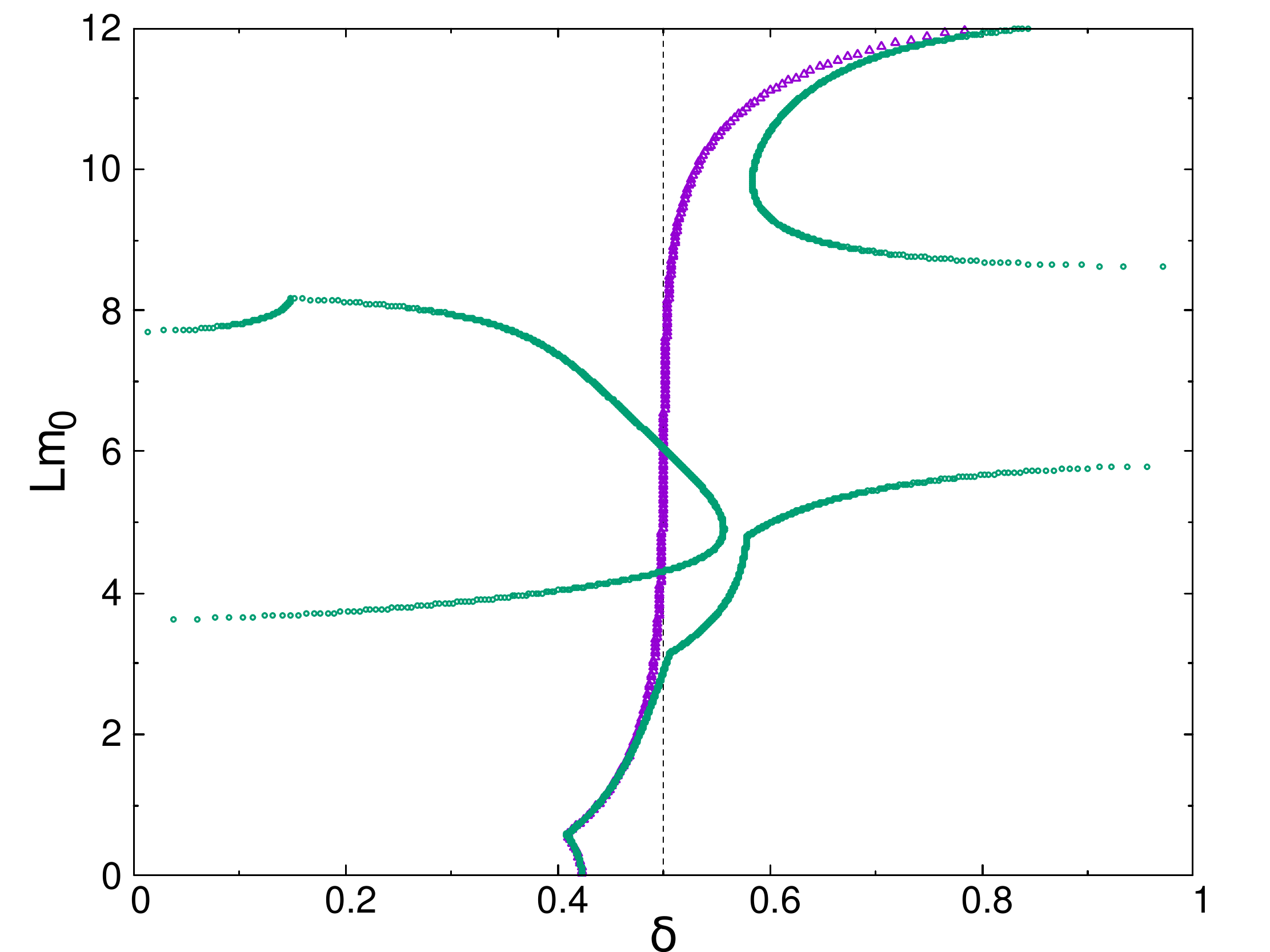}
      \put(20,16){\scriptsize \sfbtcpg{Rep.}{Rep.}}
      \put(70,16){\scriptsize \sfbtcpg{Att.}{Att.}}
       \put(20,34){\scriptsize \sfbtcpg{Rep.}{Att.}}
      \put(70,48){\scriptsize \sfbtcpg{Att.}{Rep.}}
      \put(20,62){\scriptsize \sfbtcpg{Rep.}{Rep.}}
      \put(70,62){\scriptsize \sfbtcpg{Att.}{Att.}}
     \end{overpic}
     {\footnotesize $T/m_0 = 0.1$}
    \end{center}
   \end{minipage}
  \end{tabular}
  \vspace{-1em}
  \caption{Sign-flip boundaries of the pressure on a $\delta$-$L$ plane in 2 dimensions: the purple and green curves denote $\mu/m_0=0.0$ and $\mu/m_0=0.7$ respectively.
  (Rep. or Att.)/(Rep. or Att.) indicates whether the pressure is repulsive or attractive, at $\mu/m_0=0.0$ on the left and $\mu/m_0=0.7$ on the right.}
  \label{fig:sign-flip-boundary:2d}
 \end{center}
\end{figure}

\begin{figure}[H]
 \begin{center}
  \begin{tabular}{cc}
   \begin{minipage}{0.48\hsize}
    \begin{center}
     \begin{overpic}[width=1\hsize,clip]
      {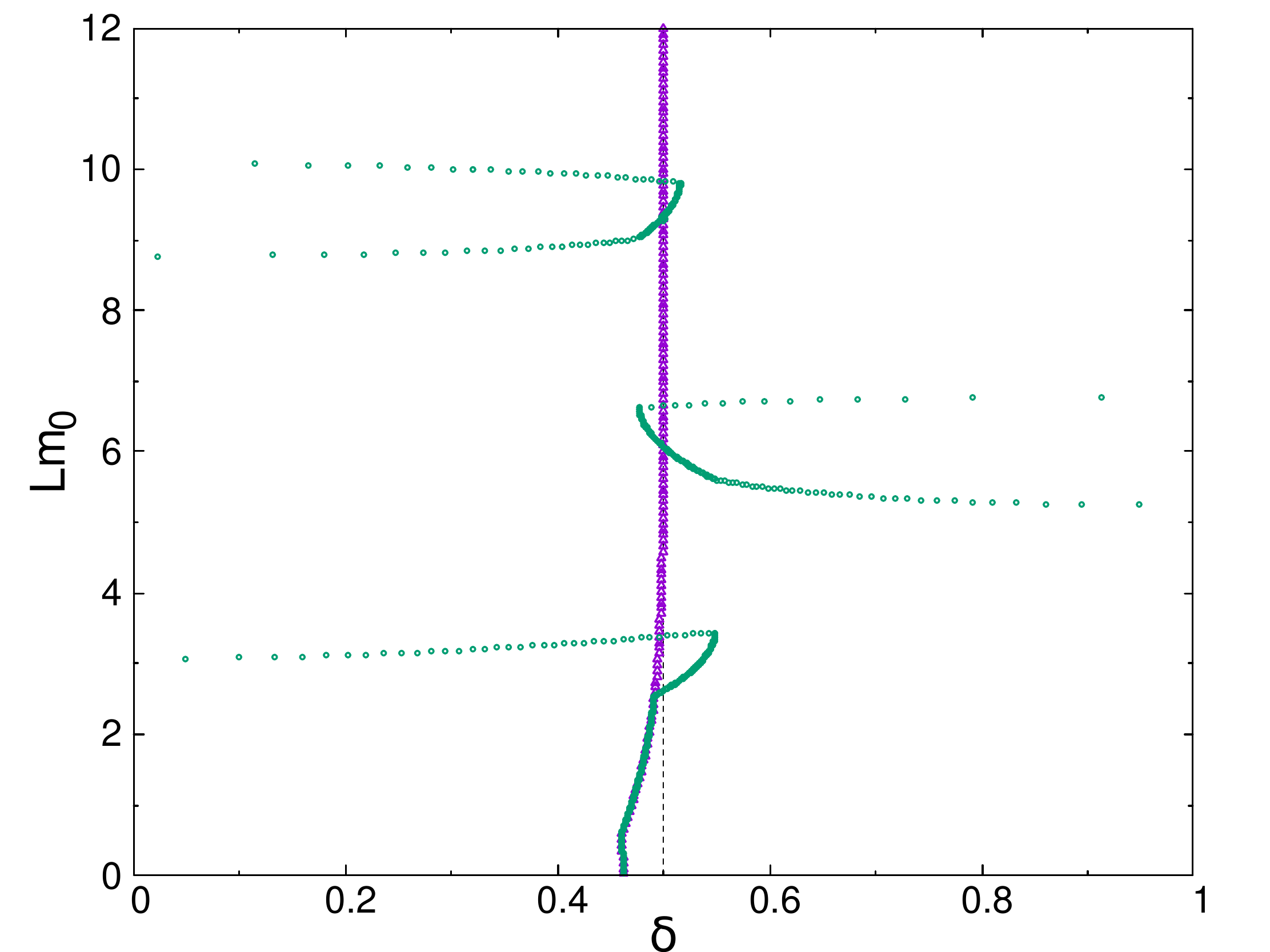}
      \put(20,16){\scriptsize \sfbtcpg{Rep.}{Rep.}}
      \put(70,16){\scriptsize \sfbtcpg{Att.}{Att.}}
      \put(20,38){\scriptsize \sfbtcpg{Rep.}{Att.}}
      \put(70,38){\scriptsize \sfbtcpg{Att.}{Rep.}}
      \put(20,58){\scriptsize \sfbtcpg{Rep.}{Rep.}}
      \put(70,58){\scriptsize \sfbtcpg{Att.}{Att.}}
     \end{overpic}
     {\footnotesize $T/m_0 = 0.005$}
    \end{center}
   \end{minipage}
   &
   \begin{minipage}{0.48\hsize}
    \begin{center}
     \begin{overpic}[width=1\hsize,clip]
      {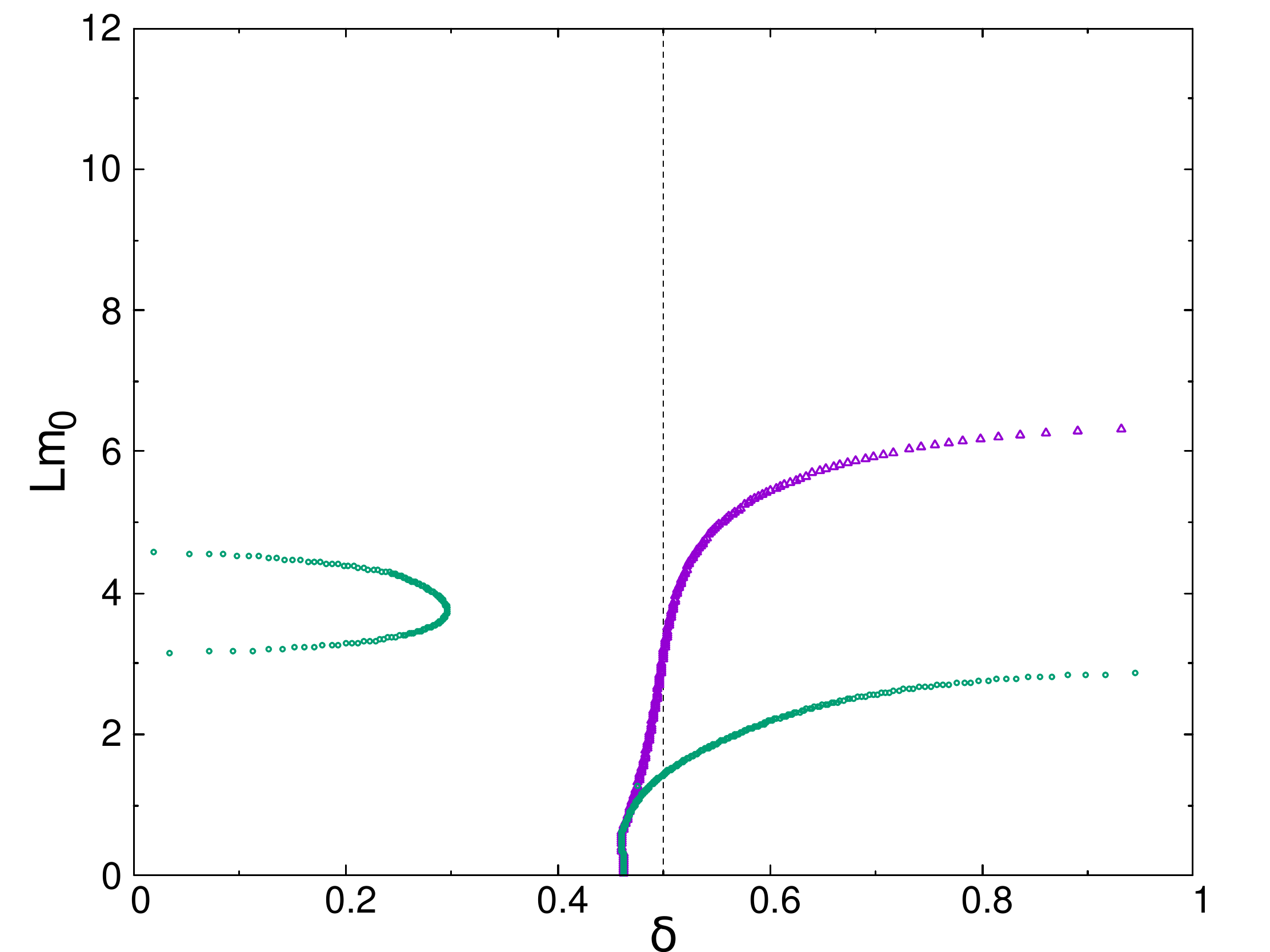}
      \put(70,16){\scriptsize \sfbtcpg{Att.}{Att.}}
      \put(14,26){\scriptsize \sfbtcpg{Rep.}{Att.}}
      \put(70,26){\scriptsize \sfbtcpg{Att.}{Rep.}}
      \put(43,58){\scriptsize \sfbtcpg{Rep.}{Rep.}}
     \end{overpic}
     {\footnotesize $T/m_0 = 0.2$}
    \end{center}
   \end{minipage}
  \end{tabular}
  \vspace{-1em}
  \caption{Sign-flip boundaries of pressure on a $\delta$-$L$ plane in 3 dimensions: the purple and green curves are $\mu/m_0=0.0$ and $\mu/m_0=1.0$ respectively.
  (Rep. or Att.)/(Rep. or Att.) indicates whether the pressure is repulsive or attractive, at $\mu/m_0=0.0$ on the left and $\mu/m_0=0.7$ on the right.}
  \label{fig:sign-flip-boundary:3d}
 \end{center}
\end{figure}
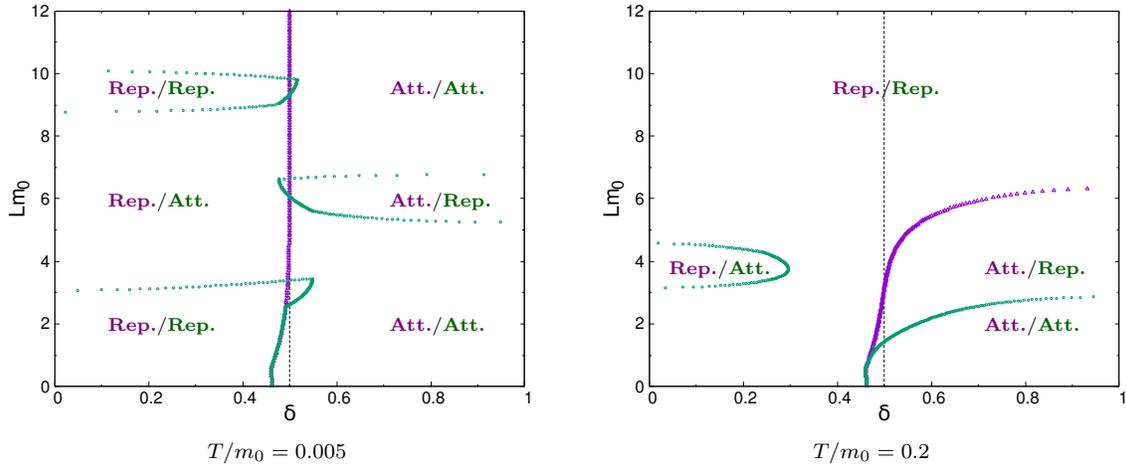

At a stable size, the sign of the pressure flips from positive to negative as the size increases.
The sign-flip points of the pressure appear at the local and global minima of the grand potential.
In Figs. \ref{fig:sign-flip-boundary:2d} and \ref{fig:sign-flip-boundary:3d} the behavior of the sigh-flip boundaries is shown on a $\delta$-$L$ plane.
At low temperature and zero chemical potential (purple curve on the left), the sign-flip boundary approaches $\delta=0.5$ with increasing size; in addition, the pressure is repulsive and attractive around the periodic and antiperiodic boundary conditions, respectively.
The finite size effect is suppressed for a larger $L$ and the thermal fluctuations induce a repulsive pressure.
Thus, the repulsive pressure is favored for a larger $L$ at $T/m_0=0.1 (D=2)$ and $T/m_0=0.2 (D=3)$.
For a finite chemical potential (green curves) the complex behavior of the sign-flip boundaries are observed on the $\delta$-$L$ plane.
At $\mu=0$ the stable size is found only for $Lm_0<1.0$ on the purple curves with a negative slope in Figs. \ref{fig:sign-flip-boundary:2d} and \ref{fig:sign-flip-boundary:3d}.
For a finite chemical potential, the stable size is found for a wide range of $\delta$.
For example, the stable size exists on the green curves near $Lm_0 \sim 4.0$  for $\delta \lesssim 0.5$ and a metastable size near $Lm_0 \sim 8.0$ for $\delta \gtrsim 0.5$ in Fig. \ref{fig:sign-flip-boundary:2d}.

The finite size effect is described by the second line in Eq. \eqref{eq:eff-pot:evaluable} at $T = \mu = 0$.
This term dominates the potential energy for a small $L$.
The term is negative, and the energy decreases via the finite size effect for $1/2 < \delta \leq 1$.
Thus, an attractive force is induced.
On the other hand, the term is positive for $0 \leq \delta < 1/3$ at the small $L$ limit and a repulsive force is induced.
Therefore, the sign of the force is flipped between $\delta=1/3$ and $1/2$.
The integrand in the second line of Eq. \eqref{eq:eff-pot:evaluable} is rewritten as, $\Re[q^{D-2}\ln(1-e^{-L\sqrt{q^2+\sigma^2}+i\pi\delta})]$.
The term is understood as an ansatz of the Fermi--Dirac $(\delta=1.0)$ and Bose--Einstein $(\delta=0.0)$ distributions in finite temperature systems.
This difference of the distribution induces an opposite contribution to the thermodynamic potential.
The phase $\delta$ is regarded as an imaginary chemical potential \cite{Roberge:1986mm, Kashiwa:2019ihm}.

\section{Conclusions}\label{sec:conclusions}
We have studied the dynamical chiral symmetry breaking in the four-fermion interaction model on $\sphere{1}$ and $\Rspace{} \times \sphere{1}$ at a finite temperature and a finite chemical potential.
We have started from the broken phase at $L \to \infty, T \to 0, \mu=0$ and introduced the finite size and thermal effects.
The $\mathrm{U}(1)$-valued boundary condition has been assigned for a compactified space.
Assuming a homogeneous chiral condensate, we have obtained the explicit expression of the effective potential in the leading order of the $1/N$ expansion.

The phase structure of the system has been evaluated by observing the extrema of the effective potential.
We have found the boundaries at which the number of the extrema and the position of the minimum change and shown the precise phase structure on $\mu$-$T$ and $L$-$T$ planes.
The finite-size effect induces complex behavior near the critical chemical potential at low temperature, and the broken and the symmetric phases alternate with increasing size of the system.
The behavior of the dynamical mass has been shown as $\delta$ and $L$ vary.

We define the grand potential whose zero-point is located at $L\to\infty$, $T\to 0$ and $\mu=0$.
The stable size, i.e. the global minimum of the grand potential, is observed for a finite chemical potential around the periodic boundary condition.
The grand potential is not bounded below around the antiperiodic boundary condition.
From the analysis of the particle number density, we have shown the trade-off relationship between the dynamical mass and the particle number density.
The mass and density gaps appear at the same $\delta$ and $L$.
The phase structure of the system is also determined by observing the pressure.
We have shown that the critical value of the second-order phase transition is fixed from the bends for the pressure.
Through the analysis of the thermodynamic quantities, it is considered that the complex behavior of the boundaries on the phase diagrams is caused by the competition among the dynamical mass, $m$, the inverse of the size, $2\pi/L$, and the chemical potential, $\mu$.

Calculating the pressure provides us alternative information about the system.
If the sign of the pressure changes from positive to negative with enlarging $L$, the sign-flip point is metastable.
We have plotted the sign-flip boundaries on the $\delta$-$L$ plane.
We have found that the contribution of a finite chemical potential generates additional metastable sizes.

In considering more realistic situations, it is interesting to introduce fermion flavors, current mass, and an electromagnetic field.
In some systems, we can not avoid considering the possibility of the inhomogeneous chiral condensate \cite{Lenz:2020bxk}.
We continue to study such situations and hope to report them in the future.

\section*{Acknowledgements}
Discussions during the KEK theory center workshop on ``Thermal Quantum Field Theories and Their Applications'' were useful to complete this work.

{\small
\bibliography{references}
\bibliographystyle{utcaps-modified}}
\end{document}